\documentclass[longauth]{aaEC}

\usepackage{graphicx}
\usepackage{natbib}
\usepackage{scalerel}
\usepackage{csquotes}
\usepackage{lastpage}

\usepackage[table]{xcolor}
\usepackage{lscape} 

\usepackage{txfonts}
\usepackage[pdfencoding=auto,psdextra]{hyperref}
\hypersetup{
    colorlinks=true,
    linkcolor=blue,
    filecolor=magenta,      
    urlcolor=blue,
    citecolor=blue
}
\makeatletter
\renewcommand*\aa@pageof{, page \thepage{} of \pageref*{LastPage}}
\makeatother

\usepackage[utf8]{inputenc}

\usepackage[switch, modulo]{lineno}
\renewcommand{\linenumbers}[0]{}

\usepackage{euclid}

\begin{document}
%
%




\title{\Euclid: Quick Data Release (Q1) -- A photometric search for ultracool dwarfs in the Euclid Deep Fields\thanks{This paper is published on
       behalf of the Euclid Consortium}}


   

\newcommand{\orcid}[1]{} 
\author{M.~{\v Z}erjal\orcid{0000-0001-6023-4974}\inst{\ref{aff1},\ref{aff2}}
\and C.~Dominguez-Tagle\inst{\ref{aff3},\ref{aff2}}
\and N.~Vitas\orcid{0000-0003-2150-0787}\inst{\ref{aff3}}
\and N.~Sedighi\inst{\ref{aff3},\ref{aff2}}
\and E.~L.~Mart\'in\orcid{0000-0002-1208-4833}\thanks{\email{ege@iac.es}}\inst{\ref{aff1},\ref{aff2}}
\and M.~R.~Zapatero~Osorio\orcid{0000-0001-5664-2852}\inst{\ref{aff4}}
\and J.~Olivares\orcid{0000-0003-0316-2956}\inst{\ref{aff5}}
\and S.~Mu\~{n}oz~Torres\orcid{0000-0003-4269-4779}\inst{\ref{aff3}}
\and S.~Tsilia\orcid{0000-0002-5378-2240}\inst{\ref{aff1},\ref{aff2}}
\and J.-Y.~Zhang\orcid{0000-0001-5392-2701}\inst{\ref{aff1},\ref{aff2}}
\and D.~Barrado\orcid{0000-0002-5971-9242}\inst{\ref{aff4}}
\and V.~J.~S.~B\'ejar\orcid{0000-0002-5086-4232}\inst{\ref{aff1},\ref{aff2}}
\and H.~Bouy\orcid{0000-0002-7084-487X}\inst{\ref{aff6},\ref{aff7}}
\and A.~Burgasser\orcid{0000-0002-6523-9536}\inst{\ref{aff8}}
\and P.~Cruz\orcid{0000-0003-1793-200X}\inst{\ref{aff4}}
\and N.~Lodieu\orcid{0000-0002-3612-8968}\inst{\ref{aff3},\ref{aff2}}
\and P.~Mas~Buitrago\orcid{0000-0001-8055-7949}\inst{\ref{aff4}}
\and N.~Phan-Bao\inst{\ref{aff9},\ref{aff10}}
\and E.~Solano\orcid{0000-0003-1885-5130}\inst{\ref{aff4}}
\and R.~Tata\inst{\ref{aff11}}
\and B.~Goldman\orcid{0000-0002-2729-7276}\inst{\ref{aff12},\ref{aff13}}
\and A.~Mohandasan\orcid{0000-0001-5182-0330}\inst{\ref{aff14}}
\and C.~Reyl\'e\orcid{0000-0003-2258-2403}\inst{\ref{aff14}}
\and R.~L.~Smart\orcid{0000-0002-4424-4766}\inst{\ref{aff15},\ref{aff16}}
\and N.~Aghanim\orcid{0000-0002-6688-8992}\inst{\ref{aff17}}
\and B.~Altieri\orcid{0000-0003-3936-0284}\inst{\ref{aff18}}
\and A.~Amara\inst{\ref{aff19}}
\and S.~Andreon\orcid{0000-0002-2041-8784}\inst{\ref{aff20}}
\and N.~Auricchio\orcid{0000-0003-4444-8651}\inst{\ref{aff21}}
\and C.~Baccigalupi\orcid{0000-0002-8211-1630}\inst{\ref{aff22},\ref{aff23},\ref{aff24},\ref{aff25}}
\and M.~Baldi\orcid{0000-0003-4145-1943}\inst{\ref{aff26},\ref{aff21},\ref{aff27}}
\and A.~Balestra\orcid{0000-0002-6967-261X}\inst{\ref{aff28}}
\and S.~Bardelli\orcid{0000-0002-8900-0298}\inst{\ref{aff21}}
\and P.~Battaglia\orcid{0000-0002-7337-5909}\inst{\ref{aff21}}
\and A.~Biviano\orcid{0000-0002-0857-0732}\inst{\ref{aff23},\ref{aff22}}
\and A.~Bonchi\orcid{0000-0002-2667-5482}\inst{\ref{aff29}}
\and E.~Branchini\orcid{0000-0002-0808-6908}\inst{\ref{aff30},\ref{aff31},\ref{aff20}}
\and M.~Brescia\orcid{0000-0001-9506-5680}\inst{\ref{aff32},\ref{aff33}}
\and J.~Brinchmann\orcid{0000-0003-4359-8797}\inst{\ref{aff34},\ref{aff35}}
\and S.~Camera\orcid{0000-0003-3399-3574}\inst{\ref{aff36},\ref{aff37},\ref{aff15}}
\and G.~Ca\~nas-Herrera\orcid{0000-0003-2796-2149}\inst{\ref{aff38},\ref{aff39},\ref{aff40}}
\and V.~Capobianco\orcid{0000-0002-3309-7692}\inst{\ref{aff15}}
\and C.~Carbone\orcid{0000-0003-0125-3563}\inst{\ref{aff41}}
\and J.~Carretero\orcid{0000-0002-3130-0204}\inst{\ref{aff42},\ref{aff43}}
\and S.~Casas\orcid{0000-0002-4751-5138}\inst{\ref{aff44}}
\and M.~Castellano\orcid{0000-0001-9875-8263}\inst{\ref{aff45}}
\and G.~Castignani\orcid{0000-0001-6831-0687}\inst{\ref{aff21}}
\and S.~Cavuoti\orcid{0000-0002-3787-4196}\inst{\ref{aff33},\ref{aff46}}
\and K.~C.~Chambers\orcid{0000-0001-6965-7789}\inst{\ref{aff47}}
\and A.~Cimatti\inst{\ref{aff48}}
\and C.~Colodro-Conde\inst{\ref{aff1}}
\and G.~Congedo\orcid{0000-0003-2508-0046}\inst{\ref{aff49}}
\and C.~J.~Conselice\orcid{0000-0003-1949-7638}\inst{\ref{aff50}}
\and L.~Conversi\orcid{0000-0002-6710-8476}\inst{\ref{aff51},\ref{aff18}}
\and Y.~Copin\orcid{0000-0002-5317-7518}\inst{\ref{aff52}}
\and F.~Courbin\orcid{0000-0003-0758-6510}\inst{\ref{aff53},\ref{aff54}}
\and H.~M.~Courtois\orcid{0000-0003-0509-1776}\inst{\ref{aff55}}
\and M.~Cropper\orcid{0000-0003-4571-9468}\inst{\ref{aff56}}
\and J.-G.~Cuby\orcid{0000-0002-8767-1442}\inst{\ref{aff57},\ref{aff58}}
\and A.~Da~Silva\orcid{0000-0002-6385-1609}\inst{\ref{aff59},\ref{aff60}}
\and H.~Degaudenzi\orcid{0000-0002-5887-6799}\inst{\ref{aff61}}
\and G.~De~Lucia\orcid{0000-0002-6220-9104}\inst{\ref{aff23}}
\and C.~Dolding\orcid{0009-0003-7199-6108}\inst{\ref{aff56}}
\and H.~Dole\orcid{0000-0002-9767-3839}\inst{\ref{aff17}}
\and M.~Douspis\orcid{0000-0003-4203-3954}\inst{\ref{aff17}}
\and F.~Dubath\orcid{0000-0002-6533-2810}\inst{\ref{aff61}}
\and X.~Dupac\inst{\ref{aff18}}
\and S.~Dusini\orcid{0000-0002-1128-0664}\inst{\ref{aff62}}
\and S.~Escoffier\orcid{0000-0002-2847-7498}\inst{\ref{aff63}}
\and M.~Farina\orcid{0000-0002-3089-7846}\inst{\ref{aff64}}
\and F.~Faustini\orcid{0000-0001-6274-5145}\inst{\ref{aff45},\ref{aff29}}
\and S.~Ferriol\inst{\ref{aff52}}
\and S.~Fotopoulou\orcid{0000-0002-9686-254X}\inst{\ref{aff65}}
\and M.~Frailis\orcid{0000-0002-7400-2135}\inst{\ref{aff23}}
\and E.~Franceschi\orcid{0000-0002-0585-6591}\inst{\ref{aff21}}
\and S.~Galeotta\orcid{0000-0002-3748-5115}\inst{\ref{aff23}}
\and K.~George\orcid{0000-0002-1734-8455}\inst{\ref{aff66}}
\and B.~Gillis\orcid{0000-0002-4478-1270}\inst{\ref{aff49}}
\and C.~Giocoli\orcid{0000-0002-9590-7961}\inst{\ref{aff21},\ref{aff27}}
\and P.~G\'omez-Alvarez\orcid{0000-0002-8594-5358}\inst{\ref{aff67},\ref{aff18}}
\and J.~Gracia-Carpio\inst{\ref{aff68}}
\and B.~R.~Granett\orcid{0000-0003-2694-9284}\inst{\ref{aff20}}
\and A.~Grazian\orcid{0000-0002-5688-0663}\inst{\ref{aff28}}
\and F.~Grupp\inst{\ref{aff68},\ref{aff66}}
\and S.~V.~H.~Haugan\orcid{0000-0001-9648-7260}\inst{\ref{aff69}}
\and J.~Hoar\inst{\ref{aff18}}
\and W.~Holmes\inst{\ref{aff70}}
\and F.~Hormuth\inst{\ref{aff71}}
\and A.~Hornstrup\orcid{0000-0002-3363-0936}\inst{\ref{aff72},\ref{aff73}}
\and K.~Jahnke\orcid{0000-0003-3804-2137}\inst{\ref{aff74}}
\and M.~Jhabvala\inst{\ref{aff75}}
\and E.~Keih\"anen\orcid{0000-0003-1804-7715}\inst{\ref{aff76}}
\and S.~Kermiche\orcid{0000-0002-0302-5735}\inst{\ref{aff63}}
\and A.~Kiessling\orcid{0000-0002-2590-1273}\inst{\ref{aff70}}
\and B.~Kubik\orcid{0009-0006-5823-4880}\inst{\ref{aff52}}
\and K.~Kuijken\orcid{0000-0002-3827-0175}\inst{\ref{aff40}}
\and M.~K\"ummel\orcid{0000-0003-2791-2117}\inst{\ref{aff66}}
\and M.~Kunz\orcid{0000-0002-3052-7394}\inst{\ref{aff77}}
\and H.~Kurki-Suonio\orcid{0000-0002-4618-3063}\inst{\ref{aff78},\ref{aff79}}
\and Q.~Le~Boulc'h\inst{\ref{aff80}}
\and A.~M.~C.~Le~Brun\orcid{0000-0002-0936-4594}\inst{\ref{aff81}}
\and S.~Ligori\orcid{0000-0003-4172-4606}\inst{\ref{aff15}}
\and P.~B.~Lilje\orcid{0000-0003-4324-7794}\inst{\ref{aff69}}
\and V.~Lindholm\orcid{0000-0003-2317-5471}\inst{\ref{aff78},\ref{aff79}}
\and I.~Lloro\orcid{0000-0001-5966-1434}\inst{\ref{aff82}}
\and G.~Mainetti\orcid{0000-0003-2384-2377}\inst{\ref{aff80}}
\and D.~Maino\inst{\ref{aff83},\ref{aff41},\ref{aff84}}
\and E.~Maiorano\orcid{0000-0003-2593-4355}\inst{\ref{aff21}}
\and O.~Mansutti\orcid{0000-0001-5758-4658}\inst{\ref{aff23}}
\and O.~Marggraf\orcid{0000-0001-7242-3852}\inst{\ref{aff85}}
\and M.~Martinelli\orcid{0000-0002-6943-7732}\inst{\ref{aff45},\ref{aff86}}
\and N.~Martinet\orcid{0000-0003-2786-7790}\inst{\ref{aff58}}
\and F.~Marulli\orcid{0000-0002-8850-0303}\inst{\ref{aff87},\ref{aff21},\ref{aff27}}
\and R.~Massey\orcid{0000-0002-6085-3780}\inst{\ref{aff88}}
\and E.~Medinaceli\orcid{0000-0002-4040-7783}\inst{\ref{aff21}}
\and S.~Mei\orcid{0000-0002-2849-559X}\inst{\ref{aff89},\ref{aff90}}
\and Y.~Mellier\inst{\ref{aff91},\ref{aff92}}
\and M.~Meneghetti\orcid{0000-0003-1225-7084}\inst{\ref{aff21},\ref{aff27}}
\and E.~Merlin\orcid{0000-0001-6870-8900}\inst{\ref{aff45}}
\and G.~Meylan\inst{\ref{aff93}}
\and A.~Mora\orcid{0000-0002-1922-8529}\inst{\ref{aff94}}
\and M.~Moresco\orcid{0000-0002-7616-7136}\inst{\ref{aff87},\ref{aff21}}
\and L.~Moscardini\orcid{0000-0002-3473-6716}\inst{\ref{aff87},\ref{aff21},\ref{aff27}}
\and R.~Nakajima\orcid{0009-0009-1213-7040}\inst{\ref{aff85}}
\and C.~Neissner\orcid{0000-0001-8524-4968}\inst{\ref{aff95},\ref{aff43}}
\and S.-M.~Niemi\orcid{0009-0005-0247-0086}\inst{\ref{aff38}}
\and C.~Padilla\orcid{0000-0001-7951-0166}\inst{\ref{aff95}}
\and S.~Paltani\orcid{0000-0002-8108-9179}\inst{\ref{aff61}}
\and F.~Pasian\orcid{0000-0002-4869-3227}\inst{\ref{aff23}}
\and K.~Pedersen\inst{\ref{aff96}}
\and W.~J.~Percival\orcid{0000-0002-0644-5727}\inst{\ref{aff97},\ref{aff98},\ref{aff99}}
\and V.~Pettorino\inst{\ref{aff38}}
\and S.~Pires\orcid{0000-0002-0249-2104}\inst{\ref{aff100}}
\and G.~Polenta\orcid{0000-0003-4067-9196}\inst{\ref{aff29}}
\and M.~Poncet\inst{\ref{aff101}}
\and L.~A.~Popa\inst{\ref{aff102}}
\and L.~Pozzetti\orcid{0000-0001-7085-0412}\inst{\ref{aff21}}
\and F.~Raison\orcid{0000-0002-7819-6918}\inst{\ref{aff68}}
\and R.~Rebolo\orcid{0000-0003-3767-7085}\inst{\ref{aff1},\ref{aff103},\ref{aff2}}
\and A.~Renzi\orcid{0000-0001-9856-1970}\inst{\ref{aff104},\ref{aff62}}
\and J.~Rhodes\orcid{0000-0002-4485-8549}\inst{\ref{aff70}}
\and G.~Riccio\inst{\ref{aff33}}
\and E.~Romelli\orcid{0000-0003-3069-9222}\inst{\ref{aff23}}
\and M.~Roncarelli\orcid{0000-0001-9587-7822}\inst{\ref{aff21}}
\and R.~Saglia\orcid{0000-0003-0378-7032}\inst{\ref{aff66},\ref{aff68}}
\and Z.~Sakr\orcid{0000-0002-4823-3757}\inst{\ref{aff105},\ref{aff106},\ref{aff107}}
\and D.~Sapone\orcid{0000-0001-7089-4503}\inst{\ref{aff108}}
\and B.~Sartoris\orcid{0000-0003-1337-5269}\inst{\ref{aff66},\ref{aff23}}
\and J.~A.~Schewtschenko\orcid{0000-0002-4913-6393}\inst{\ref{aff49}}
\and M.~Schirmer\orcid{0000-0003-2568-9994}\inst{\ref{aff74}}
\and P.~Schneider\orcid{0000-0001-8561-2679}\inst{\ref{aff85}}
\and A.~Secroun\orcid{0000-0003-0505-3710}\inst{\ref{aff63}}
\and G.~Seidel\orcid{0000-0003-2907-353X}\inst{\ref{aff74}}
\and M.~Seiffert\orcid{0000-0002-7536-9393}\inst{\ref{aff70}}
\and S.~Serrano\orcid{0000-0002-0211-2861}\inst{\ref{aff109},\ref{aff110},\ref{aff111}}
\and P.~Simon\inst{\ref{aff85}}
\and C.~Sirignano\orcid{0000-0002-0995-7146}\inst{\ref{aff104},\ref{aff62}}
\and G.~Sirri\orcid{0000-0003-2626-2853}\inst{\ref{aff27}}
\and L.~Stanco\orcid{0000-0002-9706-5104}\inst{\ref{aff62}}
\and J.~Steinwagner\orcid{0000-0001-7443-1047}\inst{\ref{aff68}}
\and P.~Tallada-Cresp\'{i}\orcid{0000-0002-1336-8328}\inst{\ref{aff42},\ref{aff43}}
\and A.~N.~Taylor\inst{\ref{aff49}}
\and I.~Tereno\orcid{0000-0002-4537-6218}\inst{\ref{aff59},\ref{aff112}}
\and S.~Toft\orcid{0000-0003-3631-7176}\inst{\ref{aff113},\ref{aff114}}
\and R.~Toledo-Moreo\orcid{0000-0002-2997-4859}\inst{\ref{aff115}}
\and F.~Torradeflot\orcid{0000-0003-1160-1517}\inst{\ref{aff43},\ref{aff42}}
\and A.~Tsyganov\inst{\ref{aff116}}
\and I.~Tutusaus\orcid{0000-0002-3199-0399}\inst{\ref{aff106}}
\and L.~Valenziano\orcid{0000-0002-1170-0104}\inst{\ref{aff21},\ref{aff117}}
\and J.~Valiviita\orcid{0000-0001-6225-3693}\inst{\ref{aff78},\ref{aff79}}
\and T.~Vassallo\orcid{0000-0001-6512-6358}\inst{\ref{aff66},\ref{aff23}}
\and G.~Verdoes~Kleijn\orcid{0000-0001-5803-2580}\inst{\ref{aff118}}
\and A.~Veropalumbo\orcid{0000-0003-2387-1194}\inst{\ref{aff20},\ref{aff31},\ref{aff30}}
\and Y.~Wang\orcid{0000-0002-4749-2984}\inst{\ref{aff119}}
\and J.~Weller\orcid{0000-0002-8282-2010}\inst{\ref{aff66},\ref{aff68}}
\and A.~Zacchei\orcid{0000-0003-0396-1192}\inst{\ref{aff23},\ref{aff22}}
\and G.~Zamorani\orcid{0000-0002-2318-301X}\inst{\ref{aff21}}
\and F.~M.~Zerbi\inst{\ref{aff20}}
\and E.~Zucca\orcid{0000-0002-5845-8132}\inst{\ref{aff21}}
\and J.~Mart\'{i}n-Fleitas\orcid{0000-0002-8594-569X}\inst{\ref{aff94}}
\and V.~Scottez\inst{\ref{aff91},\ref{aff120}}}

\institute{Instituto de Astrof\'{\i}sica de Canarias, E-38205 La Laguna, Tenerife, Spain\label{aff1}
\and
Universidad de La Laguna, Dpto. Astrof\'\i sica, E-38206 La Laguna, Tenerife, Spain\label{aff2}
\and
 Instituto de Astrof\'{\i}sica de Canarias, E-38205 La Laguna; Universidad de La Laguna, Dpto. Astrof\'\i sica, E-38206 La Laguna, Tenerife, Spain\label{aff3}
\and
Centro de Astrobiolog\'ia (CAB), CSIC-INTA, ESAC Campus, Camino Bajo del Castillo s/n, 28692 Villanueva de la Ca\~nada, Madrid, Spain\label{aff4}
\and
Departamento de Inteligencia Artificial, Universidad Nacional de Educaci\'on a Distancia (UNED), c/Juan del Rosal 16, E-28040, Madrid, Spain\label{aff5}
\and
Laboratoire d'Astrophysique de Bordeaux, CNRS and Universit\'e de Bordeaux, All\'ee Geoffroy St. Hilaire, 33165 Pessac, France\label{aff6}
\and
Institut universitaire de France (IUF), 1 rue Descartes, 75231 PARIS CEDEX 05, France\label{aff7}
\and
Department of Astronomy \& Astrophysics, University of California at San Diego, 9500 Gilman Drive, La Jolla, CA 92093, USA\label{aff8}
\and
Department of Physics, International University, Ho Chi Minh City, Vietnam\label{aff9}
\and
Vietnam National University, Ho Chi Minh City, Vietnam\label{aff10}
\and
Ohio University, Physics \& Astronomy Department,1 Ohio University, Athens, OH 45701, USA\label{aff11}
\and
International Space University, 1 rue Jean-Dominique Cassini, 67400 Illkirch-Graffenstaden, France\label{aff12}
\and
Universit\'e de Strasbourg, CNRS, Observatoire astronomique de Strasbourg, UMR 7550, 67000 Strasbourg, France\label{aff13}
\and
Universite Marie et Louis Pasteur, CNRS, Observatoire des Sciences de l'Univers THETA Franche-Comte Bourgogne, Institut UTINAM, Observatoire de Besan\c con, BP 1615, 25010 Besan\c con Cedex, France\label{aff14}
\and
INAF-Osservatorio Astrofisico di Torino, Via Osservatorio 20, 10025 Pino Torinese (TO), Italy\label{aff15}
\and
Department of Physics, Astronomy and Mathematics, University of Hertfordshire, College Lane, Hatfield AL10 9AB, UK\label{aff16}
\and
Universit\'e Paris-Saclay, CNRS, Institut d'astrophysique spatiale, 91405, Orsay, France\label{aff17}
\and
ESAC/ESA, Camino Bajo del Castillo, s/n., Urb. Villafranca del Castillo, 28692 Villanueva de la Ca\~nada, Madrid, Spain\label{aff18}
\and
School of Mathematics and Physics, University of Surrey, Guildford, Surrey, GU2 7XH, UK\label{aff19}
\and
INAF-Osservatorio Astronomico di Brera, Via Brera 28, 20122 Milano, Italy\label{aff20}
\and
INAF-Osservatorio di Astrofisica e Scienza dello Spazio di Bologna, Via Piero Gobetti 93/3, 40129 Bologna, Italy\label{aff21}
\and
IFPU, Institute for Fundamental Physics of the Universe, via Beirut 2, 34151 Trieste, Italy\label{aff22}
\and
INAF-Osservatorio Astronomico di Trieste, Via G. B. Tiepolo 11, 34143 Trieste, Italy\label{aff23}
\and
INFN, Sezione di Trieste, Via Valerio 2, 34127 Trieste TS, Italy\label{aff24}
\and
SISSA, International School for Advanced Studies, Via Bonomea 265, 34136 Trieste TS, Italy\label{aff25}
\and
Dipartimento di Fisica e Astronomia, Universit\`a di Bologna, Via Gobetti 93/2, 40129 Bologna, Italy\label{aff26}
\and
INFN-Sezione di Bologna, Viale Berti Pichat 6/2, 40127 Bologna, Italy\label{aff27}
\and
INAF-Osservatorio Astronomico di Padova, Via dell'Osservatorio 5, 35122 Padova, Italy\label{aff28}
\and
Space Science Data Center, Italian Space Agency, via del Politecnico snc, 00133 Roma, Italy\label{aff29}
\and
Dipartimento di Fisica, Universit\`a di Genova, Via Dodecaneso 33, 16146, Genova, Italy\label{aff30}
\and
INFN-Sezione di Genova, Via Dodecaneso 33, 16146, Genova, Italy\label{aff31}
\and
Department of Physics "E. Pancini", University Federico II, Via Cinthia 6, 80126, Napoli, Italy\label{aff32}
\and
INAF-Osservatorio Astronomico di Capodimonte, Via Moiariello 16, 80131 Napoli, Italy\label{aff33}
\and
Instituto de Astrof\'isica e Ci\^encias do Espa\c{c}o, Universidade do Porto, CAUP, Rua das Estrelas, PT4150-762 Porto, Portugal\label{aff34}
\and
Faculdade de Ci\^encias da Universidade do Porto, Rua do Campo de Alegre, 4150-007 Porto, Portugal\label{aff35}
\and
Dipartimento di Fisica, Universit\`a degli Studi di Torino, Via P. Giuria 1, 10125 Torino, Italy\label{aff36}
\and
INFN-Sezione di Torino, Via P. Giuria 1, 10125 Torino, Italy\label{aff37}
\and
European Space Agency/ESTEC, Keplerlaan 1, 2201 AZ Noordwijk, The Netherlands\label{aff38}
\and
Institute Lorentz, Leiden University, Niels Bohrweg 2, 2333 CA Leiden, The Netherlands\label{aff39}
\and
Leiden Observatory, Leiden University, Einsteinweg 55, 2333 CC Leiden, The Netherlands\label{aff40}
\and
INAF-IASF Milano, Via Alfonso Corti 12, 20133 Milano, Italy\label{aff41}
\and
Centro de Investigaciones Energ\'eticas, Medioambientales y Tecnol\'ogicas (CIEMAT), Avenida Complutense 40, 28040 Madrid, Spain\label{aff42}
\and
Port d'Informaci\'{o} Cient\'{i}fica, Campus UAB, C. Albareda s/n, 08193 Bellaterra (Barcelona), Spain\label{aff43}
\and
Institute for Theoretical Particle Physics and Cosmology (TTK), RWTH Aachen University, 52056 Aachen, Germany\label{aff44}
\and
INAF-Osservatorio Astronomico di Roma, Via Frascati 33, 00078 Monteporzio Catone, Italy\label{aff45}
\and
INFN section of Naples, Via Cinthia 6, 80126, Napoli, Italy\label{aff46}
\and
Institute for Astronomy, University of Hawaii, 2680 Woodlawn Drive, Honolulu, HI 96822, USA\label{aff47}
\and
Dipartimento di Fisica e Astronomia "Augusto Righi" - Alma Mater Studiorum Universit\`a di Bologna, Viale Berti Pichat 6/2, 40127 Bologna, Italy\label{aff48}
\and
Institute for Astronomy, University of Edinburgh, Royal Observatory, Blackford Hill, Edinburgh EH9 3HJ, UK\label{aff49}
\and
Jodrell Bank Centre for Astrophysics, Department of Physics and Astronomy, University of Manchester, Oxford Road, Manchester M13 9PL, UK\label{aff50}
\and
European Space Agency/ESRIN, Largo Galileo Galilei 1, 00044 Frascati, Roma, Italy\label{aff51}
\and
Universit\'e Claude Bernard Lyon 1, CNRS/IN2P3, IP2I Lyon, UMR 5822, Villeurbanne, F-69100, France\label{aff52}
\and
Institut de Ci\`{e}ncies del Cosmos (ICCUB), Universitat de Barcelona (IEEC-UB), Mart\'{i} i Franqu\`{e}s 1, 08028 Barcelona, Spain\label{aff53}
\and
Instituci\'o Catalana de Recerca i Estudis Avan\c{c}ats (ICREA), Passeig de Llu\'{\i}s Companys 23, 08010 Barcelona, Spain\label{aff54}
\and
UCB Lyon 1, CNRS/IN2P3, IUF, IP2I Lyon, 4 rue Enrico Fermi, 69622 Villeurbanne, France\label{aff55}
\and
Mullard Space Science Laboratory, University College London, Holmbury St Mary, Dorking, Surrey RH5 6NT, UK\label{aff56}
\and
Canada-France-Hawaii Telescope, 65-1238 Mamalahoa Hwy, Kamuela, HI 96743, USA\label{aff57}
\and
Aix-Marseille Universit\'e, CNRS, CNES, LAM, Marseille, France\label{aff58}
\and
Departamento de F\'isica, Faculdade de Ci\^encias, Universidade de Lisboa, Edif\'icio C8, Campo Grande, PT1749-016 Lisboa, Portugal\label{aff59}
\and
Instituto de Astrof\'isica e Ci\^encias do Espa\c{c}o, Faculdade de Ci\^encias, Universidade de Lisboa, Campo Grande, 1749-016 Lisboa, Portugal\label{aff60}
\and
Department of Astronomy, University of Geneva, ch. d'Ecogia 16, 1290 Versoix, Switzerland\label{aff61}
\and
INFN-Padova, Via Marzolo 8, 35131 Padova, Italy\label{aff62}
\and
Aix-Marseille Universit\'e, CNRS/IN2P3, CPPM, Marseille, France\label{aff63}
\and
INAF-Istituto di Astrofisica e Planetologia Spaziali, via del Fosso del Cavaliere, 100, 00100 Roma, Italy\label{aff64}
\and
School of Physics, HH Wills Physics Laboratory, University of Bristol, Tyndall Avenue, Bristol, BS8 1TL, UK\label{aff65}
\and
Universit\"ats-Sternwarte M\"unchen, Fakult\"at f\"ur Physik, Ludwig-Maximilians-Universit\"at M\"unchen, Scheinerstr.~1, 81679 M\"unchen, Germany\label{aff66}
\and
FRACTAL S.L.N.E., calle Tulip\'an 2, Portal 13 1A, 28231, Las Rozas de Madrid, Spain\label{aff67}
\and
Max Planck Institute for Extraterrestrial Physics, Giessenbachstr. 1, 85748 Garching, Germany\label{aff68}
\and
Institute of Theoretical Astrophysics, University of Oslo, P.O. Box 1029 Blindern, 0315 Oslo, Norway\label{aff69}
\and
Jet Propulsion Laboratory, California Institute of Technology, 4800 Oak Grove Drive, Pasadena, CA, 91109, USA\label{aff70}
\and
Felix Hormuth Engineering, Goethestr. 17, 69181 Leimen, Germany\label{aff71}
\and
Technical University of Denmark, Elektrovej 327, 2800 Kgs. Lyngby, Denmark\label{aff72}
\and
Cosmic Dawn Center (DAWN), Denmark\label{aff73}
\and
Max-Planck-Institut f\"ur Astronomie, K\"onigstuhl 17, 69117 Heidelberg, Germany\label{aff74}
\and
NASA Goddard Space Flight Center, Greenbelt, MD 20771, USA\label{aff75}
\and
Department of Physics and Helsinki Institute of Physics, Gustaf H\"allstr\"omin katu 2, University of Helsinki, 00014 Helsinki, Finland\label{aff76}
\and
Universit\'e de Gen\`eve, D\'epartement de Physique Th\'eorique and Centre for Astroparticle Physics, 24 quai Ernest-Ansermet, CH-1211 Gen\`eve 4, Switzerland\label{aff77}
\and
Department of Physics, P.O. Box 64, University of Helsinki, 00014 Helsinki, Finland\label{aff78}
\and
Helsinki Institute of Physics, Gustaf H{\"a}llstr{\"o}min katu 2, University of Helsinki, 00014 Helsinki, Finland\label{aff79}
\and
Centre de Calcul de l'IN2P3/CNRS, 21 avenue Pierre de Coubertin 69627 Villeurbanne Cedex, France\label{aff80}
\and
Laboratoire d'etude de l'Univers et des phenomenes eXtremes, Observatoire de Paris, Universit\'e PSL, Sorbonne Universit\'e, CNRS, 92190 Meudon, France\label{aff81}
\and
SKAO, Jodrell Bank, Lower Withington, Macclesfield SK11 9FT, UK\label{aff82}
\and
Dipartimento di Fisica "Aldo Pontremoli", Universit\`a degli Studi di Milano, Via Celoria 16, 20133 Milano, Italy\label{aff83}
\and
INFN-Sezione di Milano, Via Celoria 16, 20133 Milano, Italy\label{aff84}
\and
Universit\"at Bonn, Argelander-Institut f\"ur Astronomie, Auf dem H\"ugel 71, 53121 Bonn, Germany\label{aff85}
\and
INFN-Sezione di Roma, Piazzale Aldo Moro, 2 - c/o Dipartimento di Fisica, Edificio G. Marconi, 00185 Roma, Italy\label{aff86}
\and
Dipartimento di Fisica e Astronomia "Augusto Righi" - Alma Mater Studiorum Universit\`a di Bologna, via Piero Gobetti 93/2, 40129 Bologna, Italy\label{aff87}
\and
Department of Physics, Institute for Computational Cosmology, Durham University, South Road, Durham, DH1 3LE, UK\label{aff88}
\and
Universit\'e Paris Cit\'e, CNRS, Astroparticule et Cosmologie, 75013 Paris, France\label{aff89}
\and
CNRS-UCB International Research Laboratory, Centre Pierre Bin\'etruy, IRL2007, CPB-IN2P3, Berkeley, USA\label{aff90}
\and
Institut d'Astrophysique de Paris, 98bis Boulevard Arago, 75014, Paris, France\label{aff91}
\and
Institut d'Astrophysique de Paris, UMR 7095, CNRS, and Sorbonne Universit\'e, 98 bis boulevard Arago, 75014 Paris, France\label{aff92}
\and
Institute of Physics, Laboratory of Astrophysics, Ecole Polytechnique F\'ed\'erale de Lausanne (EPFL), Observatoire de Sauverny, 1290 Versoix, Switzerland\label{aff93}
\and
Aurora Technology for European Space Agency (ESA), Camino bajo del Castillo, s/n, Urbanizacion Villafranca del Castillo, Villanueva de la Ca\~nada, 28692 Madrid, Spain\label{aff94}
\and
Institut de F\'{i}sica d'Altes Energies (IFAE), The Barcelona Institute of Science and Technology, Campus UAB, 08193 Bellaterra (Barcelona), Spain\label{aff95}
\and
DARK, Niels Bohr Institute, University of Copenhagen, Jagtvej 155, 2200 Copenhagen, Denmark\label{aff96}
\and
Waterloo Centre for Astrophysics, University of Waterloo, Waterloo, Ontario N2L 3G1, Canada\label{aff97}
\and
Department of Physics and Astronomy, University of Waterloo, Waterloo, Ontario N2L 3G1, Canada\label{aff98}
\and
Perimeter Institute for Theoretical Physics, Waterloo, Ontario N2L 2Y5, Canada\label{aff99}
\and
Universit\'e Paris-Saclay, Universit\'e Paris Cit\'e, CEA, CNRS, AIM, 91191, Gif-sur-Yvette, France\label{aff100}
\and
Centre National d'Etudes Spatiales -- Centre spatial de Toulouse, 18 avenue Edouard Belin, 31401 Toulouse Cedex 9, France\label{aff101}
\and
Institute of Space Science, Str. Atomistilor, nr. 409 M\u{a}gurele, Ilfov, 077125, Romania\label{aff102}
\and
Consejo Superior de Investigaciones Cientificas, Calle Serrano 117, 28006 Madrid, Spain\label{aff103}
\and
Dipartimento di Fisica e Astronomia "G. Galilei", Universit\`a di Padova, Via Marzolo 8, 35131 Padova, Italy\label{aff104}
\and
Institut f\"ur Theoretische Physik, University of Heidelberg, Philosophenweg 16, 69120 Heidelberg, Germany\label{aff105}
\and
Institut de Recherche en Astrophysique et Plan\'etologie (IRAP), Universit\'e de Toulouse, CNRS, UPS, CNES, 14 Av. Edouard Belin, 31400 Toulouse, France\label{aff106}
\and
Universit\'e St Joseph; Faculty of Sciences, Beirut, Lebanon\label{aff107}
\and
Departamento de F\'isica, FCFM, Universidad de Chile, Blanco Encalada 2008, Santiago, Chile\label{aff108}
\and
Institut d'Estudis Espacials de Catalunya (IEEC),  Edifici RDIT, Campus UPC, 08860 Castelldefels, Barcelona, Spain\label{aff109}
\and
Satlantis, University Science Park, Sede Bld 48940, Leioa-Bilbao, Spain\label{aff110}
\and
Institute of Space Sciences (ICE, CSIC), Campus UAB, Carrer de Can Magrans, s/n, 08193 Barcelona, Spain\label{aff111}
\and
Instituto de Astrof\'isica e Ci\^encias do Espa\c{c}o, Faculdade de Ci\^encias, Universidade de Lisboa, Tapada da Ajuda, 1349-018 Lisboa, Portugal\label{aff112}
\and
Cosmic Dawn Center (DAWN)\label{aff113}
\and
Niels Bohr Institute, University of Copenhagen, Jagtvej 128, 2200 Copenhagen, Denmark\label{aff114}
\and
Universidad Polit\'ecnica de Cartagena, Departamento de Electr\'onica y Tecnolog\'ia de Computadoras,  Plaza del Hospital 1, 30202 Cartagena, Spain\label{aff115}
\and
Centre for Information Technology, University of Groningen, P.O. Box 11044, 9700 CA Groningen, The Netherlands\label{aff116}
\and
INFN-Bologna, Via Irnerio 46, 40126 Bologna, Italy\label{aff117}
\and
Kapteyn Astronomical Institute, University of Groningen, PO Box 800, 9700 AV Groningen, The Netherlands\label{aff118}
\and
Infrared Processing and Analysis Center, California Institute of Technology, Pasadena, CA 91125, USA\label{aff119}
\and
ICL, Junia, Universit\'e Catholique de Lille, LITL, 59000 Lille, France\label{aff120}}


%
%
%
%


%
%
\abstract{
We present a catalogue of 5306 new ultracool dwarf (UCD) candidates in the three Euclid Deep Fields in the Q1 data release. They range from late M to late T dwarfs, and include 1200 L and T dwarfs. A total of 546 objects have been spectroscopically confirmed, including 329 L dwarfs and 26 T dwarfs. We also provide empirical \Euclid colours as a function of spectral type.
Our UCD selection criteria are based only on colour ($\IE-\YE>2.5$). The combined requirement for optical detection and stringent signal-to-noise ratio threshold ensure a high purity of the sample, but at the expense of completeness, especially for T dwarfs. The detections range from magnitudes 19 and 24 in the near-infrared bands, and extend down to 26 in the optical band. 
We discuss \Euclid's capability to identify UCD candidates based on its photometric passbands. The average surface density of detected UCDs on the sky is approximately 100 objects per $\mathrm{deg}^2$, including 20 L and T dwarfs per $\mathrm{deg}^2$. This leads to an expectation of at least $1.4$\,million ultracool dwarfs in the final data release of the Euclid Wide Survey, including at least $300\,000$ L dwarfs, and more than 2600 T dwarfs, using the strict selection criteria from this work. 
}


%
%
%
%
   \titlerunning{Photometric search for ultracool dwarfs in \Euclid Q1}
   \authorrunning{M.~{\v Z}erjal et al.}
   
   \maketitle
%
%
%
%

\section{\label{sc:Intro}Introduction}
Ultracool dwarfs (UCDs), comprising the lowest-mass stars, brown dwarfs, and free-floating planetary-mass objects, represent a continuum of properties linking stellar and planetary physics. 
Their formation mechanisms remain vigorously debated: do they form like stars, via the gravitational collapse and fragmentation of molecular clouds, or like planets, through disk instabilities or core accretion followed by ejection (\citealp{2018arXiv181106833W}; \citealp{2022NatAs...6...89M}; \citealp{2024NewAR..9901711P})? Despite their intrinsic faintness, they are abundant, making up a significant fraction of the Galaxy’s population. Yet, their census remains incomplete, particularly at the faint and metal-poor ends. Building large, homogeneous catalogues and developing robust photometric and astrometric diagnostics to distinguish UCDs from background contaminants are therefore crucial steps toward constraining their formation pathways. In turn, this will shed light on the broader processes of star and planet formation, helping to advance our understanding of the lowest-mass end of the initial mass function.


The advent of major wide area imaging and spectroscopic surveys from the ground and from space, at optical and infrared wavelengths, have enabled the discovery of hundreds of ultracool dwarfs (UCDs; very low-mass stars, brown dwarfs, and planetary-mass objects). Traditionally, these faint ultracool objects are first identified by photometric selection criteria using large survey catalogues (e.g., Sloan Digital Sky Survey (SDSS; \citealp{2000AJ....120.1579Y}), Two-Micron All-Sky Survey (2MASS; \citealp{cutri2003twomasss_point_catalog,2006AJ....131.1163S}), Deep Near Infrared Survey of the Southern Sky (DENIS; \citealp{1997Msngr..87...27E}), UKIRT Infrared Deep Sky Survey (UKIDSS; \citealp{2007MNRAS.379.1599L}), Wide-field Infrared Survey Explorer (WISE; \citealp{2010AJ....140.1868W}), VISTA Hemisphere Survey (VHS; \citealp{2013Msngr.154...35M}), Panoramic Survey Telescope and Rapid Response System \citep[Pan-STARRS;][]{2016arXiv161205560C}, and Dark Energy Survey (DES; \citealp{2021ApJS..255...20A})), and later confirmed spectroscopically \citep[e.g.,][]{1999ApJ...519..802K,2000AJ....120..447K,2014ApJ...786L..18L, 
1999AJ....118.2466M,2010A&A...517A..53M,2018A&A...619L...8R}.
For example, \cite{2023MNRAS.522.1951D} extended the catalogue of \cite{2019MNRAS.489.5301C} and identified almost 20\,000 UCD candidates brighter than $z \le 23$ in DES. On the other hand, the Ultracool Sheet\footnote{\url{https://zenodo.org/records/13993077}} \citep{best_2024_13993077} -- the growing literature compilation of confirmed UCDs and imaged planets -- contains more than 4000 objects, as of December 2024.

The \Euclid space mission \citep{EuclidSkyOverview} is a wide-field space telescope equipped with high-precision optical and near-infrared imaging and slitless spectroscopy. It is designed to observe distant galaxies to explore the composition and evolution of the dark Universe. At the same time, its high sensitivity and wide coverage enable the identification of thousands of new UCDs. This is made possible primarily by its sensitive Near-Infrared Photometer and Spectrometer \citep[NISP,][]{EuclidSkyNISP}, which observes wavelengths that are otherwise partially obscured by Earth's atmosphere, such as water absorption bands. This spectral range also reveals key molecules in ultracool dwarf atmospheres, such as methane. 
\Euclid's high sensitivity makes it possible to detect fainter objects than in previous wide-field surveys. The depth provided by the Q1 data allows us to probe about 3 magnitudes deeper at optical and near-infrared wavelengths than previous ground-based surveys. The limiting sensitivity in the optical is 26.7 \citep{EuclidSkyOverview}, 
and 24.4 ($5\,\sigma$ point source) in the near-infrared, and this enables detection of faint, free-floating planetary-mass objects. For example, the first scientific results of \Euclid were derived from the Early Release Observations (ERO; \citealp{2024arXiv240513496C}), which included the photometric identification of new ultracool dwarfs in the Sigma Orionis open cluster \citep{2025A&A...697A...7M} and the LDN\,1495 region of the Taurus molecular clouds \citep{2025arXiv250216349B}. Owing to the young age of these regions, such objects have likely planetary masses, down to a few Jupiter masses. 
Upon completion of the wide (Euclid Wide Survey, EWS) and deep (Euclid Deep Survey, EDS) surveys that will take over 5 years, it has been predicted that \Euclid will have detected the largest ever number of UCDs, most of which will be present in the near-infrared images only \citep{2021MNRAS.501..281S}. The resulting large catalogue will help us better understand the populations of UCDs, their formation scenarios, and help us refine the models of their interiors and complex atmospheres.



This work presents the first study to assess the potential of \Euclid's passbands to identify ultracool dwarfs (UCDs) in its observations. By exploring the UCD parameter space using photometric data from the Euclid's first Quick Data Release, Q1 \citep{Q1cite}, we evaluated its capability to detect UCDs on a large scale, comparing it with the spectroscopic search and analysis from \citet{carlos25} and \cite{Q1-SP042}. Our goal was to understand both the strengths and limitations of the photometric data to identify UCDs.
Since the depth of the Q1 matches that of the planned EWS, this study provides a direct estimate of the expected UCD yield in the final EWS at the end of the mission -- representing a crucial step toward understanding the final survey's potential.

To ensure high reliability, we searched for UCDs detected in both near-infrared and optical bands. The resulting catalogue has a relatively low contamination rate at the cost of completeness. Fainter detections in the near-infrared EDF observations will be examined in a follow-up study and in future repeated observations, which will probe even deeper.

This paper is structured as follows.
In Sect.~\ref{sc:Data}, we present the \Euclid's Q1 data along with known UCDs from the literature, which are included in this data set and used as benchmarks. We describe the extraction of point sources from the \Euclid catalogue in Sect.~\ref{sc:Filter}. Section~\ref{sc:Selection} details the selection process for candidate UCDs in our data set. In Sect.~\ref{sc:Discussion}, we examine the properties of our candidate UCD catalogue, their spectral types and empirical colours, the catalogue's limitations, and the expected number of UCD detections in the future \Euclid observations. Finally, we draw our conclusions in Sect.~\ref{sc:Conclusions}.  We provide lists of spectroscopically confirmed benchmark sources and T dwarfs, as well as our main photometric UCD candidates, in Appendix~\ref{sec.tableappendix}.

\section{\label{sc:Data}Data}
In this paper we utilise the three catalogues of the EDFs from the Q1 data release. They cover a total area of $63\,{\rm deg}^2$, distributed approximately as follows \citep[see][]{Q1-TP001}: EDF Fornax (EDF-F); $12\,{\rm deg}^2$; EDF North (EDF-N), $23\,{\rm deg}^2$; and EDF South (EDF-S), $28\,{\rm deg}^2$. This data release includes only the first visit of these fields, but they will be continuously observed many times during the lifetime of the mission, which will deepen the detection limit by another 2 magnitudes.

We exploited the merged catalogue (MER) that consists of photometric and morphological information of 39 million objects detected in all three fields. They were observed with two instruments, VIS \citep[magnitude \IE,][]{EuclidSkyVIS}, with one wide visible filter, and NISP, 
with three photometric filters that give magnitudes \YE, \JE, and \HE. All four filters are broader than the standard ground-based photometric passbands; their widths range from 0.3 to 0.56\,\micron. Together, they cover wavelengths from about 0.5 to 2\,\micron, with almost no gaps. The passbands\footnote{Available at e.g., the \href{http://svo2.cab.inta-csic.es/theory/fps/index.php?mode=browse&gname=Euclid&asttype=\,.}{Spanish Virtual Observatory's website}.} have the following central wavelengths: 0.67\,\micron\ in \IE; 1.08\,\micron\ in \YE;  1.37\,\micron\ in \JE; and 1.77\,\micron\ in \HE \citep{Schirmer-EP18}.

The MER catalogue provides magnitudes computed in a few different ways. In this work, we adopted aperture photometry computed within twice the full width at half maximum (FWHM), with magnitudes expressed in the AB photometric system. Our choice of flux type from the MER catalogue follows the work of \cite{Q1-SP042}, where the authors discuss the optimal flux measurement for point-like sources, and compare the magnitudes with ground-based photometry.




\subsection{Reference UCDs in \Euclid fields} \label{sec.benchmarks}
Anticipating the first \Euclid observations, \citet{2024A&A...686A.171Z} analysed the existing optical and infrared photometric catalogues to search for the UCDs in the EDFs. They used Pan-STARRS release 1, 2MASS, and the AllWISE survey \citep{2014yCat.2328....0C} to search for late M, L, and T dwarfs. Their photometric selection criteria were based on the work of \citet{2019MNRAS.489.5301C}, but generalised and less strict. They found 360 M, 152 L, and three T candidate dwarfs in the three EDFs. From this sample they selected eight UCD candidates of different spectral types and obtained spectra with EMIR the Espectr{\'o}grafo Multiobjeto Infra-Rojo \citep[EMIR;][]{2022A&A...667A.107G} at the Gran Telescopio Canarias (GTC), and the Very Large Telescope (VLT)/Xshooter, which confirm their UCD nature.

\begin{figure*}
\includegraphics[width=\linewidth]{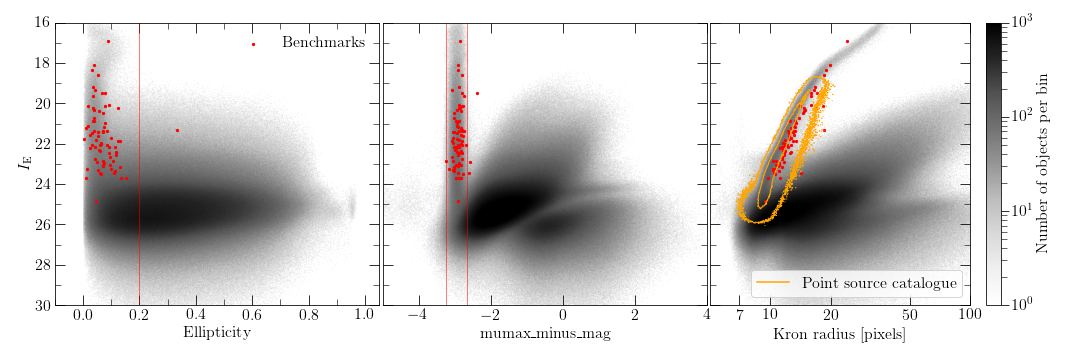} 
\caption{Morphological filters differentiate between extended sources and point sources. The filtering criteria (red lines for \texttt{ELLIPTICITY} and \texttt{MUMAX\_MINUS\_MAG}) are based on the benchmark ultracool dwarfs \citep[red dots;][]{2024A&A...686A.171Z}. Grey dots are sources from the entire \Euclid catalogue. The right panel shows that while \texttt{KRON\_RADIUS} is not used in the procedure, it correlates well with luminosity for the point sources.
} 
\label{fig.filtering}
\end{figure*}


\section{\label{sc:Filter}Selection of point sources}
The most common objects in \Euclid's catalogue are extragalactic. They are clearly extended when they are close enough, although they appear point like when very distant. Fortunately, in our parameter space of choice, there is negligible overlap between the point-like galaxies and stars, and especially substellar objects (see Sect.~\ref{sec.limitations}). 
To isolate the point-like sources, we performed a series of simple morphology and data quality filters using measurements already available\footnote{The documentation is available online in the \href{http://st-dm.pages.euclid-sgs.uk/data-product-doc/dm10/merdpd/dpcards/mer_finalcatalog.html}{Euclid SGS Data Product Description Document}.} in the Q1 tables. Our criteria were stringent, leaving out some marginal objects, such as young UCDs, which might appear extended due to a potential circumsubstellar disk and infalling material. Our goal was to produce a catalogue that is as free of contaminants as possible, rather than aiming for completeness. Our selection was defined with the help of the benchmark ultracool objects from Table~\ref{table:benchmarks}.


The most restrictive requirements were applied to the morphology of the sources, since most of the objects in \Euclid's catalogue are galaxies. Due to \texttt{POINT\_LIKE\_PROB} (probability between 0 and 1 that the source is point-like) being available only for a subset of sources, due to the strong priority placed on purity in the MER catalogue \citep{Q1-TP004}, we designed our own filters that are a little less restrictive and reach fainter magnitudes. To isolate point-like objects, we used the \texttt{ELLIPTICITY} and \texttt{MUMAX\_MINUS\_MAG} parameters. \texttt{ELLIPTICITY} is defined as $1-B/A$ where $A$ and $B$ are the semi-major and semi-minor axes of an elliptical object, respectively, and are computed directly by \texttt{SourceExtractor} \citep{1996A&AS..117..393B}. \texttt{MUMAX\_MINUS\_MAG} is a proxy for the better known \texttt{SPREAD\_MODEL} parameter \citep{10.1117/12.926785,2012ApJ...757...83D,2013A&A...554A.101B}, and is directly used for star/galaxy separation. It is defined as the difference between the peak surface brightness above the background (\texttt{MUMAX} in $\mathrm{mag\,arcsec^{-2}}$) and the magnitude \citep{Q1-TP004}. Essentially, it compares the concentration of the light at the peak and the total magnitude.
This difference distinguishes extended and point sources very efficiently, as demonstrated in Fig.~\ref{fig.filtering}. We limited this parameter to values between $-3.25$ and $-2.65\,\mathrm{mag\,arcsec^{-2}}$, and $\texttt{ELLIPTICITY}$ to $<0.2$. 

The distribution of FWHM ranges from about \ang{;;1.1} to \ang{;;1.5} for the majority of sources. Because one of the benchmarks is an outlier in this parameter (its \texttt{det\_quality\_flag} is 386, which means bad data quality) and lies beyond that limit, we formally introduced a $\textrm{FWHM} < \ang{;;1.5}$ filter to exclude it. This filter removed less than 1\% of sources from the entire sample.



\begin{figure*}
\includegraphics[width=\linewidth]{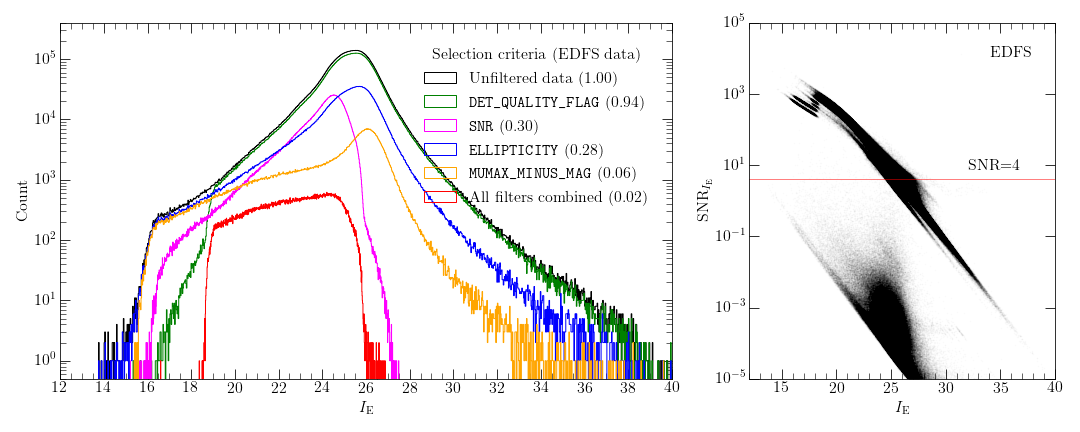} 
\caption{\textit{Left}: Magnitude distributions for each point-source selection filter (we do not plot the FWHM filter). The most limiting is the \texttt{S/N} requirement that raises the completeness and detection limits to ensure a high quality of the point-source catalogue.
\textit{Right}: Signal-to-noise ratio is correlated with luminosity. 
} 
\label{fig.magnitude_distributions_each_filtering_step}
\end{figure*}
\begin{figure}
\includegraphics[width=\linewidth]{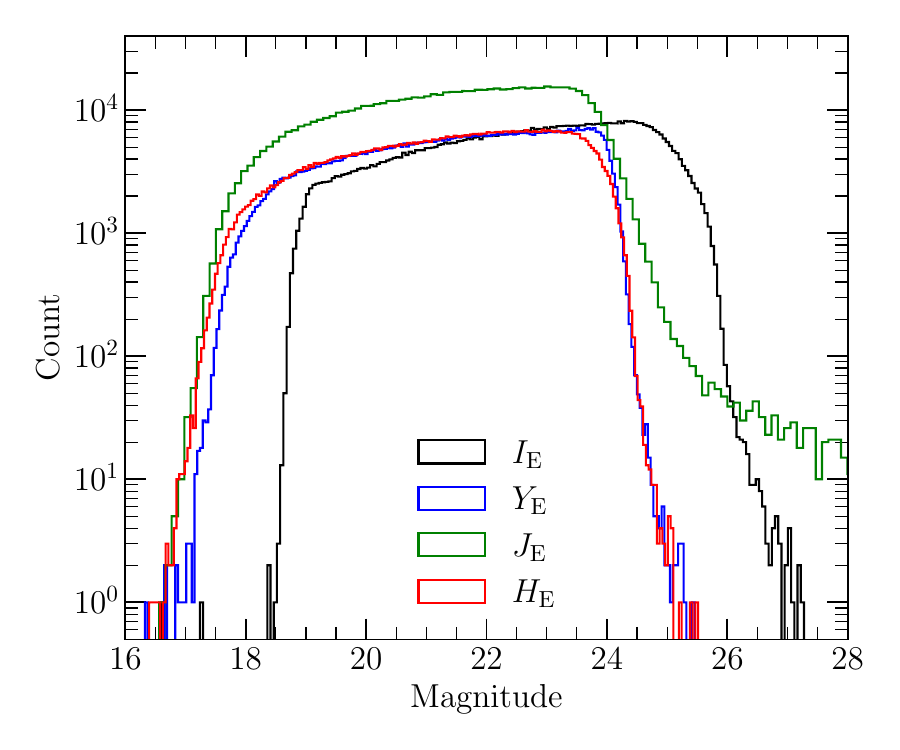} 
\caption{Magnitude distributions in the point-source catalogue. Completeness limits are approximately 23.5 for the NISP bands and 24.5 for VIS. The filtering procedure was not applied to the \JE band, since it was not used in the candidate selection from the colour-colour diagram.} 
\label{fig.magnitude_distribution}
\end{figure}

Saturated, blended, and contaminated sources, and those with similar problems, were excluded with the \texttt{DET\_QUALITY\_FLAG} smaller than 3. 
To limit ourselves only to the measurements of high quality, we eliminated all catalogue entries with signal-to-noise ratio (S/N; defined as $\mathrm{flux/flux\,error}$) less than 4. This is the most restrictive selection criterion, which consequently shifts the completeness by 1--1.5\,mag, and the limiting magnitude by several magnitudes, as demonstrated in Fig.~\ref{fig.magnitude_distributions_each_filtering_step}. 
The completeness limit in our resulting point-source catalogue, after all applied cuts, is 23.5 for the near-infrared bands and 24.5 in the visible, as shown in Fig.~\ref{fig.magnitude_distribution}. We impose such a strict constraint in order to produce as clean a sample as possible, at the expense of the fainter sources, which will be explored in a follow-up work.
We applied this filtering only to the \IE, \YE, and \HE bands that are essential for our UCD science. The \JE data remained unfiltered, because our UCD selection criteria, described below, are not based on the \JE magnitude. This means that some of our sources might have bad pixels in the \JE band, but since this magnitude is not used, it does not affect our results.



We present the set of filters that we use in Table~\ref{tab.filters}.
This simple but effective filtering resulted in a catalogue of 688\,957 point sources, representing just under 2\% of the initial sample. 

We estimated magnitude and colour error bars for the point source catalogue by sampling 1000 flux values from a normal distribution centered on the measured flux, using the flux error as the standard deviation. These sampled fluxes were then converted to magnitudes, and the 16th and 84th percentiles of the resulting distribution were taken as the lower and upper bounds of the uncertainty.


\begin{table}
\caption{Filters applied to isolate point sources.}
\label{tab.filters}
\centering
\begin{tabular}{lc}
\hline\hline \noalign{\vskip 2pt}
Parameter & Value \\
\hline\noalign{\vskip 1pt}
\texttt{ELLIPTICITY} & $<$\,0.2 \\
\texttt{MUMAX\_MINUS\_MAG} & $-3.25$ to $-2.65\,\mathrm{mag\,arcsec^{-2}}$ \\
\texttt{FWHM} & $<\ang{;;1.5}$ \\
\texttt{S/N} (\IE, \YE, and \HE) & $>$\,4 \\
\texttt{DET\_QUALITY\_FLAG} & $<$\,3 \\
\hline
\end{tabular}
\end{table}

\subsection{Extinction estimation}
\begin{figure}
\includegraphics[width=\linewidth]{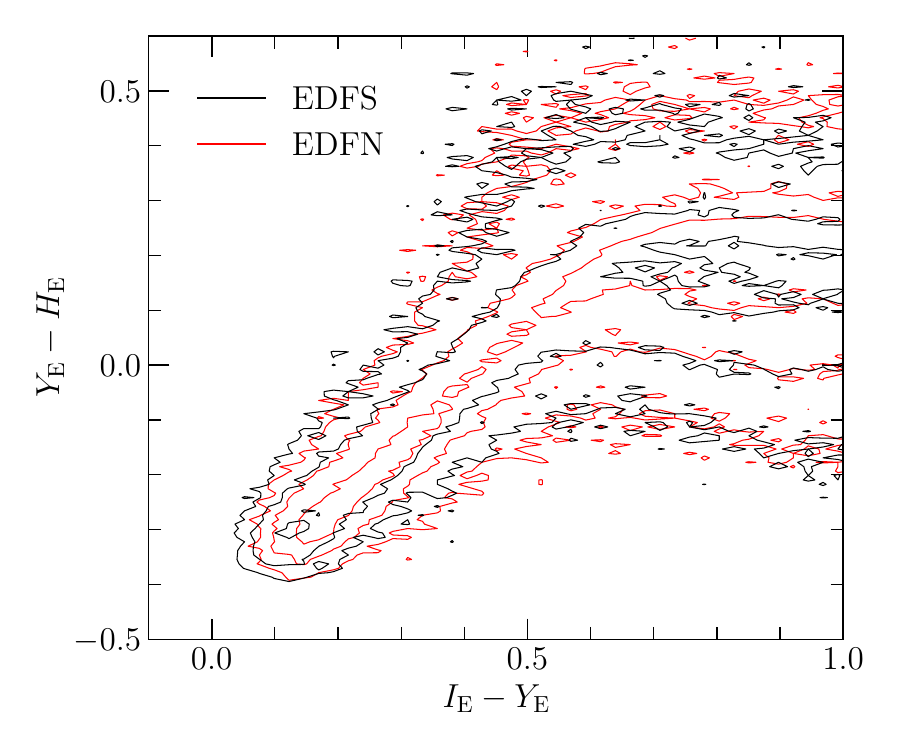} 
\caption{\Euclid colour-colour diagram.  The offset between EDF-N and EDF-S is most prominent in the solar-like region at $0.1<\IE-\YE<0.4$.} 
\label{fig.bluebottle_offset_north_south}
\end{figure}

We found an apparent relative offset between the EDF-N and the two southern fields in our $\IE-\YE$ versus $\YE-\HE$ diagrams.  This is shown, for example in Fig.~\ref{fig.bluebottle_offset_north_south}, where we plot data from EDF-N and EDF-S. The difference is particularly noticeable in the densest regions, representing solar-like stars. The relative offsets of the EDF-N with respect to EDF-S 
are 0.04 in $\IE-\YE$ and 0.02 in $\YE-\HE$. 
This might be coming from differences in interstellar extinction. While the EDFs were carefully selected to be extinction free, the average estimated\footnote{Reddening in \Euclid was estimated from the \textit{Planck} dust map \citep{2014A&A...571A..11P}, 
which is two-dimensional, providing only upper limits for objects within our Galaxy \citep[see][for more details]{Q1-TP004}.} $E(B-V)$ in the northern field (0.056) is more than 3 times higher than in the EDF-S (0.017). 
This corresponds to median extinction values of $A_{\IE}=0.13$, $A_{\YE}=0.06$, and $A_{\HE}=0.03$ in EDF-N, and $A_{\IE}=0.08$, $A_{\YE}=0.02$, and $A_{\HE}=0.009$ in EDF-S. They are typically 2--3-times larger in EDF-N, depending on the wavelength. This difference could produce the relative offsets that we detected in the data. 

However, the reddening values are upper limits, and the actual values for the UCDs, which are very close to us, are smaller. Additionally, the offsets between EDF-S and EDF-N are much smaller than the typical error bars for individual UCDs ($\pm 0.12$ 
and $\pm 0.06$, respectively). For this reason, we did not apply any corrections.

\section{\label{sc:Selection} A catalogue of photometric UCD candidates }
\begin{figure*}
\includegraphics[width=\linewidth]{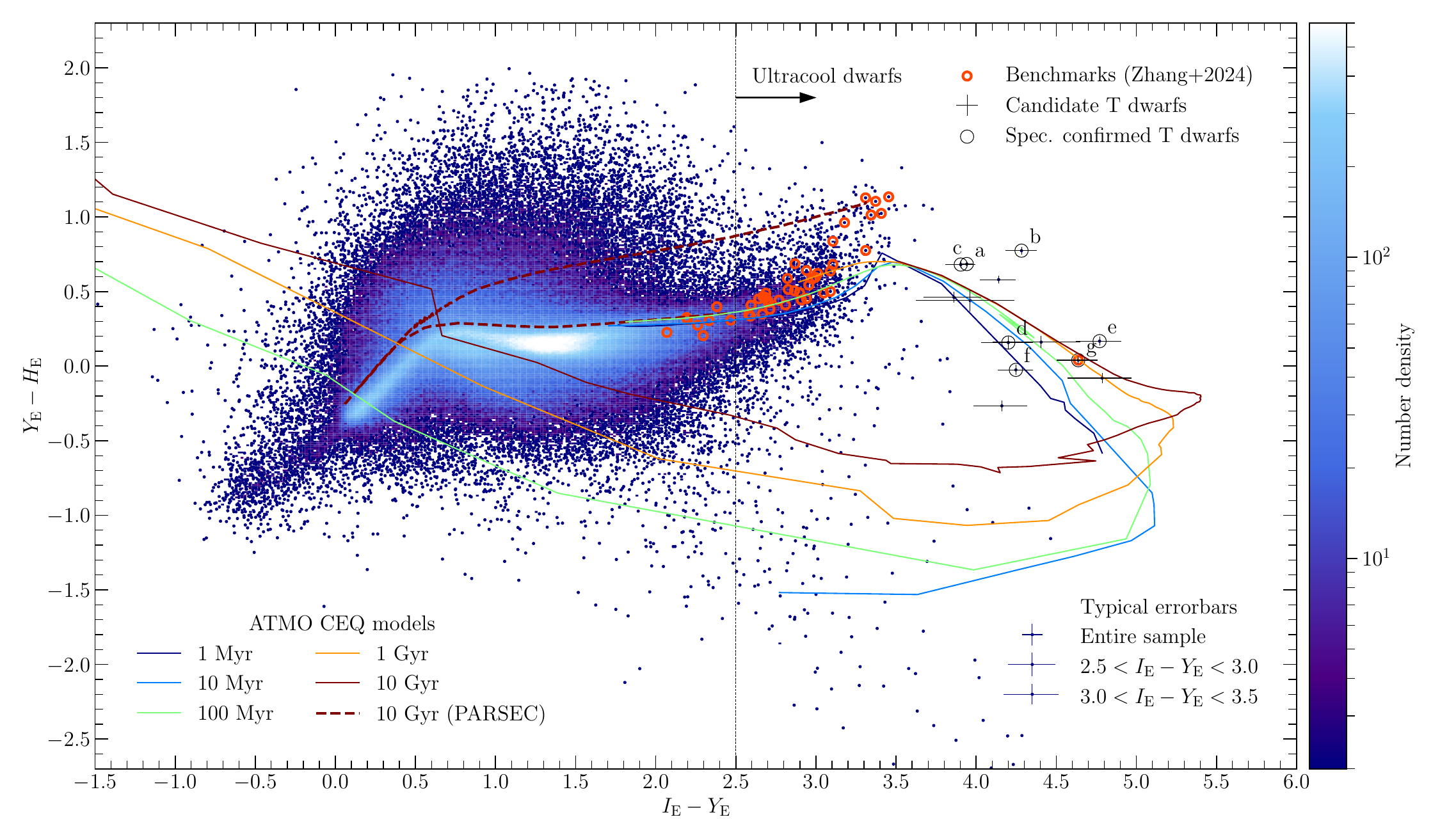} 
\caption{The `bluebottle' diagram of point sources. The central sequence starts with white dwarfs below $\IE-\YE<0$, continues with stars, and ends in the cool tip with ultracool dwarfs at $\IE-\YE \gtrsim 2.5$. Their nature is confirmed with the benchmark UCDs from \citet{2024A&A...686A.171Z}. We overplotted both the new photometric candidate T dwarfs from this work (symbols with error bars), and those that were spectroscopically confirmed (open circles), as listed in Table~\ref{tab.tdwarfs}. \texttt{ATMO} models \citep{2020Phillips}  indicate the UCD parameter space, while the \texttt{PARSEC} model traces main sequence and evolved stars with redder $\YE-\HE$. 
} 
\label{fig.bluebottle}
\end{figure*}

We visualise our point-source catalogue in a colour-colour diagram in Fig.~\ref{fig.bluebottle}. We followed \citet{2025A&A...697A...7M} and chose a combination of $\IE-\YE$ and $\YE-\HE$ that brings out the UCDs most clearly. The use of the optical band helped us to construct a high-quality catalogue of UCDs with minimal contamination, but at the expense of completeness, especially for the reddest objects (T dwarfs). We provide more details about this topic in Sect.~\ref{sec.limitations}.

\begin{figure*}
\includegraphics[width=\linewidth]{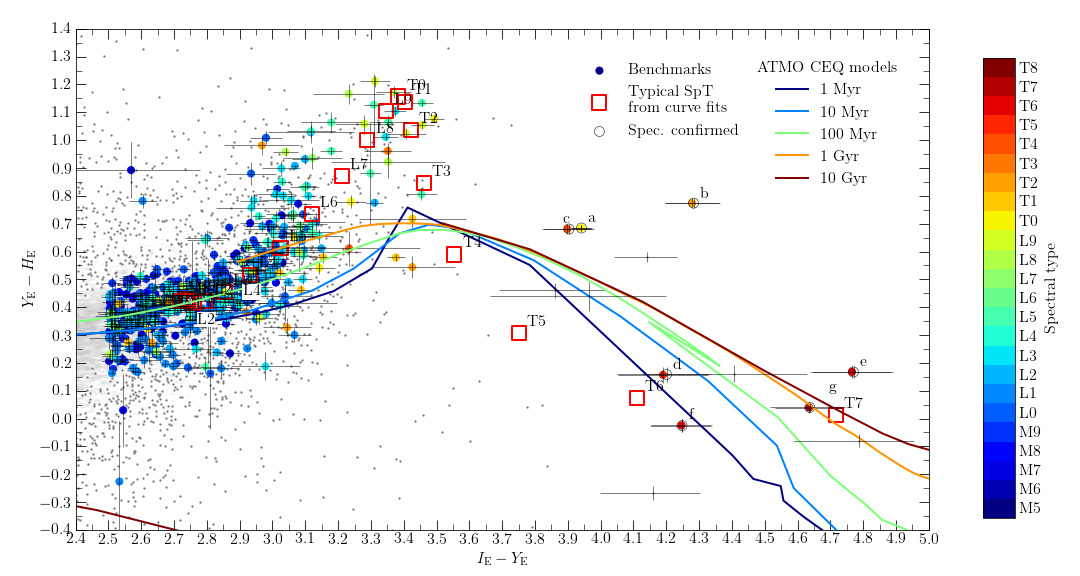} 
\caption{Zoom into the cool tip of the bluebottle diagram showing the benchmark objects (coloured dots with error bars). The spectral types are from \citet{carlos25}. We overplotted typical positions for each spectral type from the curve fits (red squares) and \texttt{ATMO} models for a range of ages. 
} 
\label{fig.bluebottle_UCD}
\end{figure*}

The data in our chosen parameter space form a shape that resembles a bluebottle\footnote{The bluebottle is a stinging marine animal, similar to a jellyfish, from the genus \textit{Physalia}.} with the main body, sail and tentacles. Therefore, we have named this colour-colour plot the `bluebottle' diagram. Its central sequence starts with white dwarfs below $\IE-\YE<0$, continues with stars, and ends in the cool tip with UCDs at $\IE-\YE \gtrapprox 2.5$.
To identify the region occupied by UCDs in Fig.~\ref{fig.bluebottle}, we overplotted the known UCDs from \citet{2024A&A...686A.171Z} that have been spectroscopically confirmed by \citet{carlos25} using \Euclid spectra. For clarity, Fig.~\ref{fig.bluebottle_UCD} shows an expanded area of the bluebottle diagram, including the cool tip and the spectroscopically confirmed UCDs. They perfectly overlap with the bluebottle, and demonstrate that this parameter space encompasses UCDs from late M dwarfs down to late T dwarfs.


\begin{figure}
\includegraphics[width=\linewidth]{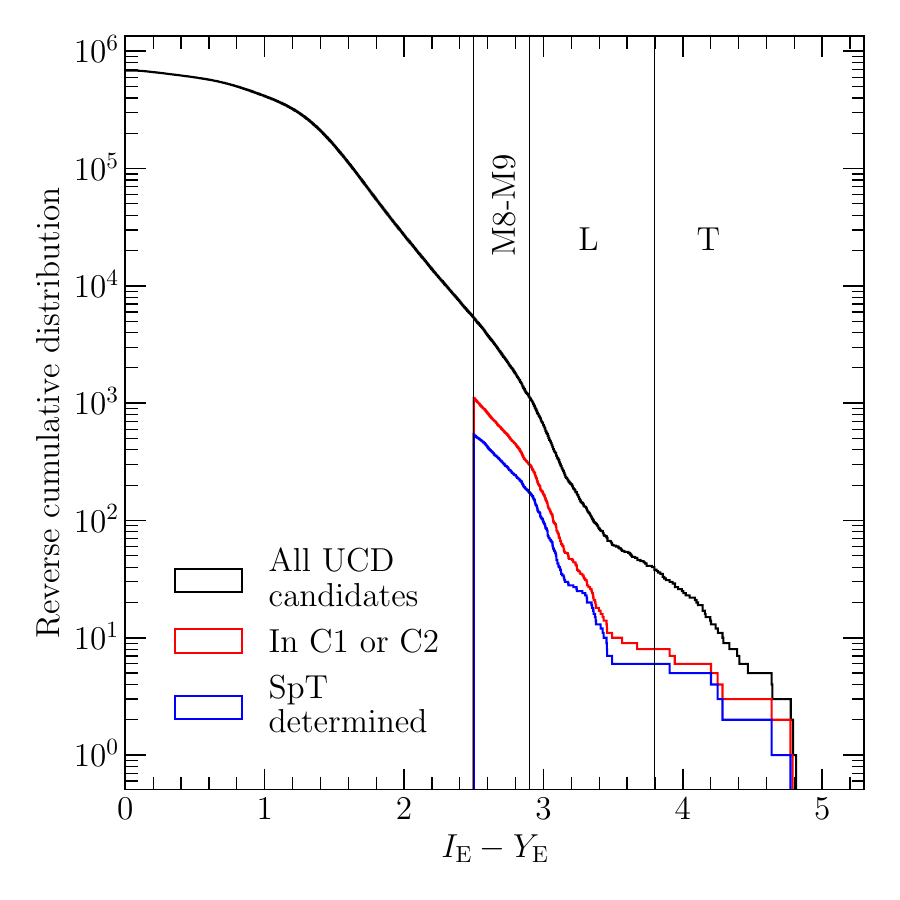} 
\caption{Reverse cumulative distribution, showing how many objects are redder than the selected colour. There are 5306 objects redder than 2.5, which correspond to late M dwarfs, and 1200 objects redder than 2.9, which roughly correspond to L0 dwarfs. The red curve corresponds to objects whose spectra exhibit UCD features, whereas the blue line corresponds to objects with assigned spectral types.
} 
\label{fig.reverse_cumulative_distribution}
\end{figure}

We used the spectroscopically confirmed UCDs as a guide in our candidate UCD selection. Figure~\ref{fig.bluebottle_UCD} shows that the typical $\IE-\YE$ colour of the M8 dwarfs is about 2.5. We therefore used the $\IE-\YE>2.5$ criterion to select the photometric UCD candidates in this work. Such a cut isolated 5306 objects, with the reddest $\IE-\YE$ colours ranging between 2.5 ($\sim$M8) and 5.0 (late T dwarfs).
 According to their colour distribution, shown in Fig.~\ref{fig.reverse_cumulative_distribution}, we expect to find about 1200 L dwarfs with $\IE-\YE>2.9$. Additionally, we report a list of 13 candidate T dwarfs. One of them has been previously identified \citep{2024A&A...686A.171Z}, 
and the rest are new. Seven of them have been spectroscopically confirmed using \Euclid spectra, as shown in figure~7 in \cite{carlos25}. 
These objects are listed in Table~\ref{tab.tdwarfs}, as well as in the main UCD catalogue. The number of T dwarfs is relatively small. This is because they are intrinsically faint, and due to our selection biases, discussed in more detail in Sect.~\ref{sec.limitations}.
In Table~\ref{table:catalogphot} we provide the main result of this work, a list of 5306 photometric UCD candidates. Of these, 546 have been assigned spectral type following the procedure from \citet{carlos25}.

\begin{figure*}
\includegraphics[width=\linewidth]{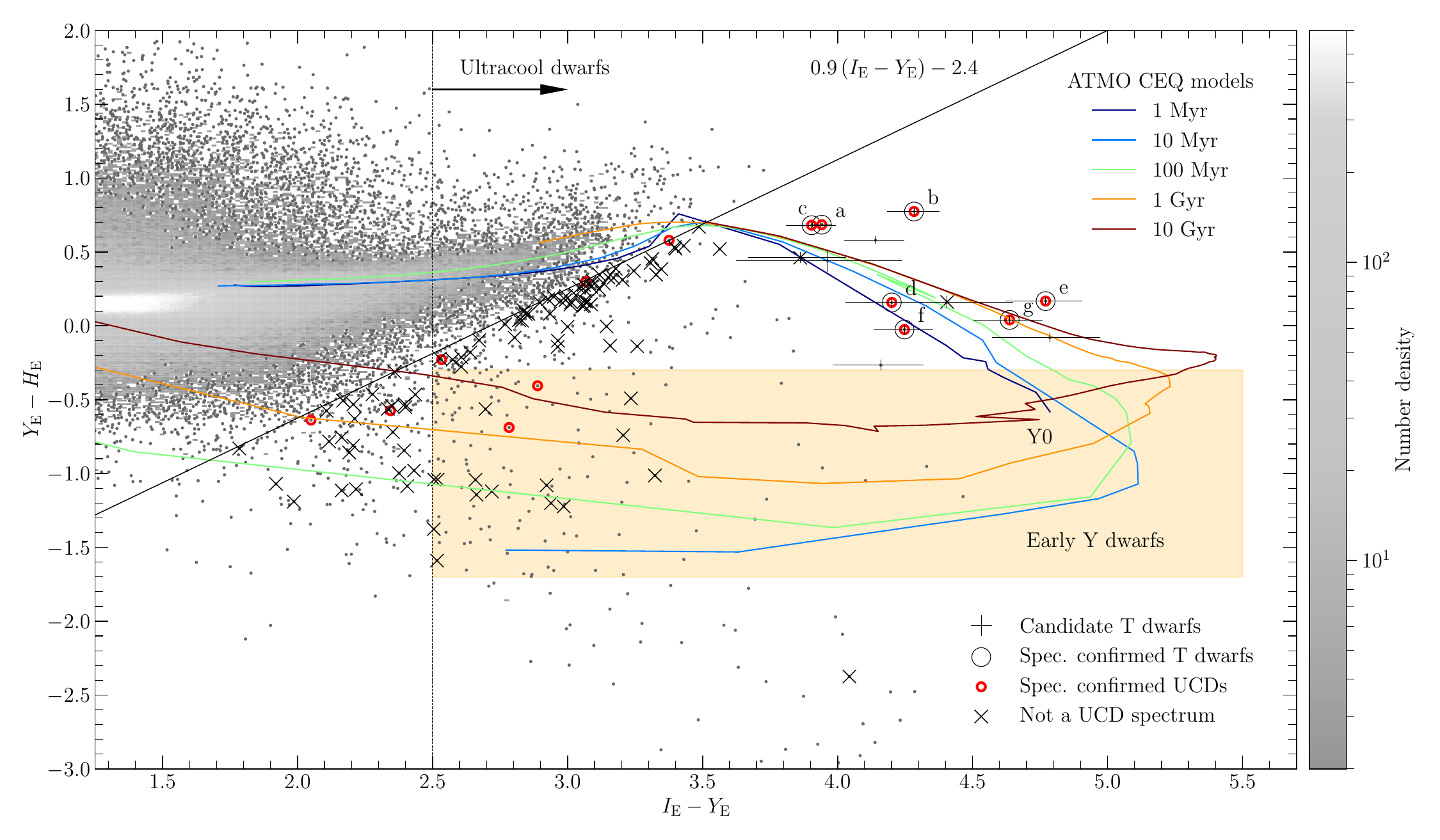} 
\caption{Bluebottle diagram focused on the ultracool dwarfs. Models predict early Y dwarfs to be located in a horizontal band below the main bluebottle body. While there are many photometric candidates found there, none of them have been spectroscopically confirmed. We manually inspected the available spectra of the objects below the black line and found that most of them are not consistent with being UCDs (black crosses). Apart from the T dwarfs, only seven other objects show UCD (or stellar) spectra.
} 
\label{fig.bluebottle_Ydwarfs_tentacles}
\end{figure*}

The \texttt{ATMO} models in Fig.~\ref{fig.bluebottle_Ydwarfs_tentacles} predict that early Y dwarfs are expected to be found in the horizontal band below the main bluebottle body, with $\YE-\HE$ colours roughly between $-1.5$ and $-0.5$. A typical $\IE-\YE$ colour of a Y0 dwarf is around 5, while we can expect objects with masses below $\sim 20 \, M_\mathrm{_{Jup}}$ 
at $\IE-\YE \simeq 2.5$. 
There are a number of objects in this region in the figure. We manually inspected their spectra (although they are not available for all objects) and could not confirm any Y dwarfs. 
On the other hand, based on colours of known Y dwarfs (e.g., \citealp{2024ApJS..271...55K}), we expect Y dwarfs to have $\IE-\YE$ colours that are even redder than T dwarfs, and $\YE-\HE$ colours that are bluer than T dwarfs; i.e., we expect them to extend the T dwarf sequence towards the bottom right in the bluebottle diagram. There are no objects in our catalogue with such colours. 

The absence of detected Y dwarfs is likely due to our strict filtering criteria. Namely, the absolute \IE magnitude of an Y0 object at 10\,Myr is around 26.2, which is already beyond the completeness limit (see Fig.~\ref{fig.magnitude_distribution}). 
Conversely, \cite{carlos25} performed an independent search for UCDs in the entire \Euclid spectral database, and did not find any Y dwarfs. This might be due to the fact that in the Q1 data release, spectra are only available for $\HE<22.5$ objects. This will change in the forthcoming data releases. 
In the future, a development of search criteria based only on NISP, supplemented with the ground-based photometry, might yield a detection of Y dwarfs in \Euclid, as discussed in Sect.~\ref{sec.limitations}.




\section{\label{sc:spec_confirm} A catalogue of spectroscopically confirmed UCDs}
To assess the nature of the selected candidates, we examined the available spectra for these objects \citep{EuclidSkyNISP,Q1-TP003}. Out of 5306 candidates, 4682 have spectra in the \textit{Euclid} Q1 release (almost 90\%). These data vary in quality, quantified by the associated \texttt{QUALITY} flags. 

As a first step, we filtered out spectra with more than 20\% of flux points having \texttt{QUALITY}~$ < 0.2$, as well as spectra with a mean quality below 0.5. This removed 255 objects from further consideration. To identify additional outliers, we computed the Mahalanobis distance $M$ \citep{bishop2006pattern}, selecting spectra with $M > 30$, which eliminated another 210 objects. These 465 outliers are labeled with `O' in Table~\ref{table:catalogphot}.

While these criteria are to some extent subjective, they are necessary to exclude severely degraded or noisy spectra that would hinder further analysis. Visual inspection of the flagged outliers revealed 16 objects that exhibit features characteristic of the UCD spectra, despite being partially corrupted. 

To emphasize the spectral morphology, we removed local spikes and applied cubic spline smoothing. Full details of the denoising procedure will be presented in a forthcoming publication. The cleaned spectra were then used in analysis.

We performed a visual comparison of each denoised spectrum against a set of spectral type templates \citep{SPEX, SPLAT}. Based on similarity to the templates and the presence of spectral features typical for UCDs in the Euclid NISP red-grism wavelength range (the shape of the H$_2$O absorption bands starting from 1.33~$\mu$m, K~I doublet at 1.25~$\mu$m, H~band continuum shape around 1.6~$\mu$m), we assigned each spectrum to one of the following classes:

\begin{itemize}
    \item C1: Spectra that clearly match UCD templates;
    \item C2: Spectra that exhibit some UCD features but do not match any template closely;
    \item C3: Spectra that contain visible signal but lack UCD-specific features;
    \item C4: Spectra that are too noisy or corrupted to be classified.
\end{itemize}

\begin{figure}
\includegraphics[width=\linewidth]{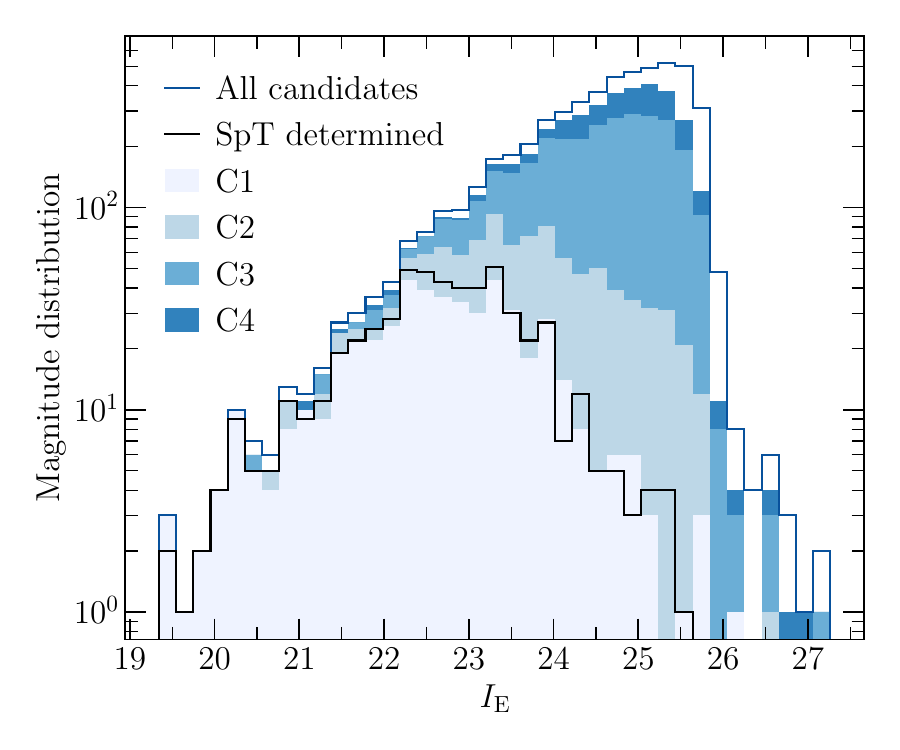} 
\caption{Magnitude distribution of the candidate UCDs with spectra (stacked histogram). Spectral types could be determined for 50\% 
of the entire sample, limited by the SNR of the spectra.
While spectral typing is limited at low SNR, spectroscopic confirmation is still possible nearly down to the VIS detection limit.
} 
\label{fig.spec_confirmed}
\end{figure}

This information is available in Table~\ref{table:catalogphot}.
There are 496 objects in the C1 category, and 615 in C2. It is possible that many more UCDs are present in our photometric candidate list, but their spectra are currently too noisy or unavailable. In the future, once repeated observations of the EDFs become available, we expect to confirm many more additional candidates. Figure~\ref{fig.spec_confirmed} shows that nearly all objects brighter than $\IE=22$ belong to the C1 class, and that we detected objects with some UCD features (class C2) nearly down to the \IE detection limit. The contamination rate is negligible for $\IE<22$, and likely remains low until $\IE=24$ where extragalactic objects start to dominate, as shown in Fig.~\ref{fig.filtering}. A contamination estimate based on the number of objects classified into classes C1, C2, and C3 shows that the contamination rate, defined as $N_\mathrm{C3} / N_\mathrm{C1+C2+C3}$, where $N_\mathrm{C3}$ and $N_\mathrm{C1+C2+C3}$ are the numbers of objects in these classes, is 67\%, consistent with the sharp increase seen at the faint end in Fig.~\ref{fig.filtering}. This is, however, a conservative estimate, as the classification into C1-C4 groups was performed manually, and some spectra assigned to the C3 class may still correspond to UCDs whose spectra are simply too heavily degraded to confirm their nature.


\subsection{Spectral typing}
We considered objects from categories C1 and C2 for spectral classification. This was done by a comparison with the SpeX standard templates \citep{SPEX, SPLAT}.  We used two different methods, as described in \cite{carlos25}. The first method used $\chi^2$ minimisation over the full wavelength range. The second one was based on minimizing the residual from the difference between the spectra and the standard templates over four wavelength ranges defined by the NISP wavelength range and the telluric absorption bands (i.e. only present in the templates). The spectral type (SpT) is obtained after weighting each method by the quality of the spectrum (see \citealp{{carlos25}} for further details).
The typical uncertainty of the SpT is $\pm$ 1 subtype. A colon `:' was added to the spectral subtypes with larger uncertainty, and a `p' for peculiar spectra. Objects with very low SNR were not classified.

In total, 546 spectra had sufficient quality to allow for a reliable determination of spectral types beyond M7. A total of 26 of them are T dwarfs, 329 are L dwarfs, and the rest are late M dwarfs. This means that 10\% of our UCD candidates are spectroscopically confirmed and have their SpT determined. The remaining objects from the C1 and C2 groups exhibit some spectral features consistent with UCD classification, e.g. the water drop at $\sim$1.35~microns, but the data quality is too limited for definitive spectral typing.
We include spectral types in Table~\ref{table:catalogphot}.


A comparison with \cite{carlos25} who published the SpT of 178 UCDs, reveals 27 objects with SpT that are not in our candidate list. Six of them were found by the spectral index search described in that paper. The remaining 21 are confirmed candidates from \cite{2024A&A...686A.171Z}.  A close inspection shows that these 21 objects do not meet our point-source and quality requirements (mostly due to low SNR; Sect.~\ref{sc:Filter}) or have $\IE-\YE < 2.5$. 
This comparison highlights the importance of conducting both photometric and spectroscopic searches independently, particularly because the latter led to the discovery of five T dwarfs that were not identified through photometry alone \citep{carlos25}.

\section{\label{sc:Discussion} Discussion}

In this section, we describe the photometric properties of UCDs that have their spectral type determined (Sect.~\ref{sec.empirical}). Then, we address the contamination rate and the detection limitations of the sample in Sect.~\ref{sec.limitations}, and the prospects for detecting metal-poor UCDs in Sect. \ref{sec.metal_poor}, compare the dataset with the UCD candidates from the literature (Sect.~\ref{sec.des}), and end with the estimated numbers of future UCD detections in the EDS and EWS, based on the results of this work.

\subsection{Empirical photometric sequence\label{sec.empirical}}
We determined the typical \Euclid colours for each UCD spectral type and compared them with the \texttt{ATMO} models used in this work \citep{2020Phillips}. Although our catalogue does not contain any widely studied brown dwarf, its large size allowed us to rely on statistical analysis for late M and L objects. We applied the same analysis to the T dwarfs as well, but the results were less robust due to their scarcity.

\begin{figure*}
\includegraphics[width=\linewidth]{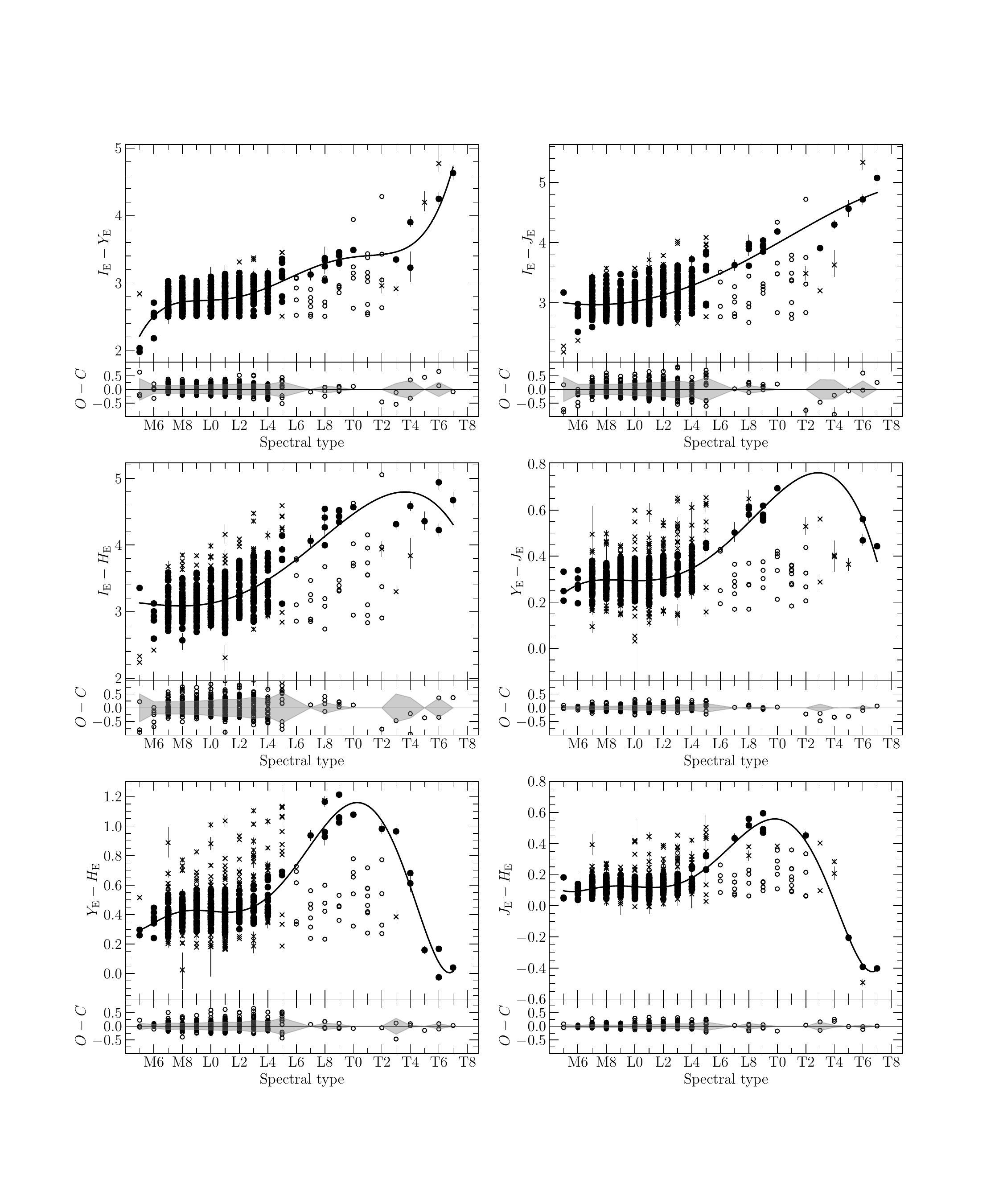} 
\caption{Typical colours for each spectral type. Black dots represent individual objects, crosses mark objects rejected during the sigma-clipping fitting, and open circles indicate objects excluded from the fit due to suspected variability. The black line shows a polynomial fit to the relation. The grey area in the $O-C$ plots represents the standard deviation of the residuals.
} 
\label{fig.spt2colour}
\end{figure*}

Figure~\ref{fig.spt2colour} shows the relations between the spectral types and the colours. The $\IE-\JE$ relation is well behaved, which is important since UCDs are brightest in the \JE band. Although $\IE-\JE$ would thus be a preferable selection criterion to $\IE-\YE$, the main limitation remains the inclusion of the optical band. For consistency with \citealp{2025A&A...697A...7M}, we therefore retain $\IE-\YE$.

Some spectral types exhibit greater scatter in colour than others. The spread is particularly pronounced for L6–T2 objects, especially in the NISP colours ($\YE-\JE$, $\YE-\HE$, and $\JE-\HE$). This increased scatter is consistent with the known variability during the L/T transition, which is attributed to changes in cloud opacity at near-infrared wavelengths \citep{2014ApJ...793...75R,2014ApJ...797..120R}. Consequently, we observed a discrepancy between the data and the \texttt{ATMO} models in Fig.~\ref{fig.bluebottle_UCD} for late L and early T dwarfs.
For this reason, we excluded L6–T2 objects with $\YE-\HE < 0.8$ from this analysis, as Fig.~\ref{fig.bluebottle_UCD} indicates that their typical $\YE-\HE$ colour is greater than 0.9.
Additionally, we excluded a small number of photometric outliers whose colours differ by more than 1 magnitude from the typical values of objects with similar spectral types, despite using a sigma-clipping approach in the fit. We provide more detail about this discrepancy in Sect.~\ref{sec.limitations}.

We fit a polynomial to each spectral type-colour relation, where spectral types were encoded numerically (e.g. 70 for L0, 75 for L5, 80 for T0, etc.). The order of the polynomial varies depending on the colour, selected to best capture the shape of the relation while avoiding overfitting. To reduce edge artefacts commonly introduced by polynomial fitting, we included a small number of M5 and M6 objects at the lower end of the spectral type range. 

Using these polynomial fits, shown in Fig.~\ref{fig.spt2colour}, we derived a typical colour for each spectral type, as listed in Table~\ref{tab.typical_colour}. The resulting $\IE-\YE$ vs. $\YE-\HE$ relation aligns well with the bluebottle diagram up to the tip of the sequence ($\IE-\YE \sim 3.4$), but begins to break down for T dwarfs. This discrepancy is likely due to the limited number of T dwarfs in the current dataset, and we expect it to improve with future \Euclid\ data releases.


\begin{table}
\caption{Empirical \Euclid colours for different spectral types.}
\label{tab.typical_colour}
\centering
\begin{tabular}{l@{\hskip 2mm}c@{\hskip 2mm}c@{\hskip 2mm}c@{\hskip 2mm}c@{\hskip 2mm}c@{\hskip 2mm}r}
\hline\hline
\noalign{\vskip 2pt}
SpT & $\IE-\YE$ & $\IE-\JE$ & $\IE-\HE$ & $\YE-\JE$ & $\YE-\HE$ & $\JE-\HE$ \\
  \hline
  \noalign{\vskip 2pt}
M7 & 2.65 & 2.97 & 3.09 & 0.29 & 0.39 & 0.11 \\
M8 & 2.72 & 2.97 & 3.08 & 0.30 & 0.42 & 0.12 \\
M9 & 2.74 & 2.99 & 3.09 & 0.30 & 0.43 & 0.13 \\
L0 & 2.74 & 3.02 & 3.12 & 0.29 & 0.42 & 0.12 \\
L1 & 2.76 & 3.07 & 3.18 & 0.30 & 0.42 & 0.12 \\
L2 & 2.80 & 3.13 & 3.25 & 0.30 & 0.42 & 0.12 \\
L3 & 2.85 & 3.20 & 3.36 & 0.32 & 0.45 & 0.14 \\
L4 & 2.93 & 3.29 & 3.48 & 0.35 & 0.52 & 0.18 \\
L5 & 3.02 & 3.38 & 3.62 & 0.38 & 0.61 & 0.24 \\
L6 & 3.12 & 3.49 & 3.78 & 0.43 & 0.73 & 0.32 \\
L7 & 3.21 & 3.60 & 3.95 & 0.49 & 0.87 & 0.41 \\
L8 & 3.29 & 3.73 & 4.13 & 0.55 & 1.00 & 0.48 \\
L9 & 3.35 & 3.86 & 4.30 & 0.61 & 1.10 & 0.54 \\
T0 & 3.38 & 3.99 & 4.47 & 0.67 & 1.16 & 0.56 \\
T1 & 3.40 & 4.12 & 4.61 & 0.72 & 1.14 & 0.52 \\
T2 & 3.42 & 4.26 & 4.72 & 0.75 & 1.04 & 0.41 \\
T3 & 3.46 & 4.39 & 4.78 & 0.76 & 0.85 & 0.25 \\
T4 & 3.55 & 4.51 & 4.79 & 0.74 & 0.59 & 0.03 \\
T5 & 3.75 & 4.63 & 4.72 & 0.68 & 0.31 & $-0.20$ \\
T6 & 4.11 & 4.74 & 4.57 & 0.56 & 0.08 & $-0.38$ \\
T7 & 4.72 & 4.83 & 4.31 & 0.38 & 0.01 & $-0.41$ \\
\hline
\end{tabular}
\end{table}

\subsection{Limitations and contamination of the catalogue \label{sec.limitations}}
Here we discuss the detection limits of the instruments, the quality of the sample, and the contamination of the UCD catalogue with evolved stars and extragalactic objects.

\begin{figure*}
\includegraphics[width=\linewidth]
{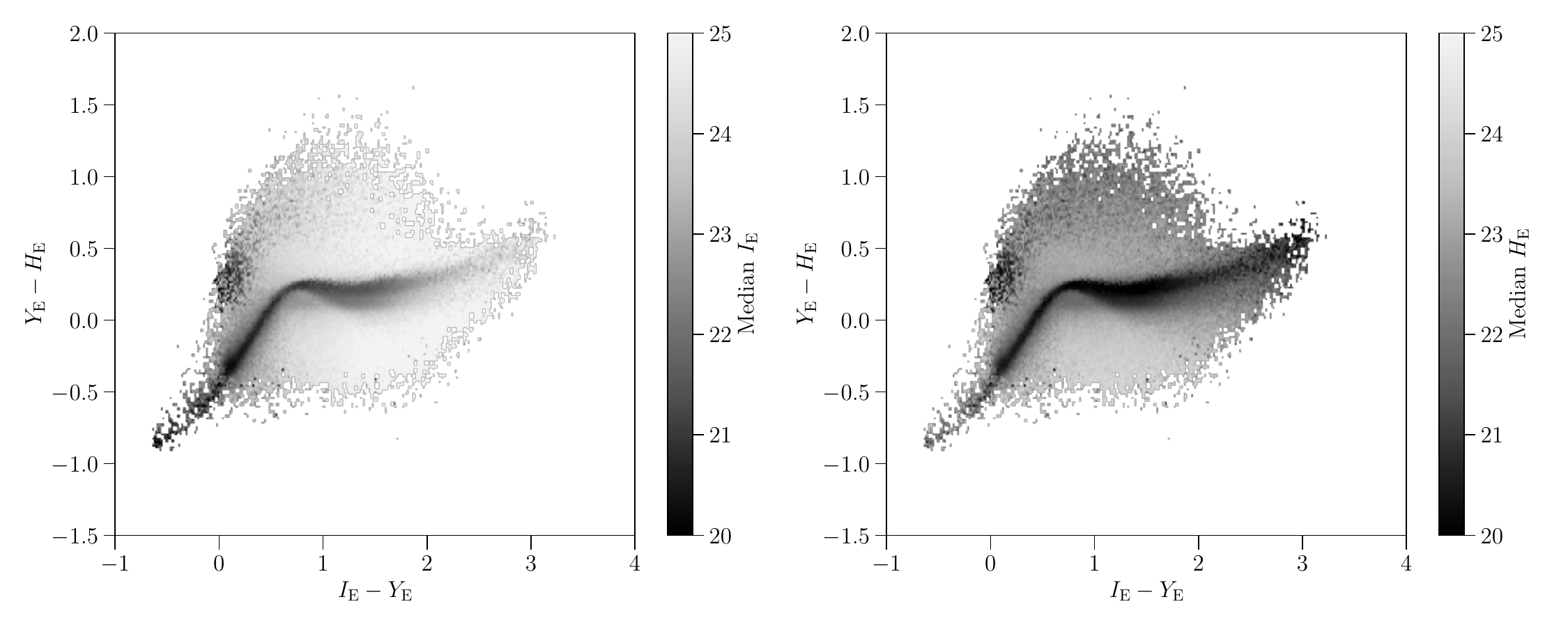}
\includegraphics[width=\linewidth]
{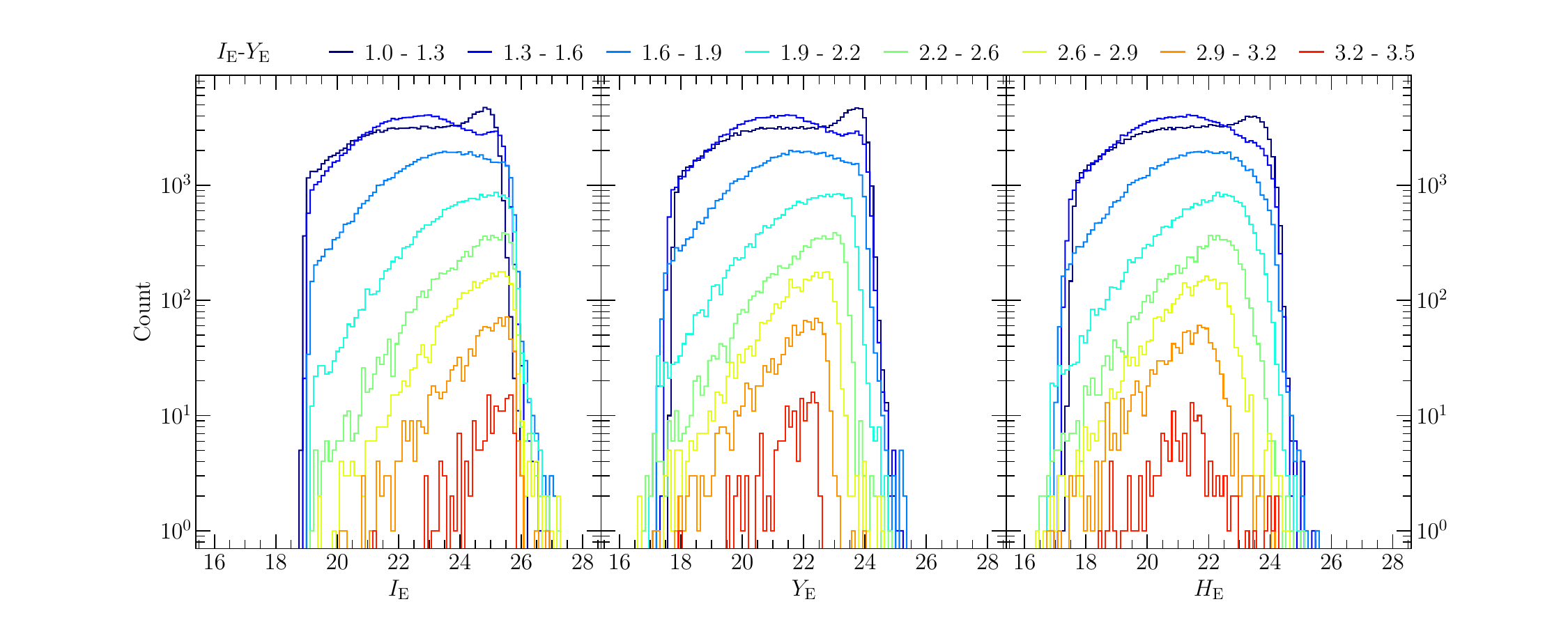} 
\caption{Completeness for the colour bins in the bluebottle diagram. Despite its very high sensitivity, VIS is the most limiting instrument for the detection of ultracool dwarfs in \Euclid. 
{\color{black} The top panels show a group of bright objects at $\approx (0,0)$, which are quasars.}
} 
\label{fig.completeness}
\end{figure*}

The most discriminating parameter in our search for UCDs was the $\IE-\YE$ colour, which is based on the measurements of two different instruments. Their sensitivities differ: the NISP instrument detects objects with magnitudes between 19 and 24; and the VIS range is broader and extends about 2 magnitudes deeper in our point source catalogue (Fig.~\ref{fig.completeness}). Since the UCDs are very red (e.g., the T dwarfs can surpass $\IE-\YE=4.5$), their luminosities in the optical are low. For example, a T dwarf with $\YE=24$ would be fainter than \IE=28. For this reason, our catalogue of UCD candidates only contains NISP sources brighter than $\YE \simeq 22.5$. The advantage of this limitation is the higher quality of the sample and a higher percentage of objects with good quality spectroscopy being available. 

We left the photometric selection of the faint NISP sources for our follow-up work. On the other hand, \cite{carlos25} complemented our current catalogue with a spectroscopic search for UCDs, which is not limited by the VIS instrument. They searched for substellar objects in the entire spectroscopic database in \Euclid, and found 27 UCD spectra of objects that are not in our candidate list because they did not meet our filtering criteria, mostly due to poor SNR. Similarly, \cite{Q1-SP042} used spectroscopic templates to search for UCDs and confirmed 33 new objects, ranging from spectral types M7 to T1.

As discussed in Sect.~\ref{sec.empirical}, approximately 75\% of objects in the L/T transition region exhibit colours that deviate significantly from those expected for their spectral types, particularly in the near-infrared. Due to the known variability and changes in cloud opacity during this transition, these objects can display L-like colours despite having T-type spectra, as shown in Fig.~\ref{fig.bluebottle_UCD}. As a result, our photometric selection revealed only 13 T dwarf candidates -- seven of which have been spectroscopically confirmed -- since only those appear within the bluebottle parameter space where T dwarfs are expected to lie.
In addition to photometry, incorporating spectroscopic searches will thus be crucial for future efforts to identify extremely low-mass UCDs (including planetary-mass objects), which models predict to overlap with the blue end of the bluebottle diagram with their $\IE-\YE$ colours between $\sim 0$ and 1.5 (see Fig.~\ref{fig.bluebottle}).



\begin{figure}
\includegraphics[width=\linewidth]{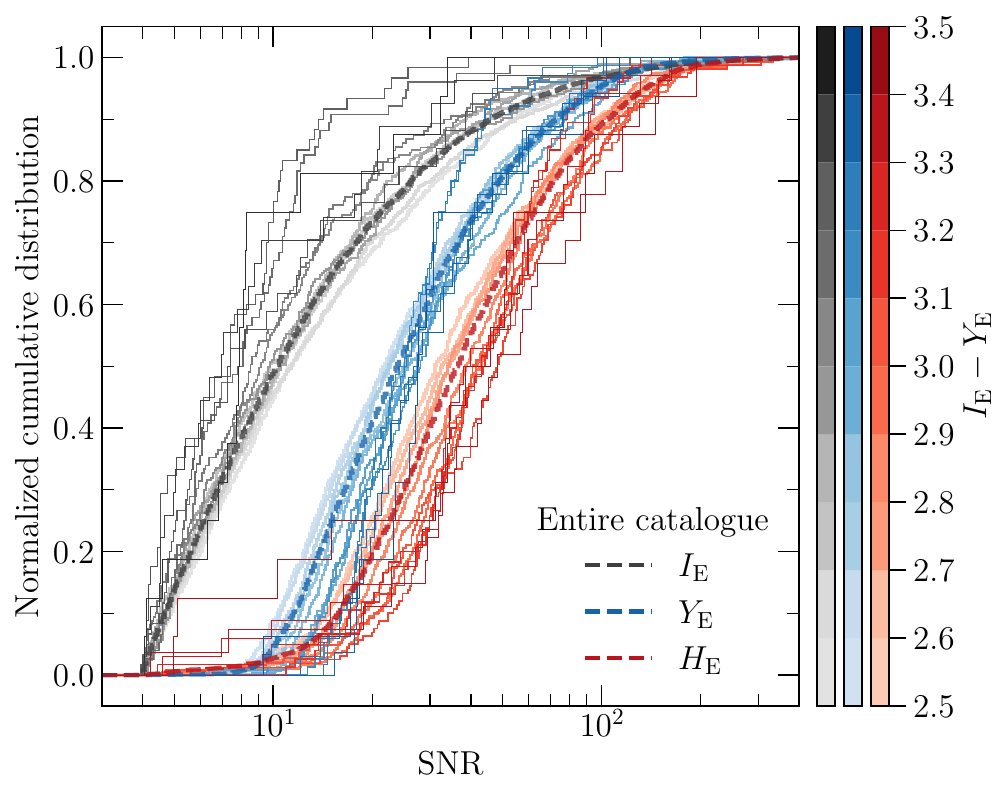} 
\caption{Signal-to-noise ratio (S/N) distribution for \IE, \YE, and \HE channels in the ultracool catalogue. The median S/N values are 10, 24, and 37, respectively. Dashed lines are distributions for the entire catalogue of ultracool candidates. Solid lines trace the variation of the S/N distribution with the $\IE-\YE$ colour. In NISP, redder (cooler) objects tend to have higher S/N on average, while in the VIS band, the opposite is true and the difference between the red and blue is more pronounced. 
} 
\label{fig.snr}
\end{figure}

Our UCD catalogue consists of candidates with good S/N, as explained in Sect.~\ref{sc:Filter}. The median values are 10, 24, and 37 for the \IE, \YE, and \HE bands, respectively. As shown in Fig.~\ref{fig.snr}, there is also a variation of S/N with $\IE-\YE$ colour, due to the aforementioned detector sensitivity bias. However, the quality criteria in the preparation of the point-source catalogue were relatively relaxed. We requested $\texttt{DET\_QUALITY\_FLAG}<3$, which keeps objects with binary flags 1 (bad pixels or contamination by close neighbours) or 2 (blended sources) in the catalogue.
Our experience with the benchmarks showed that UCD objects with bad pixels are typically outliers in the bluebottle diagram, often found far away from the main body. The `tentacles' are one such example; the majority of the benchmark objects (rejected for other reasons) with bad pixels were found there. Because a filter on bad pixels would reject candidates with otherwise adequate morphology and S/N, and a relatively large fraction of the (rejected) benchmarks were affected by this issue, we decided to ignore the bad pixel flags, with the hope that at least some of those objects then meet our UCD selection criteria in the bluebottle diagram. We manually checked the majority of the tentacle objects, and many of them had photometric issues (bad pixels, spikes from nearby bright stars, etc.). Since the tentacles extend into the region where we expect extremely ultracool dwarfs (e.g., Y dwarfs) 
according to the \texttt{ATMO} models (e.g., Fig.~\ref{fig.bluebottle}), we opted for a relaxed filtering that would not exclude potential good UCD candidates.
The issue with bad pixels will be mitigated with repeated \Euclid observations of the EDFs in the future.

\begin{figure*}
\includegraphics[width=\linewidth]{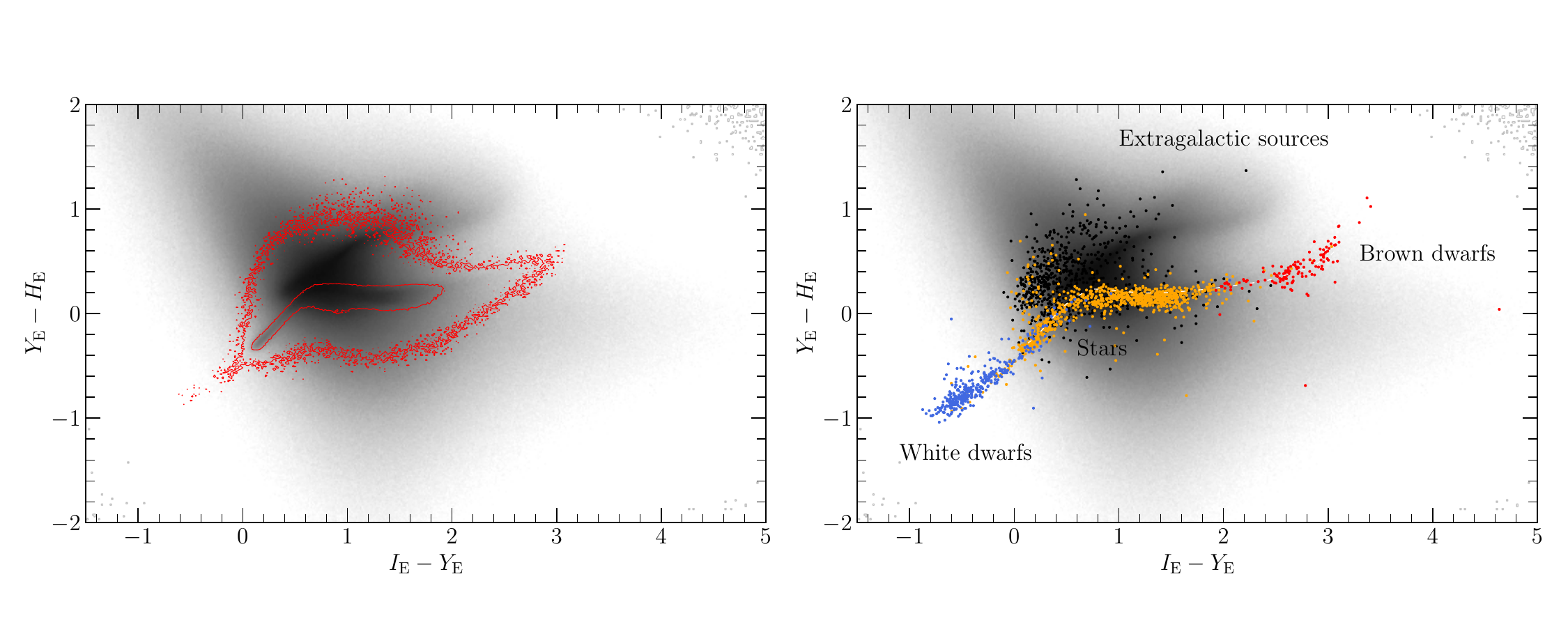} 
\caption{Contamination with extragalactic sources is negligible in the UCD parameter space of the bluebottle diagram. 
Both plots show a distribution of the entire (unfiltered) Q1 catalogue.
\textit{Left}: Red contours mark the shape of the bluebottle diagram. \textit{Right}: Objects in common between \Euclid and Simbad: black dots are QSOs, AGNs, supernovae, or galaxies; and yellow dots are stars. Brown dwarfs (red dots) are from \citet{2024A&A...686A.171Z}.
} 
\label{fig.contamination}
\end{figure*}

The majority of objects in the Q1 catalogue are extragalactic sources. Fortunately, these objects occupy a different part of the bluebottle diagram than substellar objects, as shown in Fig.~\ref{fig.contamination}. Our cross-match with the Simbad database placed galaxies (e.g., \citealp{2020yCat.1350....0G}, \citealp{2016ApJ...830...51S}, \citealp{2010A&A...512A..12B}, \citealp{2011ApJ...743..146C}, and \citealp{2004A&A...428.1043L}), quasars (e.g., \citealp{2017AJ....153..107T}, \citealp{2020yCat.1350....0G}, and \citealp{2016ApJS..224...15X}), other active galactic nuclei (e.g., \citealp{2013ApJS..207...37S}, \citealp{2021ApJS..255...20A}, and \citealp{2020A&A...634A..50P}), and supernovae (\citealp{2020yCat.1350....0G}, \citealp{2015A&A...584A..62C}, \citealp{2005A&A...430...83C}, and \citealp{2013ApJ...771...97L}) on the extragalactic branch that overlaps with the bluebottle `knee' (late K and early M dwarfs), but extends above the stellar main sequence. 
On the other hand, \citet{2025arXiv250116648T}
report the photometric and spectroscopic similarity in the near-infrared between the L- and T-type brown dwarfs, high-redshift galaxies, and `little red dots' (a recently discovered new type of object hypothesised to represent faint and/or highly reddened active galactic nuclei at high redshift; \citealp{2024ApJ...963..129M}). They found good agreement between their candidate brown dwarf spectra and the models, and realised their extragalactic nature only when their inferred distances placed them more than a few kiloparsecs away. While our spectroscopic classification indicates a contamination rate of up to 67\% beyond $\IE=24$ (Sect. \ref{sc:spec_confirm}), this value should be considered an upper limit due to the low quality of spectra at such faint magnitudes. A more reliable contamination estimate will only be possible after future repeated visits of the EDFs, when proper-motion measurements become available.

The 10-Gyr \texttt{PARSEC} isochrone \citep{2012MNRAS.427..127B, 2020MNRAS.498.3283P} shows that the asymptotic giant branch (AGB) reaches the cool tip of the bluebottle diagram (late L dwarfs; see Fig.~\ref{fig.bluebottle}). However, since the lifetime of the AGB stars is only a few Myr (e.g., \citealp{2018MNRAS.475.2282V}) these stars are very rare. It is thus unlikely that our sample of UCD candidates is contaminated with such evolved stars. 



\subsection{\label{sec.metal_poor} Metal-poor ultracool dwarfs}
Our UCD selection is based on the $\IE-\YE$ colour of the solar-metallicity objects. To investigate how this criterion affects the detection of metal-poor UCDs, we prepared a set of solar-metallicity and metal-poor models in the following way. 

The public \texttt{ATMO2020} isochrones are available only for solar metallicity. To explore the variation of the expected theoretical distribution in the $\YE - \HE = f(\IE - \YE)$ diagram of ultracool objects with decreasing metallicity, we turned to the Sonora Bobcat grid of cloudless models \citep{2021ApJ...920...85M}.

For $[\mathrm{Fe}/\mathrm{H}] = -0.5$ and $0.0$, for all values of temperature in the grid (200 to 2400\,K) and for three values of the surface gravity ($\log g \, [\mathrm{cm\,s^{-2}}]= 3, \,4$, and 5), we computed \Euclid equivalent magnitudes in each of the filters ($X$ = \IE, \YE, or \HE) by evaluation of the integral
\begin{equation}
X_\textrm{E} = \int_{\lambda_\mathrm{min}}^{\lambda_\mathrm{max}} F(\lambda)\, R_X(\lambda)\,\mathrm{d}\lambda,
\end{equation}
where $R_X(\lambda)$ is the transmission profile of a \Euclid filter \citep[provided by the Filter Profile Service,][]{2024A&A...689A..93R}, $F(\lambda)$ is the Sonora spectral flux expressed as a wavelength-dependent quantity, and $\lambda_\mathrm{min}$ and $\lambda_\mathrm{max}$ are the end wavelengths of the Sonora spectrum.

\begin{figure*}
\includegraphics[width=\linewidth]{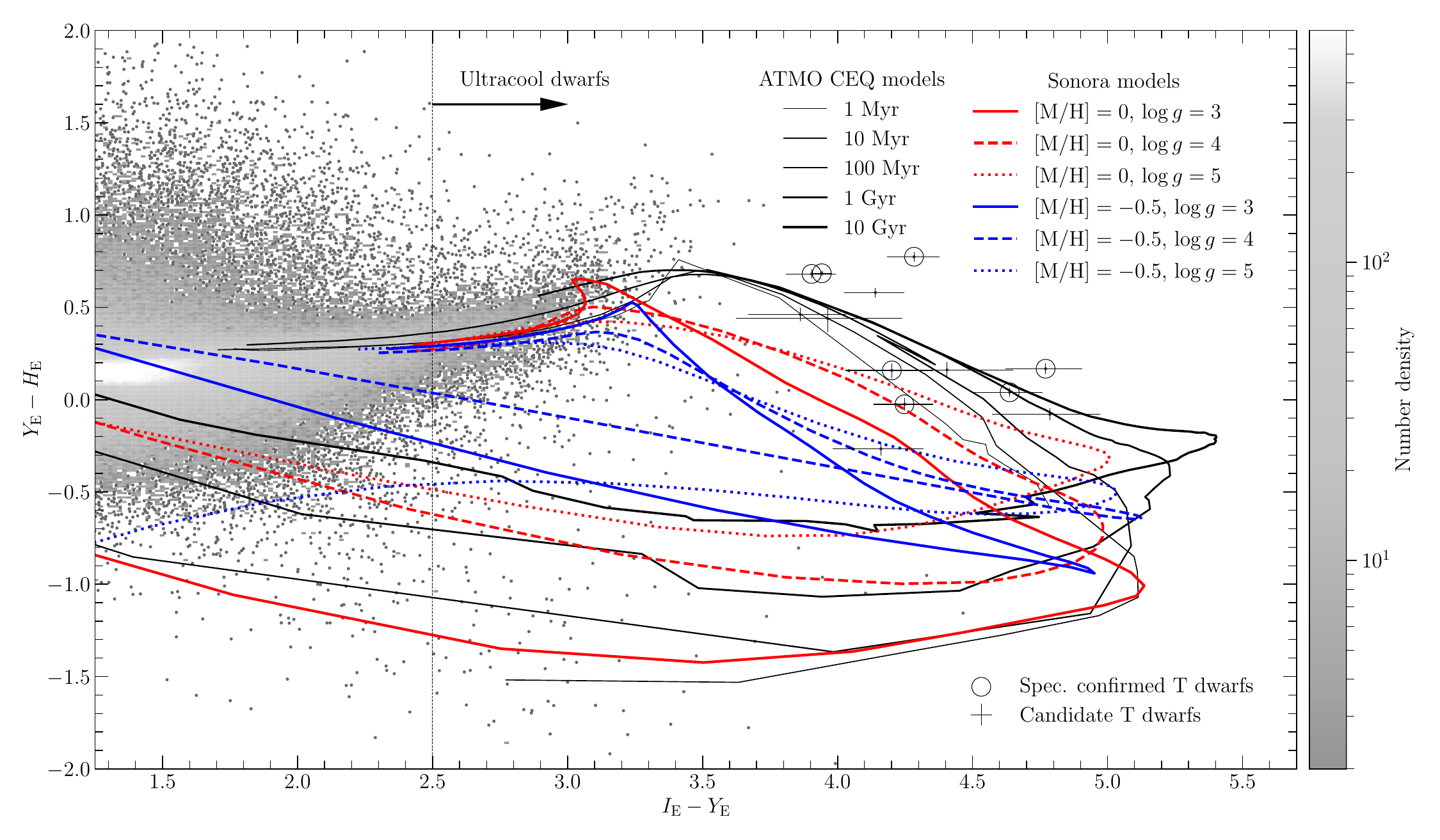} 
\caption{Bluebottle diagram with the \texttt{ATMO} and Sonora models. The latter are provided for two sets of metallicities: solar (red curves); and $\mathrm{[M/H]}=-0.5$. While we cannot isolate metal-poor M and L dwarfs, the models predict a notable colour spread with metallicity for T dwarfs in this diagram.
} 
\label{fig.metalpoor}
\end{figure*}

We compare the models in Fig.~\ref{fig.metalpoor}. They more or less overlap in the parameter space of M and L UCDs. We therefore cannot trace metal-poor objects in that part of the bluebottle diagram. On the other hand, the models predict considerably bluer $\YE-\HE$ colours for T dwarfs. There is one candidate T dwarf (\texttt{OBJECT\_ID}=$-603476608509828998$) that lies below the rest of the sample, but it has not been spectroscopically confirmed yet to be a UCD.

\subsection{Comparison with the Dark Energy Survey}\label{sec.des}
\begin{figure*}
\includegraphics[width=\linewidth]{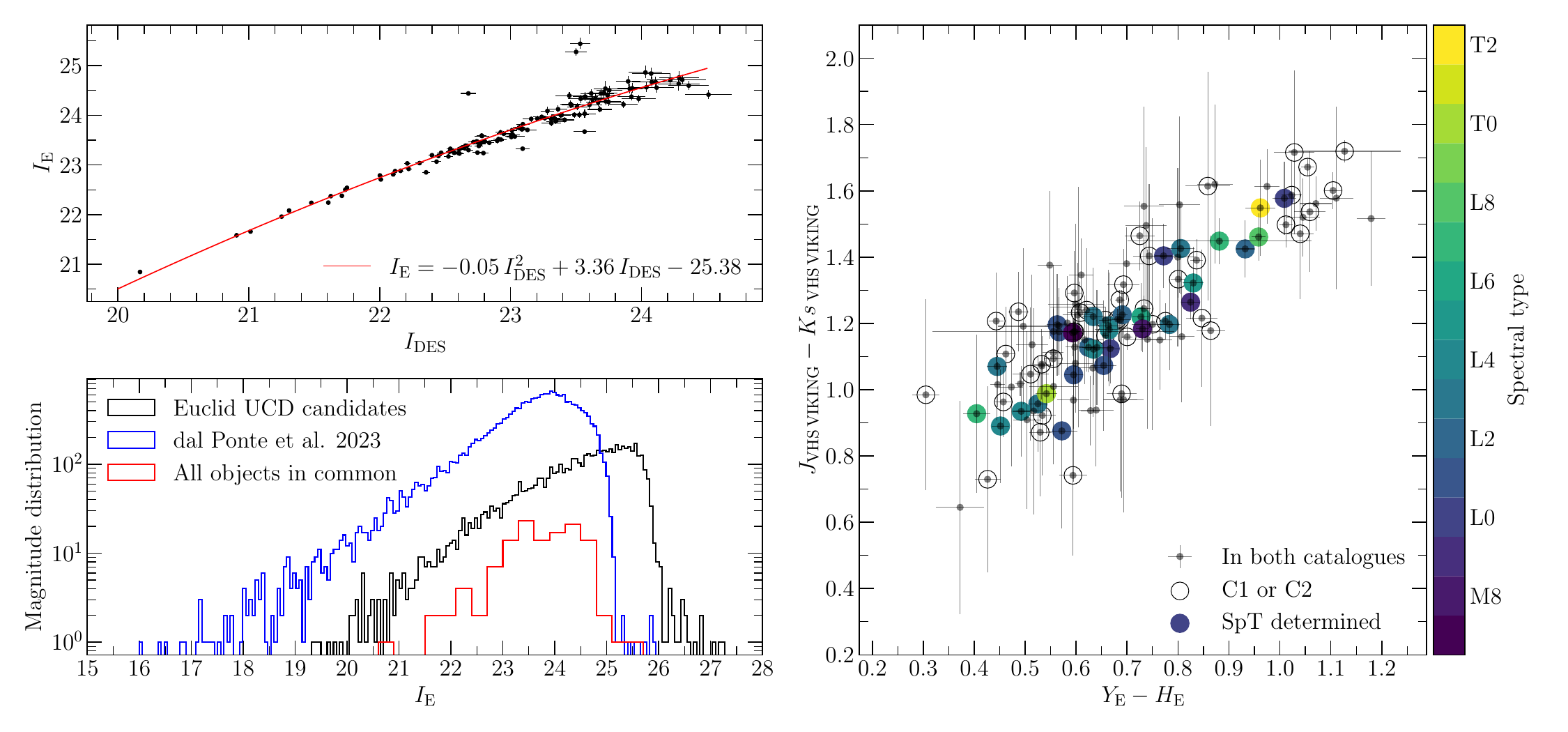} 
\caption{A cross-match with the photometric UCD candidates from the Dark Energy Survey \citep{2023MNRAS.522.1951D} reveals 125 objects in common. \textit{Left:} The relation between \IE and $I_\mathrm{DES}$ is used to translate the DES data onto the \Euclid scale and compare the magnitude distributions. Because DES covers about 100 times larger area of the sky than Q1 (EDF-N is not covered), the objects shared between the two surveys tend to fall within the magnitude range where most of the overall objects are found -- between magnitudes 22 and 25.
\textit{Right:} A tight correlation between the $\YE-\HE$ and $J_\mathrm{VHS \, VIKING}-Ks\,_\mathrm{VHS \, VIKING}$ colours (available in the \citealp{2023MNRAS.522.1951D} catalogue) supports a reliable crossmatch. More than 60\% of them were spectroscopically classified into C1 and C2 groups (black circles), and 25\% have spectral type determined. 
} 
\label{fig.dalponte}
\end{figure*}

Known UCDs in the literature are generally much brighter than our candidates. However, the crossmatch with more than 19\,000 photometric candidate UCDs from the Dark Energy Survey (DES; \citealp{2023MNRAS.522.1951D}) yielded 125 UCD candidates in common in EDF-S and EDF-F. In total, there are 188 DES candidates in the Q1 sky regions, but 56 of them do not meet our filtering criteria. 
The rest of remaining 7 were not found within the 1\,arcsec cross-matching radius, and might potentially have high proper motions.
Most of the 125 objects in common have their spectra available in \Euclid and belong to the C1 and C2 groups; many of them have their spectral types determined.
A relation between $I_\mathrm{DES}$ and \IE was used to convert the DES magnitudes into the \Euclid system and directly compare their distributions.
Figure~\ref{fig.dalponte} shows that most of them have magnitudes between $\IE=22$ and 25, which is about the detection limit in DES. This distribution lies in the fainter half of the C1 and C2 groups and strengthens the case that these objects are true UCDs, as the candidates were selected photometrically using two independent methods and surveys.

\subsection{Expected detections of UCDs in the forthcoming \Euclid data releases}


This work demonstrates the potential of \Euclid to systematically detect numerous field UCDs. While they cover the EDF regions, the Q1 catalogues have essentially the same depth of data as the planned EWS. The results of this paper can therefore be directly used to estimate the number of expected ultracool detections over 14\,000\,deg$^2$ of the sky that \Euclid is going to visit, and which corresponds to about one-third of the entire celestial sphere.

Our detection of 5306 candidate UCDs in 63\,$\mathrm{deg}^2$ of Q1 data gives an average\footnote{The depth of the images varies across the field due to dithering (e.g., \citealp{Q1-TP002}).} density of approximately $100$\,UCDs per $\mathrm{deg}^2$, and 20 L and T dwarfs per $\mathrm{deg}^2$. 
For comparison, the density in DES that detected 20\,000 UCDs in 5000\,$\mathrm{deg}^2$, is 4 objects per $\mathrm{deg}^2$ (and about 3 L and T dwarfs per $\mathrm{deg}^2$; \citealp{2023MNRAS.522.1951D}).
The density in the Galactic plane would be larger, but these regions will not be observed by \Euclid. We 
can therefore expect to detect about 1.4\,million ultracool dwarfs in the final data release of the EWS. Of these, about 300\,000 are expected to be L and T dwarfs. Similarly, we can anticipate the detection of more than 2600 T dwarfs with the method used in this paper. These are lower limits, since the present catalogue is not yet complete and does not fully utilise the potential of the data.

These overall numbers are consistent with theoretical predictions. For example, \citet{2021MNRAS.501..281S} estimated that \Euclid was going to photometrically detect around 1 million objects in the NISP bands in the EWS. However, they estimated that the \JE band, which was not used in our study, is the most sensitive to UCDs. It should be able to observe and detect about 2 million L dwarfs, 1 million T dwarfs, and a handful of Y dwarfs in the thin disc, as well as many objects in the thick disc, and even some halo objects. Conversely, the estimates for the VIS detector were about 100 times lower.


This particular catalogue of UCDs from the Q1 data release is based only on one visit of the EDFs, while \Euclid is going to repeatedly observe them during the lifetime of the mission, and thus increase the detection limit by approximately 2\,mag in each band. Our catalogue will therefore be improved in future, both for new and deeper observations, and the UCD selection methods.



\section{\label{sc:Conclusions}Conclusions}
This work presents a first catalogue of 5306 photometric ultracool dwarf candidates (UCDs) in the deep fields of \Euclid's Q1 data release. It includes approximately 1200 L and T dwarf candidates.
Out of the full sample, 546 candidate UCDs were spectroscopically confirmed; 26 of them are T dwarfs and 329 are L dwarfs.
The catalogue is limited to objects with high quality photometry and prioritises low contamination over completeness. Future work will call for a complementary search that explores fainter sources in the NISP data.

We provide empirical $\IE-\YE$, $\IE-\JE$, $\IE-\HE$, $\YE-\JE$, $\YE-\HE$, and $\JE-\HE$ colours for each UCD spectral type. A comparison with the metal-poor Sonora Bobcat models shows that we cannot easily isolate metal-poor L dwarfs in the bluebottle diagram (i.e., the $\IE-\YE$ versus $\YE-\HE$ colour-colour diagram). However, metal-poor T dwarfs are expected to have colours different from their solar-metallicity counterparts.

We assessed \Euclid's capability to identify UCD candidates based on its photometric passbands, in comparison with the spectroscopic detections reported by \cite{carlos25}. We outline the strengths and limitations of the photometric approach for selecting UCD candidates. As the analysis is based on Q1 data release -- which reaches the same depth as the planned Euclid Wide Survey -- our results provide a direct projection of the number and types of UCDs expected to be detected over the full 5-year mission.




\begin{acknowledgements}
We thank the anonymous referee for their comments, which helped to improve this work.
Funding for M{\v Z}, CDT, NS, ST, NV, JYZ, and ELM was provided by the European Union (ERC Advanced Grant, SUBSTELLAR, project number 101054354).
ELM, NL, VB and JYZ acknowledge support from the Agencia Estatal de Investigaci\'on del Ministerio de Ciencia, Innovación y Universidades under grant PID2022-137241NB-C41.
MRZO acknowledges funding support from the project PID2022-137241NB-C42 by the Spanish "Ministerio de Ciencia, Innovación y Universidades".
PC, DB, PMB, and ES acknowledge financial support from the Agencia Estatal de Investigaci\'on (AEI/10.13039/501100011033) of the Ministerio de Ciencia e Innovaci\'on through project PID2020-112949GB-I00 (Spanish Virtual Observatory \url{https://svo.cab.inta-csic.es}).
PMB acknowledges financial support from the Instituto Nacional de T\'ecnica Aeroespacial through grant PRE-OVE.
DB has been funded by grant No.\ PID2019-107061GB-C61 and PID2023-150468NB-I00 by the Spain Ministry of Science, Innovation/State Agency of Research MCIN/AEI/ 10.13039/501100011033.
NPB is funded by Vietnam National Foundation for Science and Technology Development (NAFOSTED) under grant number 103.99-2020.63.
The authors wish to acknowledge the contribution of the IAC High-Performance Computing support team and hardware facilities to the results 
of this research.
\AckQone
\AckEC  
This publication makes use of VOSA, developed under the Spanish Virtual Observatory (https://svo.cab.inta-csic.es) project funded by MCIN/AEI/10.13039/501100011033/ through grant PID2020-112949GB-I00.
VOSA has been partially updated by using funding from the European Union's Horizon 2020 Research and Innovation Programme, under Grant Agreement nº 776403 (EXOPLANETS-A).
This research has made use of the Simbad and Vizier databases, and the Aladin sky atlas operated at the centre de Donn\'ees Astronomiques de Strasbourg (CDS), and of NASA's Astrophysics Data System Bibliographic Services (ADS). 
This research makes use of ESA Datalabs (\cite{2024sdm..book....1N}; datalabs.esa.int), an initiative by ESA’s Data Science and Archives Division in the Science and Operations Department, Directorate of Science.

Software: \texttt{astropy} \citep{astropy:2018}, \texttt{NumPy} \citep{harris2020array}, \texttt{IPython} \citep{doi:10.1109/MCSE.2007.53}, \texttt{TOPCAT} \citep{2005ASPC..347...29T} and \texttt{matplotlib} \citep{doi:10.1109/MCSE.2007.55}. 
\end{acknowledgements}

%
%

\bibliography{Euclid, Q1, my} 

@ARTICLE{Q1-TP001,
       author = {{Euclid Collaboration: Aussel}, H. and {Tereno}, I. and {Schirmer}, M. and others},
        title = "{Euclid Quick Data Release (Q1) - Data release overview}",
      journal = {A\&A, submitted (Euclid Q1 SI)},
         year = 2025,
        month = mar,
          eid = {arXiv:2503.15302},
        pages = {arXiv:2503.15302},
archivePrefix = {arXiv},
       eprint = {2503.15302},
 primaryClass = {astro-ph.GA},
       adsurl = {https://ui.adsabs.harvard.edu/abs/2025arXiv250315302E}
}

@ARTICLE{Q1-TP002,
       author = {{Euclid Collaboration: McCracken}, H.~J. and {Benson}, K. and {Dolding}, C. and others},
        title = "{Euclid Quick Data Release (Q1): VIS processing and data products}",
      journal = {A\&A, accepted (Euclid Q1 SI)},
         year = 2025,
        month = mar,
          eid = {arXiv:2503.15303},
        pages = {arXiv:2503.15303},
archivePrefix = {arXiv},
       eprint = {2503.15303},
 primaryClass = {astro-ph.IM},
       adsurl = {https://ui.adsabs.harvard.edu/abs/2025arXiv250315303E}
}

@ARTICLE{Q1-TP003,
       author = {{Euclid Collaboration: Polenta}, G. and {Frailis}, M. and {Alavi}, A. and others},
        title = "{Euclid Quick Data Release (Q1). NIR processing and data products}",
      journal = {A\&A, accepted (Euclid Q1 SI)},
         year = 2025,
        month = mar,
          eid = {arXiv:2503.15304},
        pages = {arXiv:2503.15304},
archivePrefix = {arXiv},
       eprint = {2503.15304},
 primaryClass = {astro-ph.CO},
       adsurl = {https://ui.adsabs.harvard.edu/abs/2025arXiv250315304E}
}

@ARTICLE{Q1-TP004,
       author = {{Euclid Collaboration: Romelli}, E. and {K\"ummel}, M. and {Dole}, H. and others},
        title = "{Euclid Quick Data Release (Q1). From images to multiwavelength catalogues: the Euclid MERge Processing Function}",
      journal = {A\&A, in press (Euclid Q1 SI), \url{https://doi.org/10.1051/0004-6361/202554586}},
         year = 2025,
        month = mar,
          eid = {arXiv:2503.15305},
        pages = {arXiv:2503.15305},
archivePrefix = {arXiv},
       eprint = {2503.15305},
 primaryClass = {astro-ph.CO},
       adsurl = {https://ui.adsabs.harvard.edu/abs/2025arXiv250315305E}
}

@ARTICLE{Q1-SP042,
       author = {{Mohandasan}, A. and {Smart}, R.~L. and {Reyl\'e}, C. and others},
       title = "{Euclid Quick Data Release (Q1) Ultracool dwarfs in the Euclid Deep Field North}",
      journal = {A\&A, submitted (Euclid Q1 SI)},
         year = 2025,
        month = mar,
          eid = {arXiv:2503.22559},
        pages = {arXiv:2503.22559},
archivePrefix = {arXiv},
       eprint = {2503.22559},
 primaryClass = {astro-ph.SR},
       adsurl = {https://ui.adsabs.harvard.edu/abs/2025arXiv250322559M}
}

@ARTICLE{EuclidSkyOverview,
author = {{Euclid Collaboration: Mellier}, Y. and {Abdurro'uf} and {Acevedo~Barroso}, J.A. and others},
	title = {Euclid - I. Overview of the Euclid mission},
	DOI= "10.1051/0004-6361/202450810",
	url= "https://doi.org/10.1051/0004-6361/202450810",
	journal = {A\&A},
	year = 2025,
	volume = 697,
	pages = "A1",
}

@ARTICLE{EuclidSkyVIS,
author = {{Euclid Collaboration: Cropper}, M. and {Al-Bahlawan}, A. and {Amiaux}, J. and others},
	title = {Euclid - II. The VIS instrument},
	DOI= "10.1051/0004-6361/202450996",
	url= "https://doi.org/10.1051/0004-6361/202450996",
	journal = {A\&A},
	year = 2025,
	volume = 697,
	pages = "A2",
}

@ARTICLE{EuclidSkyNISP,
author = {{Euclid Collaboration: Jahnke}, K. and {Gillard}, W. and {Schirmer}, M. and others},
	title = {Euclid - III. The NISP Instrument},
	DOI= "10.1051/0004-6361/202450786",
	url= "https://doi.org/10.1051/0004-6361/202450786",
	journal = {A\&A},
	year = 2025,
	volume = 697,
	pages = "A3",
}

@misc{Q1cite,
author = "{Euclid Quick Release Q1}",
howpublished = "\url{https://doi.org/10.57780/esa-2853f3b}",
year = 2025
}

@ARTICLE{Schirmer-EP18,
       author = {{Euclid Collaboration: Schirmer}, M. and {Jahnke}, K. and {Seidel}, G. and others},
        title = "{Euclid preparation. XVIII. The NISP photometric system}",
      journal = {\aap},
         year = 2022,
        month = jun,
       volume = {662},
          eid = {A92},
        pages = {A92},
          doi = {10.1051/0004-6361/202142897},
archivePrefix = {arXiv},
       eprint = {2203.01650},
 primaryClass = {astro-ph.IM},
       adsurl = {https://ui.adsabs.harvard.edu/abs/2022A&A...662A..92E}
}

@INPROCEEDINGS{SPEX,
       author = {{Burgasser}, Adam J.},
        title = "{The SpeX Prism Library: 1000+ low-resolution, near-infrared spectra of ultracool M, L, T and Y dwarfs}",
    booktitle = {Astronomical Society of India Conference Series},
         year = 2014,
       series = {Astronomical Society of India Conference Series},
       volume = {11},
        month = jan,
        pages = {7-16},
          doi = {10.48550/arXiv.1406.4887},
archivePrefix = {arXiv},
       eprint = {1406.4887},
 primaryClass = {astro-ph.SR},
       adsurl = {https://ui.adsabs.harvard.edu/abs/2014ASInC..11....7B}
}

@INPROCEEDINGS{SPLAT,
       author = {{Burgasser}, A.~J. and {Splat Development Team}},
        title = "{The SpeX Prism Library Analysis Toolkit (SPLAT): A Data Curation Model}",
    booktitle = {Astronomical Society of India Conference Series},
         year = 2017,
       series = {Astronomical Society of India Conference Series},
       volume = {14},
        month = jan,
        pages = {7-12},
          doi = {10.48550/arXiv.1707.00062},
archivePrefix = {arXiv},
       eprint = {1707.00062},
 primaryClass = {astro-ph.SR},
       adsurl = {https://ui.adsabs.harvard.edu/abs/2017ASInC..14....7B}
}

@ARTICLE{carlos25,
       author = {{Dominguez-Tagle}, C. and {{\v{Z}}erjal}, M. and {Sedighi}, N. and {Mas-Buitrago}, P. and {Martin}, E.~L. and {Zhang}, J.-Y. and {Vitas}, N. and {B{\'e}jar}, V.~J.~S. and {Tsilia}, S. and {Mu{\~n}oz Torres}, S. and {Lodieu}, N. and {Barrado}, D. and {Solano}, E. and {Cruz}, P. and {Tata}, R. and {Phan-Bao}, N. and {Burgasser}, A.},
        title = "{Euclid Quick Data Release (Q1){\textemdash}Spectroscopic Search, Classification, and Analysis of Ultracool Dwarfs in the Deep Fields}",
      journal = {\apj},
         year = 2025,
        month = sep,
       volume = {991},
       number = {1},
          eid = {84},
        pages = {84},
          doi = {10.3847/1538-4357/adf72d},
archivePrefix = {arXiv},
       eprint = {2503.22442},
 primaryClass = {astro-ph.SR},
       adsurl = {https://ui.adsabs.harvard.edu/abs/2025ApJ...991...84D}
}

@ARTICLE{2021MNRAS.501..281S,
       author = {{Solano}, E. and {G{\'a}lvez-Ortiz}, M.~C. and {Mart{\'\i}n}, E.~L. and {G{\'o}mez Mu{\~n}oz}, I.~M. and {Rodrigo}, C. and {Burgasser}, A.~J. and {Lodieu}, N. and {B{\'e}jar}, V.~J.~S. and {Hu{\'e}lamo}, N. and {Morales-Calder{\'o}n}, M. and {Bouy}, H.},
        title = "{Ultracool dwarfs in deep extragalactic surveys using the virtual observatory: ALHAMBRA and COSMOS}",
      journal = {\mnras},
         year = 2021,
        month = jan,
       volume = {501},
       number = {1},
        pages = {281-290},
          doi = {10.1093/mnras/staa3423},
archivePrefix = {arXiv},
       eprint = {2010.16392},
 primaryClass = {astro-ph.SR},
       adsurl = {https://ui.adsabs.harvard.edu/abs/2021MNRAS.501..281S}
}

@article{astropy:2018,
Adsurl = {https://ui.adsabs.harvard.edu/#abs/2018AJ....156..123T},
Author = {{Price-Whelan}, A.~M. and {Sip{\H{o}}cz}, B.~M. and {G{\"u}nther}, H.~M. and {Lim}, P.~L. and {Crawford}, S.~M. and {Conseil}, S. and {Shupe}, D.~L. and {Craig}, M.~W. and {Dencheva}, N. and {Ginsburg}, A. and {VanderPlas}, J.~T. and {Bradley}, L.~D. and {P{\'e}rez-Su{\'a}rez}, D. and {de Val-Borro}, M. and {Paper Contributors}, (Primary and {Aldcroft}, T.~L. and {Cruz}, K.~L. and {Robitaille}, T.~P. and {Tollerud}, E.~J. and {Coordination Committee}, (Astropy and {Ardelean}, C. and {Babej}, T. and {Bach}, Y.~P. and {Bachetti}, M. and {Bakanov}, A.~V. and {Bamford}, S.~P. and {Barentsen}, G. and {Barmby}, P. and {Baumbach}, A. and {Berry}, K.~L. and {Biscani}, F. and {Boquien}, M. and {Bostroem}, K.~A. and {Bouma}, L.~G. and {Brammer}, G.~B. and {Bray}, E.~M. and {Breytenbach}, H. and {Buddelmeijer}, H. and {Burke}, D.~J. and {Calderone}, G. and {Cano Rodr{\'\i}guez}, J.~L. and {Cara}, M. and {Cardoso}, J.~V.~M. and {Cheedella}, S. and {Copin}, Y. and {Corrales}, L. and {Crichton}, D. and {D{\textquoteright}Avella}, D. and {Deil}, C. and {Depagne}, {\'E}. and {Dietrich}, J.~P. and {Donath}, A. and {Droettboom}, M. and {Earl}, N. and {Erben}, T. and {Fabbro}, S. and {Ferreira}, L.~A. and {Finethy}, T. and {Fox}, R.~T. and {Garrison}, L.~H. and {Gibbons}, S.~L.~J. and {Goldstein}, D.~A. and {Gommers}, R. and {Greco}, J.~P. and {Greenfield}, P. and {Groener}, A.~M. and {Grollier}, F. and {Hagen}, A. and {Hirst}, P. and {Homeier}, D. and {Horton}, A.~J. and {Hosseinzadeh}, G. and {Hu}, L. and {Hunkeler}, J.~S. and {Ivezi{\'c}}, {\v{Z}}. and {Jain}, A. and {Jenness}, T. and {Kanarek}, G. and {Kendrew}, S. and {Kern}, N.~S. and {Kerzendorf}, W.~E. and {Khvalko}, A. and {King}, J. and {Kirkby}, D. and {Kulkarni}, A.~M. and {Kumar}, A. and {Lee}, A. and {Lenz}, D. and {Littlefair}, S.~P. and {Ma}, Z. and {Macleod}, D.~M. and {Mastropietro}, M. and {McCully}, C. and {Montagnac}, S. and {Morris}, B.~M. and {Mueller}, M. and {Mumford}, S.~J. and {Muna}, D. and {Murphy}, N.~A. and {Nelson}, S. and {Nguyen}, G.~H. and {Ninan}, J.~P. and {N{\"o}the}, M. and {Ogaz}, S. and {Oh}, S. and {Parejko}, J.~K. and {Parley}, N. and {Pascual}, S. and {Patil}, R. and {Patil}, A.~A. and {Plunkett}, A.~L. and {Prochaska}, J.~X. and {Rastogi}, T. and {Reddy Janga}, V. and {Sabater}, J. and {Sakurikar}, P. and {Seifert}, M. and {Sherbert}, L.~E. and {Sherwood-Taylor}, H. and {Shih}, A.~Y. and {Sick}, J. and {Silbiger}, M.~T. and {Singanamalla}, S. and {Singer}, L.~P. and {Sladen}, P.~H. and {Sooley}, K.~A. and {Sornarajah}, S. and {Streicher}, O. and {Teuben}, P. and {Thomas}, S.~W. and {Tremblay}, G.~R. and {Turner}, J.~E.~H. and {Terr{\'o}n}, V. and {van Kerkwijk}, M.~H. and {de la Vega}, A. and {Watkins}, L.~L. and {Weaver}, B.~A. and {Whitmore}, J.~B. and {Woillez}, J. and {Zabalza}, V. and {Contributors}, (Astropy},
Doi = {10.3847/1538-3881/aabc4f},
Eid = {123},
Journal = {\aj},
Month = Sep,
Pages = {123},
Primaryclass = {astro-ph.IM},
Title = {{The Astropy Project: Building an Open-science Project and Status of the v2.0 Core Package}},
Volume = {156},
Year = 2018}

@article{doi:10.1109/MCSE.2007.53,
author = {P\'erez,Fernando  and Granger,Brian E. },
title = {IPython: A System for Interactive Scientific Computing},
journal = {Computing in Science \& Engineering},
volume = {9},
number = {3},
pages = {21-29},
year = {2007},
doi = {10.1109/MCSE.2007.53},
URL = { 
        https://aip.scitation.org/doi/abs/10.1109/MCSE.2007.53
},
eprint = { 
        https://aip.scitation.org/doi/pdf/10.1109/MCSE.2007.53
}
}

@article{doi:10.1109/MCSE.2007.55,
author = {Hunter,John D. },
title = {Matplotlib: A 2D Graphics Environment},
journal = {Computing in Science \& Engineering},
volume = {9},
number = {3},
pages = {90-95},
year = {2007},
doi = {10.1109/MCSE.2007.55},
URL = { 
        https://aip.scitation.org/doi/abs/10.1109/MCSE.2007.55
},
eprint = { 
        https://aip.scitation.org/doi/pdf/10.1109/MCSE.2007.55
}
}

@Article{         harris2020array,
 title         = {Array programming with {NumPy}},
 author        = {Charles R. Harris and K. Jarrod Millman and St{'{e}}fan J.
                 van der Walt and Ralf Gommers and Pauli Virtanen and David
                 Cournapeau and Eric Wieser and Julian Taylor and Sebastian
                 Berg and Nathaniel J. Smith and Robert Kern and Matti Picus
                 and Stephan Hoyer and Marten H. van Kerkwijk and Matthew
                 Brett and Allan Haldane and Jaime Fern{'{a}}ndez del
                 R{'{\i}}o and Mark Wiebe and Pearu Peterson and Pierre
                 G{'{e}}rard-Marchant and Kevin Sheppard and Tyler Reddy and
                 Warren Weckesser and Hameer Abbasi and Christoph Gohlke and
                 Travis E. Oliphant},
 year          = {2020},
 month         = sep,
 journal       = {Nature},
 volume        = {585},
 number        = {7825},
 pages         = {357--362},
 doi           = {10.1038/s41586-020-2649-2},
 publisher     = {Springer Science and Business Media {LLC}},
 url           = {https://doi.org/10.1038/s41586-020-2649-2}
}

@INPROCEEDINGS{2005ASPC..347...29T,
       author = {{Taylor}, M.~B.},
        title = "{TOPCAT \& STIL: Starlink Table/VOTable Processing Software}",
    booktitle = {Astronomical Data Analysis Software and Systems XIV},
         year = 2005,
       editor = {{Shopbell}, P. and {Britton}, M. and {Ebert}, R.},
       series = {Astronomical Society of the Pacific Conference Series},
       volume = {347},
        month = dec,
        pages = {29},
       adsurl = {https://ui.adsabs.harvard.edu/abs/2005ASPC..347...29T}
}

@ARTICLE{2024A&A...686A.171Z,
       author = {{Zhang}, J. -Y. and {Lodieu}, N. and {Mart{\'\i}n}, E.~L.},
        title = "{Reconnaissance ultracool spectra in the Euclid Deep Fields}",
      journal = {\aap},
         year = 2024,
        month = jun,
       volume = {686},
          eid = {A171},
        pages = {A171},
          doi = {10.1051/0004-6361/202348769},
archivePrefix = {arXiv},
       eprint = {2403.15288},
 primaryClass = {astro-ph.SR},
       adsurl = {https://ui.adsabs.harvard.edu/abs/2024A&A...686A.171Z}
}

@ARTICLE{2016arXiv161205560C,
       author = {{Chambers}, K.~C. and {Magnier}, E.~A. and {Metcalfe}, N. and {Flewelling}, H.~A. and {Huber}, M.~E. and {Waters}, C.~Z. and {Denneau}, L. and {Draper}, P.~W. and {Farrow}, D. and {Finkbeiner}, D.~P. and {Holmberg}, C. and {Koppenhoefer}, J. and {Price}, P.~A. and {Rest}, A. and {Saglia}, R.~P. and {Schlafly}, E.~F. and {Smartt}, S.~J. and {Sweeney}, W. and {Wainscoat}, R.~J. and {Burgett}, W.~S. and {Chastel}, S. and {Grav}, T. and {Heasley}, J.~N. and {Hodapp}, K.~W. and {Jedicke}, R. and {Kaiser}, N. and {Kudritzki}, R. -P. and {Luppino}, G.~A. and {Lupton}, R.~H. and {Monet}, D.~G. and {Morgan}, J.~S. and {Onaka}, P.~M. and {Shiao}, B. and {Stubbs}, C.~W. and {Tonry}, J.~L. and {White}, R. and {Ba{\~n}ados}, E. and {Bell}, E.~F. and {Bender}, R. and {Bernard}, E.~J. and {Boegner}, M. and {Boffi}, F. and {Botticella}, M.~T. and {Calamida}, A. and {Casertano}, S. and {Chen}, W. -P. and {Chen}, X. and {Cole}, S. and {Deacon}, N. and {Frenk}, C. and {Fitzsimmons}, A. and {Gezari}, S. and {Gibbs}, V. and {Goessl}, C. and {Goggia}, T. and {Gourgue}, R. and {Goldman}, B. and {Grant}, P. and {Grebel}, E.~K. and {Hambly}, N.~C. and {Hasinger}, G. and {Heavens}, A.~F. and {Heckman}, T.~M. and {Henderson}, R. and {Henning}, T. and {Holman}, M. and {Hopp}, U. and {Ip}, W. -H. and {Isani}, S. and {Jackson}, M. and {Keyes}, C.~D. and {Koekemoer}, A.~M. and {Kotak}, R. and {Le}, D. and {Liska}, D. and {Long}, K.~S. and {Lucey}, J.~R. and {Liu}, M. and {Martin}, N.~F. and {Masci}, G. and {McLean}, B. and {Mindel}, E. and {Misra}, P. and {Morganson}, E. and {Murphy}, D.~N.~A. and {Obaika}, A. and {Narayan}, G. and {Nieto-Santisteban}, M.~A. and {Norberg}, P. and {Peacock}, J.~A. and {Pier}, E.~A. and {Postman}, M. and {Primak}, N. and {Rae}, C. and {Rai}, A. and {Riess}, A. and {Riffeser}, A. and {Rix}, H.~W. and {R{\"o}ser}, S. and {Russel}, R. and {Rutz}, L. and {Schilbach}, E. and {Schultz}, A.~S.~B. and {Scolnic}, D. and {Strolger}, L. and {Szalay}, A. and {Seitz}, S. and {Small}, E. and {Smith}, K.~W. and {Soderblom}, D.~R. and {Taylor}, P. and {Thomson}, R. and {Taylor}, A.~N. and {Thakar}, A.~R. and {Thiel}, J. and {Thilker}, D. and {Unger}, D. and {Urata}, Y. and {Valenti}, J. and {Wagner}, J. and {Walder}, T. and {Walter}, F. and {Watters}, S.~P. and {Werner}, S. and {Wood-Vasey}, W.~M. and {Wyse}, R.},
        title = "{The Pan-STARRS1 Surveys}",
      journal = {arXiv e-prints},
         year = 2016,
        month = dec,
          eid = {arXiv:1612.05560},
        pages = {arXiv:1612.05560},
          doi = {10.48550/arXiv.1612.05560},
archivePrefix = {arXiv},
       eprint = {1612.05560},
 primaryClass = {astro-ph.IM},
       adsurl = {https://ui.adsabs.harvard.edu/abs/2016arXiv161205560C}
}

@ARTICLE{cutri2003twomasss_point_catalog,
       author = {{Cutri}, R.~M. and {Skrutskie}, M.~F. and {van Dyk}, S. and {Beichman}, C.~A. and {Carpenter}, J.~M. and {Chester}, T. and {Cambresy}, L. and {Evans}, T. and {Fowler}, J. and {Gizis}, J. and {Howard}, E. and {Huchra}, J. and {Jarrett}, T. and {Kopan}, E.~L. and {Kirkpatrick}, J.~D. and {Light}, R.~M. and {Marsh}, K.~A. and {McCallon}, H. and {Schneider}, S. and {Stiening}, R. and {Sykes}, M. and {Weinberg}, M. and {Wheaton}, W.~A. and {Wheelock}, S. and {Zacarias}, N.},
        title = "{VizieR Online Data Catalog: 2MASS All-Sky Catalog of Point Sources (Cutri+ 2003)}",
      journal = {VizieR Online Data Catalog},
         year = 2003,
        month = jun,
          eid = {II/246},
        pages = {II/246},
       adsurl = {https://ui.adsabs.harvard.edu/abs/2003yCat.2246....0C}
}

@ARTICLE{2006AJ....131.1163S,
       author = {{Skrutskie}, M.~F. and {Cutri}, R.~M. and {Stiening}, R. and {Weinberg}, M.~D. and {Schneider}, S. and {Carpenter}, J.~M. and {Beichman}, C. and {Capps}, R. and {Chester}, T. and {Elias}, J. and {Huchra}, J. and {Liebert}, J. and {Lonsdale}, C. and {Monet}, D.~G. and {Price}, S. and {Seitzer}, P. and {Jarrett}, T. and {Kirkpatrick}, J.~D. and {Gizis}, J.~E. and {Howard}, E. and {Evans}, T. and {Fowler}, J. and {Fullmer}, L. and {Hurt}, R. and {Light}, R. and {Kopan}, E.~L. and {Marsh}, K.~A. and {McCallon}, H.~L. and {Tam}, R. and {Van Dyk}, S. and {Wheelock}, S.},
        title = "{The Two Micron All Sky Survey (2MASS)}",
      journal = {\aj},
         year = 2006,
        month = feb,
       volume = {131},
       number = {2},
        pages = {1163-1183},
          doi = {10.1086/498708},
       adsurl = {https://ui.adsabs.harvard.edu/abs/2006AJ....131.1163S}
}

@misc{2014yCat.2328....0C,
       author = {{Cutri}, R.~M. and {Wright}, E.~L. and {Conrow}, T. and {Fowler}, J.~W. and {Eisenhardt}, P.~R.~M. and {Grillmair}, C. and {Kirkpatrick}, J.~D. and {Masci}, F. and {McCallon}, H.~L. and {Wheelock}, S.~L. and {Fajardo-Acosta}, S. and {Yan}, L. and {Benford}, D. and {Harbut}, M. and {Jarrett}, T. and {Lake}, S. and {Leisawitz}, D. and {Ressler}, M.~E. and {Stanford}, S.~A. and {Tsai}, C. -W. and {Liu}, F. and {Helou}, G. and {Mainzer}, A. and {Gettngs}, D. and {Gonzalez}, A. and {Hoffman}, D. and {Marsh}, K.~A. and {Padgett}, D. and {Skrutskie}, M.~F. and {Beck}, R. and {Papin}, M. and {Wittman}, M.},
        title = "{VizieR Online Data Catalog: AllWISE Data Release (Cutri+ 2013)}",
 howpublished = {VizieR On-line Data Catalog: II/328.  Originally published in: IPAC/Caltech (2013)},
         year = 2021,
        month = feb,
          eid = {II/328},
       adsurl = {https://ui.adsabs.harvard.edu/abs/2014yCat.2328....0C}
}

@ARTICLE{2019MNRAS.489.5301C,
       author = {{Carnero Rosell}, A. and {Santiago}, B. and {dal Ponte}, M. and {Burningham}, B. and {da Costa}, L.~N. and {James}, D.~J. and {Marshall}, J.~L. and {McMahon}, R.~G. and {Bechtol}, K. and {De Paris}, L. and {Li}, T. and {Pieres}, A. and {Garc{\'\i}a-Bellido}, J. and {Abbott}, T.~M.~C. and {Annis}, J. and {Avila}, S. and {Bernstein}, G.~M. and {Brooks}, D. and {Burke}, D.~L. and {Carrasco Kind}, M. and {Carretero}, J. and {De Vicente}, J. and {Drlica-Wagner}, A. and {Fosalba}, P. and {Frieman}, J. and {Gaztanaga}, E. and {Gruendl}, R.~A. and {Gschwend}, J. and {Gutierrez}, G. and {Hollowood}, D.~L. and {Maia}, M.~A.~G. and {Menanteau}, F. and {Miquel}, R. and {Plazas}, A.~A. and {Roodman}, A. and {Sanchez}, E. and {Scarpine}, V. and {Schindler}, R. and {Serrano}, S. and {Sevilla-Noarbe}, I. and {Smith}, M. and {Sobreira}, F. and {Soares-Santos}, M. and {Suchyta}, E. and {Swanson}, M.~E.~C. and {Tarle}, G. and {Vikram}, V. and {Walker}, A.~R. and {DES Collaboration}},
        title = "{Brown dwarf census with the Dark Energy Survey year 3 data and the thin disc scale height of early L types}",
      journal = {\mnras},
         year = 2019,
        month = nov,
       volume = {489},
       number = {4},
        pages = {5301-5325},
          doi = {10.1093/mnras/stz2398},
archivePrefix = {arXiv},
       eprint = {1903.10806},
 primaryClass = {astro-ph.SR},
       adsurl = {https://ui.adsabs.harvard.edu/abs/2019MNRAS.489.5301C}
}

@ARTICLE{2024ApJ...963..129M,
       author = {{Matthee}, Jorryt and {Naidu}, Rohan P. and {Brammer}, Gabriel and {Chisholm}, John and {Eilers}, Anna-Christina and {Goulding}, Andy and {Greene}, Jenny and {Kashino}, Daichi and {Labbe}, Ivo and {Lilly}, Simon J. and {Mackenzie}, Ruari and {Oesch}, Pascal A. and {Weibel}, Andrea and {Wuyts}, Stijn and {Xiao}, Mengyuan and {Bordoloi}, Rongmon and {Bouwens}, Rychard and {van Dokkum}, Pieter and {Illingworth}, Garth and {Kramarenko}, Ivan and {Maseda}, Michael V. and {Mason}, Charlotte and {Meyer}, Romain A. and {Nelson}, Erica J. and {Reddy}, Naveen A. and {Shivaei}, Irene and {Simcoe}, Robert A. and {Yue}, Minghao},
        title = "{Little Red Dots: An Abundant Population of Faint Active Galactic Nuclei at z {\ensuremath{\sim}} 5 Revealed by the EIGER and FRESCO JWST Surveys}",
      journal = {\apj},
         year = 2024,
        month = mar,
       volume = {963},
       number = {2},
          eid = {129},
        pages = {129},
          doi = {10.3847/1538-4357/ad2345},
archivePrefix = {arXiv},
       eprint = {2306.05448},
 primaryClass = {astro-ph.GA},
       adsurl = {https://ui.adsabs.harvard.edu/abs/2024ApJ...963..129M}
}

@ARTICLE{2025arXiv250116648T,
       author = {{Tu}, Zhijun and {Wang}, Shu and {Chen}, Xiaodian and {Liu}, Jifeng},
        title = "{Three Brown Dwarfs Masquerading as High-Redshift Galaxies in JWST Observations}",
      journal = {arXiv e-prints},
         year = 2025,
        month = jan,
          eid = {arXiv:2501.16648},
        pages = {arXiv:2501.16648},
          doi = {10.48550/arXiv.2501.16648},
archivePrefix = {arXiv},
       eprint = {2501.16648},
 primaryClass = {astro-ph.SR},
       adsurl = {https://ui.adsabs.harvard.edu/abs/2025arXiv250116648T}
}

@ARTICLE{2020Phillips,
       author = {{Phillips}, M.~W. and {Tremblin}, P. and {Baraffe}, I. and {Chabrier}, G. and {Allard}, N.~F. and {Spiegelman}, F. and {Goyal}, J.~M. and {Drummond}, B. and {H{\'e}brard}, E.},
        title = "{A new set of atmosphere and evolution models for cool T-Y brown dwarfs and giant exoplanets}",
      journal = {\aap},
         year = 2020,
        month = may,
       volume = {637},
          eid = {A38},
        pages = {A38},
          doi = {10.1051/0004-6361/201937381},
archivePrefix = {arXiv},
       eprint = {2003.13717},
 primaryClass = {astro-ph.SR},
       adsurl = {https://ui.adsabs.harvard.edu/abs/2020A&A...637A..38P}
}

@ARTICLE{2012MNRAS.427..127B,
       author = {{Bressan}, Alessandro and {Marigo}, Paola and {Girardi}, L{\'e}o. and {Salasnich}, Bernardo and {Dal Cero}, Claudia and {Rubele}, Stefano and {Nanni}, Ambra},
        title = "{PARSEC: stellar tracks and isochrones with the PAdova and TRieste Stellar Evolution Code}",
      journal = {\mnras},
         year = 2012,
        month = nov,
       volume = {427},
       number = {1},
        pages = {127-145},
          doi = {10.1111/j.1365-2966.2012.21948.x},
archivePrefix = {arXiv},
       eprint = {1208.4498},
 primaryClass = {astro-ph.SR},
       adsurl = {https://ui.adsabs.harvard.edu/abs/2012MNRAS.427..127B}
}

@ARTICLE{2020MNRAS.498.3283P,
       author = {{Pastorelli}, Giada and {Marigo}, Paola and {Girardi}, L{\'e}o and {Aringer}, Bernhard and {Chen}, Yang and {Rubele}, Stefano and {Trabucchi}, Michele and {Bladh}, Sara and {Boyer}, Martha L. and {Bressan}, Alessandro and {Dalcanton}, Julianne J. and {Groenewegen}, Martin A.~T. and {Lebzelter}, Thomas and {Mowlavi}, Nami and {Chubb}, Katy L. and {Cioni}, Maria-Rosa L. and {de Grijs}, Richard and {Ivanov}, Valentin D. and {Nanni}, Ambra and {van Loon}, Jacco Th and {Zaggia}, Simone},
        title = "{Constraining the thermally pulsing asymptotic giant branch phase with resolved stellar populations in the Large Magellanic Cloud}",
      journal = {\mnras},
         year = 2020,
        month = nov,
       volume = {498},
       number = {3},
        pages = {3283-3301},
          doi = {10.1093/mnras/staa2565},
archivePrefix = {arXiv},
       eprint = {2008.08595},
 primaryClass = {astro-ph.SR},
       adsurl = {https://ui.adsabs.harvard.edu/abs/2020MNRAS.498.3283P}
}

@ARTICLE{1996A&AS..117..393B,
       author = {{Bertin}, E. and {Arnouts}, S.},
        title = "{SExtractor: Software for source extraction.}",
      journal = {\aaps},
         year = 1996,
        month = jun,
       volume = {117},
        pages = {393-404},
          doi = {10.1051/aas:1996164},
       adsurl = {https://ui.adsabs.harvard.edu/abs/1996A&AS..117..393B}
}

@ARTICLE{2025arXiv250216349B,
       author = {{Bouy}, H. and {Mart{\'\i}n}, E.~L. and {Cuillandre}, J. -C. and {Barrado}, D. and {Tamura}, M. and {Bertin}, E. and {{\v{Z}}erjal}, M. and {Points}, S. and {Olivares}, J. and {Hu{\'e}lamo}, N. and {Rodrigues}, T.},
        title = "{Free-floating planetary mass objects in LDN 1495 from Euclid Early Release Observations}",
      journal = {arXiv e-prints},
         year = 2025,
        month = feb,
          eid = {arXiv:2502.16349},
        pages = {arXiv:2502.16349},
          doi = {10.48550/arXiv.2502.16349},
archivePrefix = {arXiv},
       eprint = {2502.16349},
 primaryClass = {astro-ph.SR},
       adsurl = {https://ui.adsabs.harvard.edu/abs/2025arXiv250216349B}
}

@inproceedings{10.1117/12.926785,
author = {Joseph J. Mohr and Robert Armstrong and Emmanuel Bertin and Greg Daues and Shantanu Desai and Michelle Gower and Robert Gruendl and William Hanlon and Nikolay Kuropatkin and Huan Lin and John Marriner and Donald Petravic and Ignacio Sevilla and Molly Swanson and Todd Tomashek and Douglas Tucker and Brian Yanny},
title = {{The Dark Energy Survey data processing and calibration system}},
volume = {8451},
booktitle = {Software and Cyberinfrastructure for Astronomy II},
editor = {Nicole M. Radziwill and Gianluca Chiozzi},
organization = {International Society for Optics and Photonics},
publisher = {SPIE},
pages = {84510D},
year = {2012},
doi = {10.1117/12.926785},
URL = {https://doi.org/10.1117/12.926785}
}

@ARTICLE{2012ApJ...757...83D,
       author = {{Desai}, S. and {Armstrong}, R. and {Mohr}, J.~J. and {Semler}, D.~R. and {Liu}, J. and {Bertin}, E. and {Allam}, S.~S. and {Barkhouse}, W.~A. and {Bazin}, G. and {Buckley-Geer}, E.~J. and {Cooper}, M.~C. and {Hansen}, S.~M. and {High}, F.~W. and {Lin}, H. and {Lin}, Y. -T. and {Ngeow}, C. -C. and {Rest}, A. and {Song}, J. and {Tucker}, D. and {Zenteno}, A.},
        title = "{The Blanco Cosmology Survey: Data Acquisition, Processing, Calibration, Quality Diagnostics, and Data Release}",
      journal = {\apj},
         year = 2012,
        month = sep,
       volume = {757},
       number = {1},
          eid = {83},
        pages = {83},
          doi = {10.1088/0004-637X/757/1/83},
archivePrefix = {arXiv},
       eprint = {1204.1210},
 primaryClass = {astro-ph.CO},
       adsurl = {https://ui.adsabs.harvard.edu/abs/2012ApJ...757...83D}
}

@ARTICLE{2013A&A...554A.101B,
       author = {{Bouy}, H. and {Bertin}, E. and {Moraux}, E. and {Cuillandre}, J. -C. and {Bouvier}, J. and {Barrado}, D. and {Solano}, E. and {Bayo}, A.},
        title = "{Dynamical analysis of nearby clusters. Automated astrometry from the ground: precision proper motions over a wide field}",
      journal = {\aap},
         year = 2013,
        month = jun,
       volume = {554},
          eid = {A101},
        pages = {A101},
          doi = {10.1051/0004-6361/201220748},
archivePrefix = {arXiv},
       eprint = {1306.4446},
 primaryClass = {astro-ph.IM},
       adsurl = {https://ui.adsabs.harvard.edu/abs/2013A&A...554A.101B}
}

@ARTICLE{2018MNRAS.475.2282V,
       author = {{Ventura}, P. and {Karakas}, A. and {Dell'Agli}, F. and {Garc{\'\i}a-Hern{\'a}ndez}, D.~A. and {Guzman-Ramirez}, L.},
        title = "{Gas and dust from solar metallicity AGB stars}",
      journal = {\mnras},
         year = 2018,
        month = apr,
       volume = {475},
       number = {2},
        pages = {2282-2305},
          doi = {10.1093/mnras/stx3338},
archivePrefix = {arXiv},
       eprint = {1712.08582},
 primaryClass = {astro-ph.SR},
       adsurl = {https://ui.adsabs.harvard.edu/abs/2018MNRAS.475.2282V}
}

@ARTICLE{2014A&A...571A..11P,
       author = {{Planck Collaboration} and {Abergel}, A. and {Ade}, P.~A.~R. and {Aghanim}, N. and {Alves}, M.~I.~R. and {Aniano}, G. and {Armitage-Caplan}, C. and {Arnaud}, M. and {Ashdown}, M. and {Atrio-Barandela}, F. and {Aumont}, J. and {Baccigalupi}, C. and {Banday}, A.~J. and {Barreiro}, R.~B. and {Bartlett}, J.~G. and {Battaner}, E. and {Benabed}, K. and {Beno{\^\i}t}, A. and {Benoit-L{\'e}vy}, A. and {Bernard}, J. -P. and {Bersanelli}, M. and {Bielewicz}, P. and {Bobin}, J. and {Bock}, J.~J. and {Bonaldi}, A. and {Bond}, J.~R. and {Borrill}, J. and {Bouchet}, F.~R. and {Boulanger}, F. and {Bridges}, M. and {Bucher}, M. and {Burigana}, C. and {Butler}, R.~C. and {Cardoso}, J. -F. and {Catalano}, A. and {Chamballu}, A. and {Chary}, R. -R. and {Chiang}, H.~C. and {Chiang}, L. -Y. and {Christensen}, P.~R. and {Church}, S. and {Clemens}, M. and {Clements}, D.~L. and {Colombi}, S. and {Colombo}, L.~P.~L. and {Combet}, C. and {Couchot}, F. and {Coulais}, A. and {Crill}, B.~P. and {Curto}, A. and {Cuttaia}, F. and {Danese}, L. and {Davies}, R.~D. and {Davis}, R.~J. and {de Bernardis}, P. and {de Rosa}, A. and {de Zotti}, G. and {Delabrouille}, J. and {Delouis}, J. -M. and {D{\'e}sert}, F. -X. and {Dickinson}, C. and {Diego}, J.~M. and {Dole}, H. and {Donzelli}, S. and {Dor{\'e}}, O. and {Douspis}, M. and {Draine}, B.~T. and {Dupac}, X. and {Efstathiou}, G. and {En{\ss}lin}, T.~A. and {Eriksen}, H.~K. and {Falgarone}, E. and {Finelli}, F. and {Forni}, O. and {Frailis}, M. and {Fraisse}, A.~A. and {Franceschi}, E. and {Galeotta}, S. and {Ganga}, K. and {Ghosh}, T. and {Giard}, M. and {Giardino}, G. and {Giraud-H{\'e}raud}, Y. and {Gonz{\'a}lez-Nuevo}, J. and {G{\'o}rski}, K.~M. and {Gratton}, S. and {Gregorio}, A. and {Grenier}, I.~A. and {Gruppuso}, A. and {Guillet}, V. and {Hansen}, F.~K. and {Hanson}, D. and {Harrison}, D.~L. and {Helou}, G. and {Henrot-Versill{\'e}}, S. and {Hern{\'a}ndez-Monteagudo}, C. and {Herranz}, D. and {Hildebrandt}, S.~R. and {Hivon}, E. and {Hobson}, M. and {Holmes}, W.~A. and {Hornstrup}, A. and {Hovest}, W. and {Huffenberger}, K.~M. and {Jaffe}, A.~H. and {Jaffe}, T.~R. and {Jewell}, J. and {Joncas}, G. and {Jones}, W.~C. and {Juvela}, M. and {Keih{\"a}nen}, E. and {Keskitalo}, R. and {Kisner}, T.~S. and {Knoche}, J. and {Knox}, L. and {Kunz}, M. and {Kurki-Suonio}, H. and {Lagache}, G. and {L{\"a}hteenm{\"a}ki}, A. and {Lamarre}, J. -M. and {Lasenby}, A. and {Laureijs}, R.~J. and {Lawrence}, C.~R. and {Leonardi}, R. and {Le{\'o}n-Tavares}, J. and {Lesgourgues}, J. and {Levrier}, F. and {Liguori}, M. and {Lilje}, P.~B. and {Linden-V{\o}rnle}, M. and {L{\'o}pez-Caniego}, M. and {Lubin}, P.~M. and {Mac{\'\i}as-P{\'e}rez}, J.~F. and {Maffei}, B. and {Maino}, D. and {Mandolesi}, N. and {Maris}, M. and {Marshall}, D.~J. and {Martin}, P.~G. and {Mart{\'\i}nez-Gonz{\'a}lez}, E. and {Masi}, S. and {Massardi}, M. and {Matarrese}, S. and {Matthai}, F. and {Mazzotta}, P. and {McGehee}, P. and {Melchiorri}, A. and {Mendes}, L. and {Mennella}, A. and {Migliaccio}, M. and {Mitra}, S. and {Miville-Desch{\^e}nes}, M. -A. and {Moneti}, A. and {Montier}, L. and {Morgante}, G. and {Mortlock}, D. and {Munshi}, D. and {Murphy}, J.~A. and {Naselsky}, P. and {Nati}, F. and {Natoli}, P. and {Netterfield}, C.~B. and {N{\o}rgaard-Nielsen}, H.~U. and {Noviello}, F. and {Novikov}, D. and {Novikov}, I. and {Osborne}, S. and {Oxborrow}, C.~A. and {Paci}, F. and {Pagano}, L. and {Pajot}, F. and {Paladini}, R. and {Paoletti}, D. and {Pasian}, F. and {Patanchon}, G. and {Perdereau}, O. and {Perotto}, L. and {Perrotta}, F. and {Piacentini}, F. and {Piat}, M. and {Pierpaoli}, E. and {Pietrobon}, D. and {Plaszczynski}, S. and {Pointecouteau}, E. and {Polenta}, G. and {Ponthieu}, N. and {Popa}, L. and {Poutanen}, T. and {Pratt}, G.~W. and {Pr{\'e}zeau}, G. and {Prunet}, S. and {Puget}, J. -L. and {Rachen}, J.~P. and {Reach}, W.~T. and {Rebolo}, R. and {Reinecke}, M. and {Remazeilles}, M. and {Renault}, C. and {Ricciardi}, S. and {Riller}, T.},
        title = "{Planck 2013 results. XI. All-sky model of thermal dust emission}",
      journal = {\aap},
         year = 2014,
        month = nov,
       volume = {571},
          eid = {A11},
        pages = {A11},
          doi = {10.1051/0004-6361/201323195},
archivePrefix = {arXiv},
       eprint = {1312.1300},
 primaryClass = {astro-ph.GA},
       adsurl = {https://ui.adsabs.harvard.edu/abs/2014A&A...571A..11P}
}

@ARTICLE{2021ApJ...920...85M,
       author = {{Marley}, Mark S. and {Saumon}, Didier and {Visscher}, Channon and {Lupu}, Roxana and {Freedman}, Richard and {Morley}, Caroline and {Fortney}, Jonathan J. and {Seay}, Christopher and {Smith}, Adam J.~R.~W. and {Teal}, D.~J. and {Wang}, Ruoyan},
        title = "{The Sonora Brown Dwarf Atmosphere and Evolution Models. I. Model Description and Application to Cloudless Atmospheres in Rainout Chemical Equilibrium}",
      journal = {\apj},
         year = 2021,
        month = oct,
       volume = {920},
       number = {2},
          eid = {85},
        pages = {85},
          doi = {10.3847/1538-4357/ac141d},
archivePrefix = {arXiv},
       eprint = {2107.07434},
 primaryClass = {astro-ph.SR},
       adsurl = {https://ui.adsabs.harvard.edu/abs/2021ApJ...920...85M}
}

@ARTICLE{2024A&A...689A..93R,
       author = {{Rodrigo}, Carlos and {Cruz}, Patricia and {Aguilar}, John F. and {Aller}, Alba and {Solano}, Enrique and {G{\'a}lvez-Ortiz}, Maria Cruz and {Jim{\'e}nez-Esteban}, Francisco and {Mas-Buitrago}, Pedro and {Bayo}, Amelia and {Cort{\'e}s-Contreras}, Miriam and {Murillo-Ojeda}, Raquel and {Bonoli}, Silvia and {Cenarro}, Javier and {Dupke}, Renato and {L{\'o}pez-Sanjuan}, Carlos and {Mar{\'\i}n-Franch}, Antonio and {de Oliveira}, Claudia Mendes and {Moles}, Mariano and {Taylor}, Keith and {Varela}, Jes{\'u}s and {Rami{\'o}}, H{\'e}ctor V{\'a}zquez},
        title = "{Photometric segregation of dwarf and giant FGK stars using the SVO Filter Profile Service and photometric tools}",
      journal = {\aap},
         year = 2024,
        month = sep,
       volume = {689},
          eid = {A93},
        pages = {A93},
          doi = {10.1051/0004-6361/202449998},
archivePrefix = {arXiv},
       eprint = {2406.03310},
 primaryClass = {astro-ph.SR},
       adsurl = {https://ui.adsabs.harvard.edu/abs/2024A&A...689A..93R}
}

@ARTICLE{2000AJ....120..447K,
       author = {{Kirkpatrick}, J. Davy and {Reid}, I. Neill and {Liebert}, James and {Gizis}, John E. and {Burgasser}, Adam J. and {Monet}, David G. and {Dahn}, Conard C. and {Nelson}, Brant and {Williams}, Rik J.},
        title = "{67 Additional L Dwarfs Discovered by the Two Micron All Sky Survey}",
      journal = {\aj},
         year = 2000,
        month = jul,
       volume = {120},
       number = {1},
        pages = {447-472},
          doi = {10.1086/301427},
archivePrefix = {arXiv},
       eprint = {astro-ph/0003317},
 primaryClass = {astro-ph},
       adsurl = {https://ui.adsabs.harvard.edu/abs/2000AJ....120..447K}
}

@ARTICLE{1999ApJ...519..802K,
       author = {{Kirkpatrick}, J. Davy and {Reid}, I. Neill and {Liebert}, James and {Cutri}, Roc M. and {Nelson}, Brant and {Beichman}, Charles A. and {Dahn}, Conard C. and {Monet}, David G. and {Gizis}, John E. and {Skrutskie}, Michael F.},
        title = "{Dwarfs Cooler than ``M``: The Definition of Spectral Type ``L'' Using Discoveries from the 2 Micron All-Sky Survey (2MASS)}",
      journal = {\apj},
         year = 1999,
        month = jul,
       volume = {519},
       number = {2},
        pages = {802-833},
          doi = {10.1086/307414},
       adsurl = {https://ui.adsabs.harvard.edu/abs/1999ApJ...519..802K}
}

@ARTICLE{2017AJ....153..107T,
       author = {{Tie}, S.~S. and {Martini}, P. and {Mudd}, D. and {Ostrovski}, F. and {Reed}, S.~L. and {Lidman}, C. and {Kochanek}, C. and {Davis}, T.~M. and {Sharp}, R. and {Uddin}, S. and {King}, A. and {Wester}, W. and {Tucker}, B.~E. and {Tucker}, D.~L. and {Buckley-Geer}, E. and {Carollo}, D. and {Childress}, M. and {Glazebrook}, K. and {Hinton}, S.~R. and {Lewis}, G. and {Macaulay}, E. and {O'Neill}, C.~R. and {Abbott}, T.~M.~C. and {Abdalla}, F.~B. and {Annis}, J. and {Benoit-L{\'e}vy}, A. and {Bertin}, E. and {Brooks}, D. and {Carnero Rosell}, A. and {Carrasco Kind}, M. and {Carretero}, J. and {Cunha}, C.~E. and {da Costa}, L.~N. and {DePoy}, D.~L. and {Desai}, S. and {Doel}, P. and {Eifler}, T.~F. and {Evrard}, A.~E. and {Finley}, D.~A. and {Flaugher}, B. and {Fosalba}, P. and {Frieman}, J. and {Garc{\'\i}a-Bellido}, J. and {Gaztanaga}, E. and {Gerdes}, D.~W. and {Goldstein}, D.~A. and {Gruen}, D. and {Gruendl}, R.~A. and {Gutierrez}, G. and {Honscheid}, K. and {James}, D.~J. and {Kuehn}, K. and {Kuropatkin}, N. and {Lima}, M. and {Maia}, M.~A.~G. and {Marshall}, J.~L. and {Menanteau}, F. and {Miller}, C.~J. and {Miquel}, R. and {Nichol}, R.~C. and {Nord}, B. and {Ogando}, R. and {Plazas}, A.~A. and {Romer}, A.~K. and {Sanchez}, E. and {Santiago}, B. and {Scarpine}, V. and {Schubnell}, M. and {Sevilla-Noarbe}, I. and {Smith}, R.~C. and {Soares-Santos}, M. and {Sobreira}, F. and {Suchyta}, E. and {Swanson}, M.~E.~C. and {Tarle}, G. and {Thomas}, D. and {Walker}, A.~R. and {DES Collaboration}},
        title = "{A Study of Quasar Selection in the Supernova Fields of the Dark Energy Survey}",
      journal = {\aj},
         year = 2017,
        month = mar,
       volume = {153},
       number = {3},
          eid = {107},
        pages = {107},
          doi = {10.3847/1538-3881/aa5b8d},
archivePrefix = {arXiv},
       eprint = {1611.05456},
 primaryClass = {astro-ph.GA},
       adsurl = {https://ui.adsabs.harvard.edu/abs/2017AJ....153..107T}
}

@misc{2020yCat.1350....0G,
       author = {{Gaia Collaboration}},
        title = "{VizieR Online Data Catalog: Gaia EDR3 (Gaia Collaboration, 2020)}",
 howpublished = {VizieR On-line Data Catalog: I/350.  Originally published in: 2021A\&A...649A...1G},
         year = 2020,
        month = nov,
          eid = {I/350},
          doi = {10.26093/cds/vizier.1350},
       adsurl = {https://ui.adsabs.harvard.edu/abs/2020yCat.1350....0G}
}

@ARTICLE{2016ApJS..224...15X,
       author = {{Xue}, Y.~Q. and {Luo}, B. and {Brandt}, W.~N. and {Alexander}, D.~M. and {Bauer}, F.~E. and {Lehmer}, B.~D. and {Yang}, G.},
        title = "{The 2 Ms Chandra Deep Field-North Survey and the 250 ks Extended Chandra Deep Field-South Survey: Improved Point-source Catalogs}",
      journal = {\apjs},
         year = 2016,
        month = jun,
       volume = {224},
       number = {2},
          eid = {15},
        pages = {15},
          doi = {10.3847/0067-0049/224/2/15},
archivePrefix = {arXiv},
       eprint = {1602.06299},
 primaryClass = {astro-ph.GA},
       adsurl = {https://ui.adsabs.harvard.edu/abs/2016ApJS..224...15X}
}

@ARTICLE{2013ApJS..207...37S,
       author = {{Shim}, Hyunjin and {Im}, Myungshin and {Ko}, Jongwan and {Jeon}, Yiseul and {Karouzos}, Marios and {Kim}, Seong Jin and {Lee}, Hyung Mok and {Papovich}, Casey and {Willmer}, Christopher and {Weiner}, Benjamin J.},
        title = "{Hectospec and Hydra Spectra of Infrared Luminous Sources in the AKARI North Ecliptic Pole Survey Field}",
      journal = {\apjs},
         year = 2013,
        month = aug,
       volume = {207},
       number = {2},
          eid = {37},
        pages = {37},
          doi = {10.1088/0067-0049/207/2/37},
       adsurl = {https://ui.adsabs.harvard.edu/abs/2013ApJS..207...37S}
}

@ARTICLE{2021ApJS..255...20A,
       author = {{Abbott}, T.~M.~C. and {Adam{\'o}w}, M. and {Aguena}, M. and {Allam}, S. and {Amon}, A. and {Annis}, J. and {Avila}, S. and {Bacon}, D. and {Banerji}, M. and {Bechtol}, K. and {Becker}, M.~R. and {Bernstein}, G.~M. and {Bertin}, E. and {Bhargava}, S. and {Bridle}, S.~L. and {Brooks}, D. and {Burke}, D.~L. and {Carnero Rosell}, A. and {Carrasco Kind}, M. and {Carretero}, J. and {Castander}, F.~J. and {Cawthon}, R. and {Chang}, C. and {Choi}, A. and {Conselice}, C. and {Costanzi}, M. and {Crocce}, M. and {da Costa}, L.~N. and {Davis}, T.~M. and {De Vicente}, J. and {DeRose}, J. and {Desai}, S. and {Diehl}, H.~T. and {Dietrich}, J.~P. and {Drlica-Wagner}, A. and {Eckert}, K. and {Elvin-Poole}, J. and {Everett}, S. and {Evrard}, A.~E. and {Ferrero}, I. and {Fert{\'e}}, A. and {Flaugher}, B. and {Fosalba}, P. and {Friedel}, D. and {Frieman}, J. and {Garc{\'\i}a-Bellido}, J. and {Gaztanaga}, E. and {Gelman}, L. and {Gerdes}, D.~W. and {Giannantonio}, T. and {Gill}, M.~S.~S. and {Gruen}, D. and {Gruendl}, R.~A. and {Gschwend}, J. and {Gutierrez}, G. and {Hartley}, W.~G. and {Hinton}, S.~R. and {Hollowood}, D.~L. and {Honscheid}, K. and {Huterer}, D. and {James}, D.~J. and {Jeltema}, T. and {Johnson}, M.~D. and {Kent}, S. and {Kron}, R. and {Kuehn}, K. and {Kuropatkin}, N. and {Lahav}, O. and {Li}, T.~S. and {Lidman}, C. and {Lin}, H. and {MacCrann}, N. and {Maia}, M.~A.~G. and {Manning}, T.~A. and {Maloney}, J.~D. and {March}, M. and {Marshall}, J.~L. and {Martini}, P. and {Melchior}, P. and {Menanteau}, F. and {Miquel}, R. and {Morgan}, R. and {Myles}, J. and {Neilsen}, E. and {Ogando}, R.~L.~C. and {Palmese}, A. and {Paz-Chinch{\'o}n}, F. and {Petravick}, D. and {Pieres}, A. and {Plazas}, A.~A. and {Pond}, C. and {Rodriguez-Monroy}, M. and {Romer}, A.~K. and {Roodman}, A. and {Rykoff}, E.~S. and {Sako}, M. and {Sanchez}, E. and {Santiago}, B. and {Scarpine}, V. and {Serrano}, S. and {Sevilla-Noarbe}, I. and {Smith}, J. Allyn and {Smith}, M. and {Soares-Santos}, M. and {Suchyta}, E. and {Swanson}, M.~E.~C. and {Tarle}, G. and {Thomas}, D. and {To}, C. and {Tremblay}, P.~E. and {Troxel}, M.~A. and {Tucker}, D.~L. and {Turner}, D.~J. and {Varga}, T.~N. and {Walker}, A.~R. and {Wechsler}, R.~H. and {Weller}, J. and {Wester}, W. and {Wilkinson}, R.~D. and {Yanny}, B. and {Zhang}, Y. and {Nikutta}, R. and {Fitzpatrick}, M. and {Jacques}, A. and {Scott}, A. and {Olsen}, K. and {Huang}, L. and {Herrera}, D. and {Juneau}, S. and {Nidever}, D. and {Weaver}, B.~A. and {Adean}, C. and {Correia}, V. and {de Freitas}, M. and {Freitas}, F.~N. and {Singulani}, C. and {Vila-Verde}, G. and {Linea Science Server}},
        title = "{The Dark Energy Survey Data Release 2}",
      journal = {\apjs},
         year = 2021,
        month = aug,
       volume = {255},
       number = {2},
          eid = {20},
        pages = {20},
          doi = {10.3847/1538-4365/ac00b3},
archivePrefix = {arXiv},
       eprint = {2101.05765},
 primaryClass = {astro-ph.IM},
       adsurl = {https://ui.adsabs.harvard.edu/abs/2021ApJS..255...20A}
}

@ARTICLE{2020A&A...634A..50P,
       author = {{Poulain}, M. and {Paolillo}, M. and {De Cicco}, D. and {Brandt}, W.~N. and {Bauer}, F.~E. and {Falocco}, S. and {Vagnetti}, F. and {Grado}, A. and {Ragosta}, F. and {Botticella}, M.~T. and {Cappellaro}, E. and {Pignata}, G. and {Vaccari}, M. and {Schipani}, P. and {Covone}, G. and {Longo}, G. and {Napolitano}, N.~R.},
        title = "{Extending the variability selection of active galactic nuclei in the W-CDF-S and SERVS/SWIRE region}",
      journal = {\aap},
         year = 2020,
        month = feb,
       volume = {634},
          eid = {A50},
        pages = {A50},
          doi = {10.1051/0004-6361/201937108},
archivePrefix = {arXiv},
       eprint = {2001.02560},
 primaryClass = {astro-ph.GA},
       adsurl = {https://ui.adsabs.harvard.edu/abs/2020A&A...634A..50P}
}

@ARTICLE{2015A&A...584A..62C,
       author = {{Cappellaro}, E. and {Botticella}, M.~T. and {Pignata}, G. and {Grado}, A. and {Greggio}, L. and {Limatola}, L. and {Vaccari}, M. and {Baruffolo}, A. and {Benetti}, S. and {Bufano}, F. and {Capaccioli}, M. and {Cascone}, E. and {Covone}, G. and {De Cicco}, D. and {Falocco}, S. and {Della Valle}, M. and {Jarvis}, M. and {Marchetti}, L. and {Napolitano}, N.~R. and {Paolillo}, M. and {Pastorello}, A. and {Radovich}, M. and {Schipani}, P. and {Spiro}, S. and {Tomasella}, L. and {Turatto}, M.},
        title = "{Supernova rates from the SUDARE VST-OmegaCAM search. I. Rates per unit volume}",
      journal = {\aap},
         year = 2015,
        month = dec,
       volume = {584},
          eid = {A62},
        pages = {A62},
          doi = {10.1051/0004-6361/201526712},
archivePrefix = {arXiv},
       eprint = {1509.04496},
 primaryClass = {astro-ph.CO},
       adsurl = {https://ui.adsabs.harvard.edu/abs/2015A&A...584A..62C}
}

@ARTICLE{2005A&A...430...83C,
       author = {{Cappellaro}, E. and {Riello}, M. and {Altavilla}, G. and {Botticella}, M.~T. and {Benetti}, S. and {Clocchiatti}, A. and {Danziger}, J.~I. and {Mazzali}, P. and {Pastorello}, A. and {Patat}, F. and {Salvo}, M. and {Turatto}, M. and {Valenti}, S.},
        title = "{Death rate of massive stars at redshift {\ensuremath{\sim}}0.3}",
      journal = {\aap},
         year = 2005,
        month = jan,
       volume = {430},
        pages = {83-93},
          doi = {10.1051/0004-6361:20041256},
archivePrefix = {arXiv},
       eprint = {astro-ph/0407216},
 primaryClass = {astro-ph},
       adsurl = {https://ui.adsabs.harvard.edu/abs/2005A&A...430...83C}
}

@ARTICLE{2013ApJ...771...97L,
       author = {{Lunnan}, R. and {Chornock}, R. and {Berger}, E. and {Milisavljevic}, D. and {Drout}, M. and {Sanders}, N.~E. and {Challis}, P.~M. and {Czekala}, I. and {Foley}, R.~J. and {Fong}, W. and {Huber}, M.~E. and {Kirshner}, R.~P. and {Leibler}, C. and {Marion}, G.~H. and {McCrum}, M. and {Narayan}, G. and {Rest}, A. and {Roth}, K.~C. and {Scolnic}, D. and {Smartt}, S.~J. and {Smith}, K. and {Soderberg}, A.~M. and {Stubbs}, C.~W. and {Tonry}, J.~L. and {Burgett}, W.~S. and {Chambers}, K.~C. and {Kudritzki}, R. -P. and {Magnier}, E.~A. and {Price}, P.~A.},
        title = "{PS1-10bzj: A Fast, Hydrogen-poor Superluminous Supernova in a Metal-poor Host Galaxy}",
      journal = {\apj},
         year = 2013,
        month = jul,
       volume = {771},
       number = {2},
          eid = {97},
        pages = {97},
          doi = {10.1088/0004-637X/771/2/97},
archivePrefix = {arXiv},
       eprint = {1303.1531},
 primaryClass = {astro-ph.HE},
       adsurl = {https://ui.adsabs.harvard.edu/abs/2013ApJ...771...97L}
}

@ARTICLE{2016ApJ...830...51S,
       author = {{Straatman}, Caroline M.~S. and {Spitler}, Lee R. and {Quadri}, Ryan F. and {Labb{\'e}}, Ivo and {Glazebrook}, Karl and {Persson}, S. Eric and {Papovich}, Casey and {Tran}, Kim-Vy H. and {Brammer}, Gabriel B. and {Cowley}, Michael and {Tomczak}, Adam and {Nanayakkara}, Themiya and {Alcorn}, Leo and {Allen}, Rebecca and {Broussard}, Adam and {van Dokkum}, Pieter and {Forrest}, Ben and {van Houdt}, Josha and {Kacprzak}, Glenn G. and {Kawinwanichakij}, Lalitwadee and {Kelson}, Daniel D. and {Lee}, Janice and {McCarthy}, Patrick J. and {Mehrtens}, Nicola and {Monson}, Andrew and {Murphy}, David and {Rees}, Glen and {Tilvi}, Vithal and {Whitaker}, Katherine E.},
        title = "{The FourStar Galaxy Evolution Survey (ZFOURGE): Ultraviolet to Far-infrared Catalogs, Medium-bandwidth Photometric Redshifts with Improved Accuracy, Stellar Masses, and Confirmation of Quiescent Galaxies to z {\ensuremath{\sim}} 3.5}",
      journal = {\apj},
         year = 2016,
        month = oct,
       volume = {830},
       number = {1},
          eid = {51},
        pages = {51},
          doi = {10.3847/0004-637X/830/1/51},
archivePrefix = {arXiv},
       eprint = {1608.07579},
 primaryClass = {astro-ph.GA},
       adsurl = {https://ui.adsabs.harvard.edu/abs/2016ApJ...830...51S}
}

@ARTICLE{2010A&A...512A..12B,
       author = {{Balestra}, I. and {Mainieri}, V. and {Popesso}, P. and {Dickinson}, M. and {Nonino}, M. and {Rosati}, P. and {Teimoorinia}, H. and {Vanzella}, E. and {Cristiani}, S. and {Cesarsky}, C. and {Fosbury}, R.~A.~E. and {Kuntschner}, H. and {Rettura}, A.},
        title = "{The Great Observatories Origins Deep Survey. VLT/VIMOS spectroscopy in the GOODS-south field: Part II}",
      journal = {\aap},
         year = 2010,
        month = mar,
       volume = {512},
          eid = {A12},
        pages = {A12},
          doi = {10.1051/0004-6361/200913626},
archivePrefix = {arXiv},
       eprint = {1001.1115},
 primaryClass = {astro-ph.CO},
       adsurl = {https://ui.adsabs.harvard.edu/abs/2010A&A...512A..12B}
}

@ARTICLE{2011ApJ...743..146C,
       author = {{Cameron}, E. and {Carollo}, C.~M. and {Oesch}, P.~A. and {Bouwens}, R.~J. and {Illingworth}, G.~D. and {Trenti}, M. and {Labb{\'e}}, I. and {Magee}, D.},
        title = "{Active and Passive Galaxies at z \raisebox{-0.5ex}\textasciitilde 2: Rest-frame Optical Morphologies with WFC3}",
      journal = {\apj},
         year = 2011,
        month = dec,
       volume = {743},
       number = {2},
          eid = {146},
        pages = {146},
          doi = {10.1088/0004-637X/743/2/146},
archivePrefix = {arXiv},
       eprint = {1007.2422},
 primaryClass = {astro-ph.CO},
       adsurl = {https://ui.adsabs.harvard.edu/abs/2011ApJ...743..146C}
}

@ARTICLE{2004A&A...428.1043L,
       author = {{Le F{\`e}vre}, O. and {Vettolani}, G. and {Paltani}, S. and {Tresse}, L. and {Zamorani}, G. and {Le Brun}, V. and {Moreau}, C. and {Bottini}, D. and {Maccagni}, D. and {Picat}, J.~P. and {Scaramella}, R. and {Scodeggio}, M. and {Zanichelli}, A. and {Adami}, C. and {Arnouts}, S. and {Bardelli}, S. and {Bolzonella}, M. and {Cappi}, A. and {Charlot}, S. and {Contini}, T. and {Foucaud}, S. and {Franzetti}, P. and {Garilli}, B. and {Gavignaud}, I. and {Guzzo}, L. and {Ilbert}, O. and {Iovino}, A. and {McCracken}, H.~J. and {Mancini}, D. and {Marano}, B. and {Marinoni}, C. and {Mathez}, G. and {Mazure}, A. and {Meneux}, B. and {Merighi}, R. and {Pell{\`o}}, R. and {Pollo}, A. and {Pozzetti}, L. and {Radovich}, M. and {Zucca}, E. and {Arnaboldi}, M. and {Bondi}, M. and {Bongiorno}, A. and {Busarello}, G. and {Ciliegi}, P. and {Gregorini}, L. and {Mellier}, Y. and {Merluzzi}, P. and {Ripepi}, V. and {Rizzo}, D.},
        title = "{The VIMOS VLT Deep Survey. Public release of 1599 redshifts to I$_{AB}${\ensuremath{\leq}}24 across the Chandra Deep Field South}",
      journal = {\aap},
         year = 2004,
        month = dec,
       volume = {428},
        pages = {1043-1049},
          doi = {10.1051/0004-6361:20048072},
archivePrefix = {arXiv},
       eprint = {astro-ph/0403628},
 primaryClass = {astro-ph},
       adsurl = {https://ui.adsabs.harvard.edu/abs/2004A&A...428.1043L}
}

@ARTICLE{2024ApJS..271...55K,
       author = {{Kirkpatrick}, J. Davy and {Marocco}, Federico and {Gelino}, Christopher R. and {Raghu}, Yadukrishna and {Faherty}, Jacqueline K. and {Bardalez Gagliuffi}, Daniella C. and {Schurr}, Steven D. and {Apps}, Kevin and {Schneider}, Adam C. and {Meisner}, Aaron M. and {Kuchner}, Marc J. and {Caselden}, Dan and {Smart}, R.~L. and {Casewell}, S.~L. and {Raddi}, Roberto and {Kesseli}, Aurora and {Stevnbak Andersen}, Nikolaj and {Antonini}, Edoardo and {Beaulieu}, Paul and {Bickle}, Thomas P. and {Bilsing}, Martin and {Chieng}, Raymond and {Colin}, Guillaume and {Deen}, Sam and {Dereveanco}, Alexandru and {Doll}, Katharina and {Durantini Luca}, Hugo A. and {Frazer}, Anya and {Gantier}, Jean Marc and {Gramaize}, L{\'e}opold and {Grant}, Kristin and {Hamlet}, Leslie K. and {Higashimura}, Hiro and {Hyogo}, Michiharu and {Ja{\l}owiczor}, Peter A. and {Jonkeren}, Alexander and {Kabatnik}, Martin and {Kiwy}, Frank and {Martin}, David W. and {Michaels}, Marianne N. and {Pendrill}, William and {Pessanha Machado}, Celso and {Pumphrey}, Benjamin and {Rothermich}, Austin and {Russwurm}, Rebekah and {Sainio}, Arttu and {Sanchez}, John and {Sapelkin-Tambling}, Fyodor Theo and {Sch{\"u}mann}, J{\"o}rg and {Selg-Mann}, Karl and {Singh}, Harshdeep and {Stenner}, Andres and {Sun}, Guoyou and {Tanner}, Christopher and {Th{\'e}venot}, Melina and {Ventura}, Maurizio and {Voloshin}, Nikita V. and {Walla}, Jim and {W{\k{e}}dracki}, Zbigniew and {Adorno}, Jose I. and {Aganze}, Christian and {Allers}, Katelyn N. and {Brooks}, Hunter and {Burgasser}, Adam J. and {Calamari}, Emily and {Connor}, Thomas and {Costa}, Edgardo and {Eisenhardt}, Peter R. and {Gagn{\'e}}, Jonathan and {Gerasimov}, Roman and {Gonzales}, Eileen C. and {Hsu}, Chih-Chun and {Kiman}, Rocio and {Li}, Guodong and {Low}, Ryan and {Mamajek}, Eric and {Pantoja}, Blake M. and {Popinchalk}, Mark and {Rees}, Jon M. and {Stern}, Daniel and {Su{\'a}rez}, Genaro and {Theissen}, Christopher and {Tsai}, Chao-Wei and {Vos}, Johanna M. and {Zurek}, David and {The Backyard Worlds: Planet 9 Collaboration}},
        title = "{The Initial Mass Function Based on the Full-sky 20 pc Census of {\ensuremath{\sim}}3600 Stars and Brown Dwarfs}",
      journal = {\apjs},
         year = 2024,
        month = apr,
       volume = {271},
       number = {2},
          eid = {55},
        pages = {55},
          doi = {10.3847/1538-4365/ad24e2},
archivePrefix = {arXiv},
       eprint = {2312.03639},
 primaryClass = {astro-ph.SR},
       adsurl = {https://ui.adsabs.harvard.edu/abs/2024ApJS..271...55K}
}

@ARTICLE{2018A&A...619L...8R,
       author = {{Reyl{\'e}}, C.},
        title = "{New ultra-cool and brown dwarf candidates in Gaia DR2}",
      journal = {\aap},
         year = 2018,
        month = nov,
       volume = {619},
          eid = {L8},
        pages = {L8},
          doi = {10.1051/0004-6361/201834082},
archivePrefix = {arXiv},
       eprint = {1809.08244},
 primaryClass = {astro-ph.SR},
       adsurl = {https://ui.adsabs.harvard.edu/abs/2018A&A...619L...8R}
}

@ARTICLE{2014ApJ...786L..18L,
       author = {{Luhman}, K.~L.},
        title = "{Discovery of a \raisebox{-0.5ex}\textasciitilde250 K Brown Dwarf at 2 pc from the Sun}",
      journal = {\apjl},
         year = 2014,
        month = may,
       volume = {786},
       number = {2},
          eid = {L18},
        pages = {L18},
          doi = {10.1088/2041-8205/786/2/L18},
archivePrefix = {arXiv},
       eprint = {1404.6501},
 primaryClass = {astro-ph.GA},
       adsurl = {https://ui.adsabs.harvard.edu/abs/2014ApJ...786L..18L}
}

@ARTICLE{1999AJ....118.2466M,
       author = {{Mart{\'\i}n}, Eduardo L. and {Delfosse}, Xavier and {Basri}, Gibor and {Goldman}, Bertrand and {Forveille}, Thierry and {Zapatero Osorio}, Maria Rosa},
        title = "{Spectroscopic Classification of Late-M and L Field Dwarfs}",
      journal = {\aj},
         year = 1999,
        month = nov,
       volume = {118},
       number = {5},
        pages = {2466-2482},
          doi = {10.1086/301107},
       adsurl = {https://ui.adsabs.harvard.edu/abs/1999AJ....118.2466M}
}

@ARTICLE{2010A&A...517A..53M,
       author = {{Mart{\'\i}n}, E.~L. and {Phan-Bao}, N. and {Bessell}, M. and {Delfosse}, X. and {Forveille}, T. and {Magazz{\`u}}, A. and {Reyl{\'e}}, C. and {Bouy}, H. and {Tata}, R.},
        title = "{Spectroscopic characterization of 78 DENIS ultracool dwarf candidates in the solar neighborhood and the Upper Scorpii OB association}",
      journal = {\aap},
         year = 2010,
        month = jul,
       volume = {517},
          eid = {A53},
        pages = {A53},
          doi = {10.1051/0004-6361/201014202},
archivePrefix = {arXiv},
       eprint = {1004.1775},
 primaryClass = {astro-ph.SR},
       adsurl = {https://ui.adsabs.harvard.edu/abs/2010A&A...517A..53M}
}

@ARTICLE{2013ApJS..205....6M,
       author = {{Mace}, Gregory N. and {Kirkpatrick}, J. Davy and {Cushing}, Michael C. and {Gelino}, Christopher R. and {Griffith}, Roger L. and {Skrutskie}, Michael F. and {Marsh}, Kenneth A. and {Wright}, Edward L. and {Eisenhardt}, Peter R. and {McLean}, Ian S. and {Thompson}, Maggie A. and {Mix}, Katholeen and {Bailey}, Vanessa and {Beichman}, Charles A. and {Bloom}, Joshua S. and {Burgasser}, Adam J. and {Fortney}, Jonathan J. and {Hinz}, Philip M. and {Knox}, Russell P. and {Lowrance}, Patrick J. and {Marley}, Mark S. and {Morley}, Caroline V. and {Rodigas}, Timothy J. and {Saumon}, Didier and {Sheppard}, Scott S. and {Stock}, Nathan D.},
        title = "{A Study of the Diverse T Dwarf Population Revealed by WISE}",
      journal = {\apjs},
         year = 2013,
        month = mar,
       volume = {205},
       number = {1},
          eid = {6},
        pages = {6},
          doi = {10.1088/0067-0049/205/1/6},
archivePrefix = {arXiv},
       eprint = {1301.3913},
 primaryClass = {astro-ph.SR},
       adsurl = {https://ui.adsabs.harvard.edu/abs/2013ApJS..205....6M}
}

@ARTICLE{2022A&A...667A.107G,
       author = {{Garz{\'o}n}, F. and {Balcells}, M. and {Gallego}, J. and {Gry}, C. and {Guzm{\'a}n}, R. and {Hammersley}, P. and {Herrero}, A. and {Mu{\~n}oz-Tu{\~n}{\'o}n}, C. and {Pell{\'o}}, R. and {Prieto}, M. and {Bourrec}, {\'E}. and {Cabello}, C. and {Cardiel}, N. and {Gonz{\'a}lez-Fern{\'a}ndez}, C. and {Laporte}, N. and {Milliard}, B. and {Pascual}, S. and {Patrick}, L.~R. and {Patr{\'o}n}, J. and {Ram{\'\i}rez-Alegr{\'\i}a}, S. and {Streblyanska}, A.},
        title = "{EMIR, the near-infrared camera and multi-object spectrograph for the GTC. EMIR at GTC}",
      journal = {\aap},
         year = 2022,
        month = nov,
       volume = {667},
          eid = {A107},
        pages = {A107},
          doi = {10.1051/0004-6361/202244729},
archivePrefix = {arXiv},
       eprint = {2209.15395},
 primaryClass = {astro-ph.IM},
       adsurl = {https://ui.adsabs.harvard.edu/abs/2022A&A...667A.107G}
}

@ARTICLE{1997Msngr..87...27E,
       author = {{Epchtein}, N. and {de Batz}, B. and {Capoani}, L. and {Chevallier}, L. and {Copet}, E. and {Fouqu{\'e}}, P. and {Lacombe}, P. and {Le Bertre}, T. and {Pau}, S. and {Rouan}, D. and {Ruphy}, S. and {Simon}, G. and {Tiph{\`e}ne}, D. and {Burton}, W.~B. and {Bertin}, E. and {Deul}, E. and {Habing}, H. and {Borsenberger}, J. and {Dennefeld}, M. and {Guglielmo}, F. and {Loup}, C. and {Mamon}, G. and {Ng}, Y. and {Omont}, A. and {Provost}, L. and {Renault}, J. -C. and {Tanguy}, F. and {Kimeswenger}, S. and {Kienel}, C. and {Garzon}, F. and {Persi}, P. and {Ferrari-Toniolo}, M. and {Robin}, A. and {Paturel}, G. and {Vauglin}, I. and {Forveille}, T. and {Delfosse}, X. and {Hron}, J. and {Schultheis}, M. and {Appenzeller}, I. and {Wagner}, S. and {Balazs}, L. and {Holl}, A. and {L{\'e}pine}, J. and {Boscolo}, P. and {Picazzio}, E. and {Duc}, P. -A. and {Mennessier}, M. -O.},
        title = "{The deep near-infrared southern sky survey (DENIS).}",
      journal = {The Messenger},
         year = 1997,
        month = mar,
       volume = {87},
        pages = {27-34},
       adsurl = {https://ui.adsabs.harvard.edu/abs/1997Msngr..87...27E}
}

@ARTICLE{2007MNRAS.379.1599L,
       author = {{Lawrence}, A. and {Warren}, S.~J. and {Almaini}, O. and {Edge}, A.~C. and {Hambly}, N.~C. and {Jameson}, R.~F. and {Lucas}, P. and {Casali}, M. and {Adamson}, A. and {Dye}, S. and {Emerson}, J.~P. and {Foucaud}, S. and {Hewett}, P. and {Hirst}, P. and {Hodgkin}, S.~T. and {Irwin}, M.~J. and {Lodieu}, N. and {McMahon}, R.~G. and {Simpson}, C. and {Smail}, I. and {Mortlock}, D. and {Folger}, M.},
        title = "{The UKIRT Infrared Deep Sky Survey (UKIDSS)}",
      journal = {\mnras},
         year = 2007,
        month = aug,
       volume = {379},
       number = {4},
        pages = {1599-1617},
          doi = {10.1111/j.1365-2966.2007.12040.x},
archivePrefix = {arXiv},
       eprint = {astro-ph/0604426},
 primaryClass = {astro-ph},
       adsurl = {https://ui.adsabs.harvard.edu/abs/2007MNRAS.379.1599L}
}

@ARTICLE{2010AJ....140.1868W,
       author = {{Wright}, Edward L. and {Eisenhardt}, Peter R.~M. and {Mainzer}, Amy K. and {Ressler}, Michael E. and {Cutri}, Roc M. and {Jarrett}, Thomas and {Kirkpatrick}, J. Davy and {Padgett}, Deborah and {McMillan}, Robert S. and {Skrutskie}, Michael and {Stanford}, S.~A. and {Cohen}, Martin and {Walker}, Russell G. and {Mather}, John C. and {Leisawitz}, David and {Gautier}, III, Thomas N. and {McLean}, Ian and {Benford}, Dominic and {Lonsdale}, Carol J. and {Blain}, Andrew and {Mendez}, Bryan and {Irace}, William R. and {Duval}, Valerie and {Liu}, Fengchuan and {Royer}, Don and {Heinrichsen}, Ingolf and {Howard}, Joan and {Shannon}, Mark and {Kendall}, Martha and {Walsh}, Amy L. and {Larsen}, Mark and {Cardon}, Joel G. and {Schick}, Scott and {Schwalm}, Mark and {Abid}, Mohamed and {Fabinsky}, Beth and {Naes}, Larry and {Tsai}, Chao-Wei},
        title = "{The Wide-field Infrared Survey Explorer (WISE): Mission Description and Initial On-orbit Performance}",
      journal = {\aj},
         year = 2010,
        month = dec,
       volume = {140},
       number = {6},
        pages = {1868-1881},
          doi = {10.1088/0004-6256/140/6/1868},
archivePrefix = {arXiv},
       eprint = {1008.0031},
 primaryClass = {astro-ph.IM},
       adsurl = {https://ui.adsabs.harvard.edu/abs/2010AJ....140.1868W}
}

@ARTICLE{2013Msngr.154...35M,
       author = {{McMahon}, R.~G. and {Banerji}, M. and {Gonzalez}, E. and {Koposov}, S.~E. and {Bejar}, V.~J. and {Lodieu}, N. and {Rebolo}, R. and {VHS Collaboration}},
        title = "{First Scientific Results from the VISTA Hemisphere Survey (VHS)}",
      journal = {The Messenger},
         year = 2013,
        month = dec,
       volume = {154},
        pages = {35-37},
       adsurl = {https://ui.adsabs.harvard.edu/abs/2013Msngr.154...35M}
}

@ARTICLE{2023MNRAS.522.1951D,
       author = {{dal Ponte}, M. and {Santiago}, B. and {Carnero Rosell}, A. and {De Paris}, L. and {Pace}, A.~B. and {Bechtol}, K. and {Abbott}, T.~M.~C. and {Aguena}, M. and {Allam}, S. and {Alves}, O. and {Bacon}, D. and {Bertin}, E. and {Bocquet}, S. and {Brooks}, D. and {Burke}, D.~L. and {Carrasco Kind}, M. and {Carretero}, J. and {Conselice}, C. and {Costanzi}, M. and {Desai}, S. and {De Vicente}, J. and {Doel}, P. and {Everett}, S. and {Ferrero}, I. and {Flaugher}, B. and {Frieman}, J. and {Garc{\'\i}a-Bellido}, J. and {Gerdes}, D.~W. and {Gruendl}, R.~A. and {Gruen}, D. and {Gutierrez}, G. and {Hinton}, S.~R. and {Hollowood}, D.~L. and {James}, D.~J. and {Kuehn}, K. and {Kuropatkin}, N. and {Marshall}, J.~L. and {Mena-Fern{\'a}ndez}, J. and {Menanteau}, F. and {Miquel}, R. and {Ogando}, R.~L.~C. and {Palmese}, A. and {Paz-Chinch{\'o}n}, F. and {Pereira}, M.~E.~S. and {Plazas Malag{\'o}n}, A.~A. and {Pieres}, A. and {Raveri}, M. and {Rodriguez-Monroy}, M. and {Sanchez}, E. and {Scarpine}, V. and {Schubnell}, M. and {Sevilla-Noarbe}, I. and {Smith}, M. and {Soares-Santos}, M. and {Suchyta}, E. and {Swanson}, M.~E.~C. and {Tarle}, G. and {Thomas}, D. and {To}, C. and {Weaverdyck}, N. and {DES Collaboration}},
        title = "{Ultracool dwarfs candidates based on 6 yr of the Dark Energy Survey data}",
      journal = {\mnras},
         year = 2023,
        month = jun,
       volume = {522},
       number = {2},
        pages = {1951-1967},
          doi = {10.1093/mnras/stad955},
archivePrefix = {arXiv},
       eprint = {2303.15156},
 primaryClass = {astro-ph.SR},
       adsurl = {https://ui.adsabs.harvard.edu/abs/2023MNRAS.522.1951D}
}

@ARTICLE{2000AJ....120.1579Y,
       author = {{York}, Donald G. and {Adelman}, J. and {Anderson}, Jr., John E. and {Anderson}, Scott F. and {Annis}, James and {Bahcall}, Neta A. and {Bakken}, J.~A. and {Barkhouser}, Robert and {Bastian}, Steven and {Berman}, Eileen and {Boroski}, William N. and {Bracker}, Steve and {Briegel}, Charlie and {Briggs}, John W. and {Brinkmann}, J. and {Brunner}, Robert and {Burles}, Scott and {Carey}, Larry and {Carr}, Michael A. and {Castander}, Francisco J. and {Chen}, Bing and {Colestock}, Patrick L. and {Connolly}, A.~J. and {Crocker}, J.~H. and {Csabai}, Istv{\'a}n and {Czarapata}, Paul C. and {Davis}, John Eric and {Doi}, Mamoru and {Dombeck}, Tom and {Eisenstein}, Daniel and {Ellman}, Nancy and {Elms}, Brian R. and {Evans}, Michael L. and {Fan}, Xiaohui and {Federwitz}, Glenn R. and {Fiscelli}, Larry and {Friedman}, Scott and {Frieman}, Joshua A. and {Fukugita}, Masataka and {Gillespie}, Bruce and {Gunn}, James E. and {Gurbani}, Vijay K. and {de Haas}, Ernst and {Haldeman}, Merle and {Harris}, Frederick H. and {Hayes}, J. and {Heckman}, Timothy M. and {Hennessy}, G.~S. and {Hindsley}, Robert B. and {Holm}, Scott and {Holmgren}, Donald J. and {Huang}, Chi-hao and {Hull}, Charles and {Husby}, Don and {Ichikawa}, Shin-Ichi and {Ichikawa}, Takashi and {Ivezi{\'c}}, {\v{Z}}eljko and {Kent}, Stephen and {Kim}, Rita S.~J. and {Kinney}, E. and {Klaene}, Mark and {Kleinman}, A.~N. and {Kleinman}, S. and {Knapp}, G.~R. and {Korienek}, John and {Kron}, Richard G. and {Kunszt}, Peter Z. and {Lamb}, D.~Q. and {Lee}, B. and {Leger}, R. French and {Limmongkol}, Siriluk and {Lindenmeyer}, Carl and {Long}, Daniel C. and {Loomis}, Craig and {Loveday}, Jon and {Lucinio}, Rich and {Lupton}, Robert H. and {MacKinnon}, Bryan and {Mannery}, Edward J. and {Mantsch}, P.~M. and {Margon}, Bruce and {McGehee}, Peregrine and {McKay}, Timothy A. and {Meiksin}, Avery and {Merelli}, Aronne and {Monet}, David G. and {Munn}, Jeffrey A. and {Narayanan}, Vijay K. and {Nash}, Thomas and {Neilsen}, Eric and {Neswold}, Rich and {Newberg}, Heidi Jo and {Nichol}, R.~C. and {Nicinski}, Tom and {Nonino}, Mario and {Okada}, Norio and {Okamura}, Sadanori and {Ostriker}, Jeremiah P. and {Owen}, Russell and {Pauls}, A. George and {Peoples}, John and {Peterson}, R.~L. and {Petravick}, Donald and {Pier}, Jeffrey R. and {Pope}, Adrian and {Pordes}, Ruth and {Prosapio}, Angela and {Rechenmacher}, Ron and {Quinn}, Thomas R. and {Richards}, Gordon T. and {Richmond}, Michael W. and {Rivetta}, Claudio H. and {Rockosi}, Constance M. and {Ruthmansdorfer}, Kurt and {Sandford}, Dale and {Schlegel}, David J. and {Schneider}, Donald P. and {Sekiguchi}, Maki and {Sergey}, Gary and {Shimasaku}, Kazuhiro and {Siegmund}, Walter A. and {Smee}, Stephen and {Smith}, J. Allyn and {Snedden}, S. and {Stone}, R. and {Stoughton}, Chris and {Strauss}, Michael A. and {Stubbs}, Christopher and {SubbaRao}, Mark and {Szalay}, Alexander S. and {Szapudi}, Istvan and {Szokoly}, Gyula P. and {Thakar}, Anirudda R. and {Tremonti}, Christy and {Tucker}, Douglas L. and {Uomoto}, Alan and {Vanden Berk}, Dan and {Vogeley}, Michael S. and {Waddell}, Patrick and {Wang}, Shu-i. and {Watanabe}, Masaru and {Weinberg}, David H. and {Yanny}, Brian and {Yasuda}, Naoki and {SDSS Collaboration}},
        title = "{The Sloan Digital Sky Survey: Technical Summary}",
      journal = {\aj},
         year = 2000,
        month = sep,
       volume = {120},
       number = {3},
        pages = {1579-1587},
          doi = {10.1086/301513},
archivePrefix = {arXiv},
       eprint = {astro-ph/0006396},
 primaryClass = {astro-ph},
       adsurl = {https://ui.adsabs.harvard.edu/abs/2000AJ....120.1579Y}
}

@misc{best_2024_13993077,
  author       = {Best, William M. J. and
                  Dupuy, Trent J. and
                  Liu, Michael C. and
                  Sanghi, Aniket and
                  Siverd, Robert J. and
                  Zhang, Zhoujian},
  title        = {The UltracoolSheet: Photometry, Astrometry,
                   Spectroscopy, and Multiplicity for 4000+ Ultracool
                   Dwarfs and Imaged Exoplanets
                  },
  month        = dec,
  year         = 2024,
  publisher    = {Zenodo},
  version      = {2.0.1},
  doi          = {10.5281/zenodo.13993077},
  url          = {https://doi.org/10.5281/zenodo.13993077},
}

@ARTICLE{2024arXiv240513496C,
       author = {{Cuillandre}, J. -C. and {Bertin}, E. and {Bolzonella}, M. and {Bouy}, H. and {Gwyn}, S. and {Isani}, S. and {Kluge}, M. and {Lai}, O. and {Lan{\c{c}}on}, A. and {Lang}, D.~A. and {Laureijs}, R. and {Saifollahi}, T. and {Schirmer}, M. and {Stone}, C. and {Abdurro'uf} and {Aghanim}, N. and {Altieri}, B. and {Annibali}, F. and {Atek}, H. and {Awad}, P. and {Baes}, M. and {Ba{\~n}ados}, E. and {Barrado}, D. and {Belladitta}, S. and {Belokurov}, V. and {Boselli}, A. and {Bournaud}, F. and {Bovy}, J. and {Bowler}, R.~A.~A. and {Buenadicha}, G. and {Buitrago}, F. and {Cantiello}, M. and {Carollo}, D. and {Codis}, S. and {Collins}, M.~L.~M. and {Congedo}, G. and {Dalessandro}, E. and {de Lapparent}, V. and {De Paolis}, F. and {Diego}, J.~M. and {Dimauro}, P. and {Dinis}, J. and {Dole}, H. and {Duc}, P. -A. and {Erkal}, D. and {Ezziati}, M. and {Ferguson}, A.~M.~N. and {Ferr{\'e}-Mateu}, A. and {Franco}, A. and {Gavazzi}, R. and {George}, K. and {Gillard}, W. and {Golden-Marx}, J.~B. and {Goldman}, B. and {Gonzalez}, A.~H. and {Habas}, R. and {Hartley}, W.~G. and {Hatch}, N.~A. and {Kohley}, R. and {Hoar}, J. and {Howell}, J.~M. and {Hunt}, L.~K. and {Jablonka}, P. and {Jauzac}, M. and {Kang}, Y. and {Knapen}, J.~H. and {Kneib}, J. -P. and {Kohley}, R. and {Kuzma}, P.~B. and {Larsen}, S.~S. and {Marchal}, O. and {Mart{\'\i}n-Fleitas}, J. and {Marcos-Arenal}, P. and {Marleau}, F.~R. and {Mart{\'\i}n}, E.~L. and {Massari}, D. and {McConnachie}, A.~W. and {Meneghetti}, M. and {Miluzio}, M. and {Miro Carretero}, J. and {Miyatake}, H. and {Mondelin}, M. and {Montes}, M. and {Mora}, A. and {M{\"u}ller}, O. and {Nally}, C. and {Noeske}, K. and {Nucita}, A.~A. and {Oesch}, P.~A. and {Oguri}, M. and {Peletier}, R.~F. and {Poulain}, M. and {Quilley}, L. and {Racca}, G.~D. and {Rejkuba}, M. and {Rhodes}, J. and {Rocci}, P. -F. and {Rom{\'a}n}, J. and {Sacquegna}, S. and {Saremi}, E. and {Scaramella}, R. and {Schinnerer}, E. and {Serjeant}, S. and {Sola}, E. and {Sorce}, J.~G. and {Tarsitano}, F. and {Tereno}, I. and {Toft}, S. and {Tortora}, C. and {Urbano}, M. and {Venhola}, A. and {Voggel}, K. and {Weaver}, J.~R. and {Xu}, X. and {{\v{Z}}erjal}, M. and {Z{\"o}ller}, R. and {Andreon}, S. and {Auricchio}, N. and {Baldi}, M. and {Balestra}, A. and {Bardelli}, S. and {Basset}, A. and {Bender}, R. and {Bodendorf}, C. and {Branchini}, E. and {Brau-Nogue}, S. and {Brescia}, M. and {Brinchmann}, J. and {Camera}, S. and {Capobianco}, V. and {Carbone}, C. and {Carretero}, J. and {Casas}, S. and {Castander}, F.~J. and {Castellano}, M. and {Cavuoti}, S. and {Cimatti}, A. and {Conselice}, C.~J. and {Conversi}, L. and {Copin}, Y. and {Courbin}, F. and {Courtois}, H.~M. and {Cropper}, M. and {Cuby}, J. -G. and {Da Silva}, A. and {Degaudenzi}, H. and {Di Giorgio}, A.~M. and {Douspis}, M. and {Duncan}, C.~A.~J. and {Dupac}, X. and {Dusini}, S. and {Fabricius}, M. and {Farina}, M. and {Farrens}, S. and {Ferriol}, S. and {Fotopoulou}, S. and {Frailis}, M. and {Franceschi}, E. and {Galeotta}, S. and {Garilli}, B. and {Gillis}, B. and {Giocoli}, C. and {G{\'o}mez-Alvarez}, P. and {Grazian}, A. and {Grupp}, F. and {Guzzo}, L. and {Haugan}, S.~V.~H. and {Hoar}, J. and {Hoekstra}, H. and {Holmes}, W. and {Hook}, I. and {Hormuth}, F. and {Hornstrup}, A. and {Hudelot}, P. and {Jahnke}, K. and {Jhabvala}, M. and {Keih{\"a}nen}, E. and {Kermiche}, S. and {Kiessling}, A. and {Kilbinger}, M. and {Kitching}, T. and {Kubik}, B. and {Kuijken}, K. and {K{\"u}mmel}, M. and {Kunz}, M. and {Kurki-Suonio}, H. and {Lahav}, O. and {Ligori}, S. and {Lilje}, P.~B. and {Lindholm}, V. and {Lloro}, I. and {Maino}, D. and {Maiorano}, E. and {Mansutti}, O. and {Marggraf}, O. and {Markovic}, K. and {Martinet}, N. and {Marulli}, F. and {Massey}, R. and {Maurogordato}, S.},
        title = "{Euclid: Early Release Observations -- Programme overview and pipeline for compact- and diffuse-emission photometry}",
      journal = {arXiv e-prints},
         year = 2024,
        month = may,
          eid = {arXiv:2405.13496},
        pages = {arXiv:2405.13496},
          doi = {10.48550/arXiv.2405.13496},
archivePrefix = {arXiv},
       eprint = {2405.13496},
 primaryClass = {astro-ph.IM},
       adsurl = {https://ui.adsabs.harvard.edu/abs/2024arXiv240513496C}
}

@INCOLLECTION{2024sdm..book....1N,
       author = {{Navarro}, Vicente and {del Rio}, Sara and {Angel Diego}, Miguel and {Lopez-Caniego}, Marcos and {Marinic}, Filip and {Kruk}, Sandor and {Reerink}, Jan and {Arviset}, Christophe},
        title = "{ESA Datalabs: Digital Innovation in Space Science}",
    booktitle = {Space Data Management. Studies in Big Data},
         year = 2024,
       volume = {141},
        pages = {1-13},
          doi = {10.1007/978-981-97-0041-7_1},
       adsurl = {https://ui.adsabs.harvard.edu/abs/2024sdm..book....1N}
}

@book{bishop2006pattern,
  title     = {Pattern Recognition and Machine Learning},
  author    = {Christopher M. Bishop},
  year      = {2006},
  publisher = {Springer},
  isbn      = {978-0-387-31073-2}
}

@ARTICLE{2014ApJ...797..120R,
       author = {{Radigan}, Jacqueline},
        title = "{An Independent Analysis of the Brown Dwarf Atmosphere Monitoring (BAM) Data: Large-amplitude Variability is Rare Outside the L/T Transition}",
      journal = {\apj},
         year = 2014,
        month = dec,
       volume = {797},
       number = {2},
          eid = {120},
        pages = {120},
          doi = {10.1088/0004-637X/797/2/120},
archivePrefix = {arXiv},
       eprint = {1408.5919},
 primaryClass = {astro-ph.SR},
       adsurl = {https://ui.adsabs.harvard.edu/abs/2014ApJ...797..120R}
}

@ARTICLE{2014ApJ...793...75R,
       author = {{Radigan}, Jacqueline and {Lafreni{\`e}re}, David and {Jayawardhana}, Ray and {Artigau}, Etienne},
        title = "{Strong Brightness Variations Signal Cloudy-to-clear Transition of Brown Dwarfs}",
      journal = {\apj},
         year = 2014,
        month = oct,
       volume = {793},
       number = {2},
          eid = {75},
        pages = {75},
          doi = {10.1088/0004-637X/793/2/75},
archivePrefix = {arXiv},
       eprint = {1404.3247},
 primaryClass = {astro-ph.SR},
       adsurl = {https://ui.adsabs.harvard.edu/abs/2014ApJ...793...75R}
}

@ARTICLE{2025A&A...697A...7M,
       author = {{Mart{\'\i}n}, E.~L. and {{\v{Z}}erjal}, M. and {Bouy}, H. and {Martin-Gonzalez}, D. and {Mu{\~n}oz Torres}, S. and {Barrado}, D. and {Olivares}, J. and {P{\'e}rez-Garrido}, A. and {Mas-Buitrago}, P. and {Cruz}, P. and {Solano}, E. and {Zapatero Osorio}, M.~R. and {Lodieu}, N. and {B{\'e}jar}, V.~J.~S. and {Zhang}, J. -Y. and {del Burgo}, C. and {Hu{\'e}lamo}, N. and {Laureijs}, R. and {Mora}, A. and {Saifollahi}, T. and {Cuillandre}, J. -C. and {Schirmer}, M. and {Tata}, R. and {Points}, S. and {Phan-Bao}, N. and {Goldman}, B. and {Casewell}, S.~L. and {Reyl{\'e}}, C. and {Smart}, R.~L. and {Dominguez-Tagle}, C. and {Escobar}, A. and {Sedighi}, N. and {Tsilia}, S. and {Vitas}, N. and {Ayadi}, A. and {Aghanim}, N. and {Altieri}, B. and {Andreon}, S. and {Auricchio}, N. and {Baldi}, M. and {Balestra}, A. and {Bardelli}, S. and {Basset}, A. and {Bender}, R. and {Bonino}, D. and {Branchini}, E. and {Brescia}, M. and {Brinchmann}, J. and {Camera}, S. and {Capobianco}, V. and {Carbone}, C. and {Carretero}, J. and {Casas}, S. and {Castellano}, M. and {Cavuoti}, S. and {Cimatti}, A. and {Congedo}, G. and {Conselice}, C.~J. and {Conversi}, L. and {Copin}, Y. and {Corcione}, L. and {Courbin}, F. and {Courtois}, H.~M. and {Cropper}, M. and {Da Silva}, A. and {Degaudenzi}, H. and {Di Giorgio}, A.~M. and {Dinis}, J. and {Dubath}, F. and {Dupac}, X. and {Dusini}, S. and {Ealet}, A. and {Farina}, M. and {Farrens}, S. and {Ferriol}, S. and {Fosalba}, P. and {Frailis}, M. and {Franceschi}, E. and {Fumana}, M. and {Galeotta}, S. and {Garilli}, B. and {Gillard}, W. and {Gillis}, B. and {Giocoli}, C. and {G{\'o}mez-Alvarez}, P. and {Grazian}, A. and {Grupp}, F. and {Guzzo}, L. and {Haugan}, S.~V.~H. and {Hoar}, J. and {Hoekstra}, H. and {Holmes}, W. and {Hook}, I. and {Hormuth}, F. and {Hornstrup}, A. and {Hu}, D. and {Hudelot}, P. and {Jahnke}, K. and {Jhabvala}, M. and {Keih{\"a}nen}, E. and {Kermiche}, S. and {Kiessling}, A. and {Kilbinger}, M. and {Kitching}, T. and {Kohley}, R. and {Kubik}, B. and {K{\"u}mmel}, M. and {Kunz}, M. and {Kurki-Suonio}, H. and {Le Mignant}, D. and {Ligori}, S. and {Lilje}, P.~B. and {Lindholm}, V. and {Lloro}, I. and {Maino}, D. and {Maiorano}, E. and {Mansutti}, O. and {Marggraf}, O. and {Martinet}, N. and {Marulli}, F. and {Massey}, R. and {Medinaceli}, E. and {Mei}, S. and {Melchior}, M. and {Mellier}, Y. and {Meneghetti}, M. and {Meylan}, G. and {Mohr}, J.~J. and {Moresco}, M. and {Moscardini}, L. and {Niemi}, S. -M. and {Padilla}, C. and {Paltani}, S. and {Pasian}, F. and {Pedersen}, K. and {Percival}, W.~J. and {Pettorino}, V. and {Pires}, S. and {Polenta}, G. and {Poncet}, M. and {Popa}, L.~A. and {Pozzetti}, L. and {Racca}, G.~D. and {Raison}, F. and {Rebolo}, R. and {Renzi}, A. and {Rhodes}, J. and {Riccio}, G. and {Rix}, Hans-Walter and {Romelli}, E. and {Roncarelli}, M. and {Rossetti}, E. and {Saglia}, R. and {Sapone}, D. and {Sartoris}, B. and {Sauvage}, M. and {Scaramella}, R. and {Schneider}, P. and {Secroun}, A. and {Seidel}, G. and {Seiffert}, M. and {Serrano}, S. and {Sirignano}, C. and {Sirri}, G. and {Stanco}, L. and {Tallada-Cresp{\'\i}}, P. and {Taylor}, A.~N. and {Teplitz}, H.~I. and {Tereno}, I. and {Toledo-Moreo}, R. and {Tsyganov}, A. and {Tutusaus}, I. and {Valenziano}, L. and {Vassallo}, T. and {Verdoes Kleijn}, G. and {Wang}, Y. and {Weller}, J. and {Williams}, O.~R. and {Zucca}, E. and {Baccigalupi}, C. and {Willis}, G. and {Simon}, P. and {Mart{\'\i}n-Fleitas}, J. and {Scott}, D.},
        title = "{Euclid: Early Release Observations {\textendash} A glance at free-floating newborn planets in the {\ensuremath{\sigma}} Orionis cluster}",
      journal = {\aap},
         year = 2025,
        month = may,
       volume = {697},
          eid = {A7},
        pages = {A7},
          doi = {10.1051/0004-6361/202450793},
archivePrefix = {arXiv},
       eprint = {2405.13497},
 primaryClass = {astro-ph.EP},
       adsurl = {https://ui.adsabs.harvard.edu/abs/2025A&A...697A...7M}
}

@ARTICLE{2018arXiv181106833W,
       author = {{Whitworth}, Anthony},
        title = "{Brown Dwarf Formation: Theory}",
      journal = {arXiv e-prints},
         year = 2018,
        month = nov,
          eid = {arXiv:1811.06833},
        pages = {arXiv:1811.06833},
          doi = {10.48550/arXiv.1811.06833},
archivePrefix = {arXiv},
       eprint = {1811.06833},
 primaryClass = {astro-ph.GA},
       adsurl = {https://ui.adsabs.harvard.edu/abs/2018arXiv181106833W}
}

@ARTICLE{2022NatAs...6...89M,
       author = {{Miret-Roig}, N{\'u}ria and {Bouy}, Herv{\'e} and {Raymond}, Sean N. and {Tamura}, Motohide and {Bertin}, Emmanuel and {Barrado}, David and {Olivares}, Javier and {Galli}, Phillip A.~B. and {Cuillandre}, Jean-Charles and {Sarro}, Luis Manuel and {Berihuete}, Angel and {Hu{\'e}lamo}, Nuria},
        title = "{A rich population of free-floating planets in the Upper Scorpius young stellar association}",
      journal = {Nature Astronomy},
         year = 2022,
        month = feb,
       volume = {6},
        pages = {89-97},
          doi = {10.1038/s41550-021-01513-x},
archivePrefix = {arXiv},
       eprint = {2112.11999},
 primaryClass = {astro-ph.EP},
       adsurl = {https://ui.adsabs.harvard.edu/abs/2022NatAs...6...89M}
}

@ARTICLE{2024NewAR..9901711P,
       author = {{Palau}, Aina and {Hu{\'e}lamo}, Nuria and {Barrado}, David and {Dunham}, Michael M. and {Lee}, Chang Won},
        title = "{Observations of pre- and proto-brown dwarfs in nearby clouds: Paving the way to further constraining theories of brown dwarf formation}",
      journal = {\nar},
         year = 2024,
        month = dec,
       volume = {99},
          eid = {101711},
        pages = {101711},
          doi = {10.1016/j.newar.2024.101711},
archivePrefix = {arXiv},
       eprint = {2410.07769},
 primaryClass = {astro-ph.GA},
       adsurl = {https://ui.adsabs.harvard.edu/abs/2024NewAR..9901711P}
}

%

\begin{appendix}

\section{\label{sec.tableappendix}Tables}


In Table~\ref{table:benchmarks} we present the 60 UCDs from \citet{2024A&A...686A.171Z} that have been spectroscopically confirmed by \citet{carlos25} using \Euclid NISP data. These are used as benchmark objects in this paper. 
In Table~\ref{tab.tdwarfs} we list the seven confirmed T dwarfs and six candidates.
In Table~\ref{table:catalogphot} we list the main result of this paper, namely our catalogue of photometrically selected UCD candidates.


\renewcommand{\arraystretch}{1.3}
\begin{sidewaystable}
\centering
\caption{Benchmarks}
\label{table:benchmarks}
\scalebox{0.91}{
\begin{tabular}{@{\hskip 0mm}r@{\hskip 2mm}r@{\hskip 2mm}r@{\hskip 2mm}r@{\hskip 2mm}r@{\hskip 2mm}r@{\hskip 2mm}r@{\hskip 2mm}r@{\hskip 2mm}r@{\hskip 2mm}r@{\hskip 0mm}r@{\hskip 2mm}r@{\hskip 2mm}r@{\hskip 2mm}r@{\hskip 2mm}r@{\hskip 2mm}l@{\hskip 2mm}l@{\hskip 0mm}}
\hline\hline \noalign{\vskip 1pt}
\multicolumn{1}{c}{\Euclid ID} & \multicolumn{1}{c}{RA} & \multicolumn{1}{c}{Dec} & \multicolumn{4}{c}{\Euclid magnitude} & \multicolumn{1}{c}{$e$} & \multicolumn{1}{c}{$r_\mathrm{K}$} & \multicolumn{1}{c}{$m$} & \multicolumn{1}{c}{Flags} & \multicolumn{4}{c}{S/N} &SpT &\multicolumn{1}{c}{gaia\_id} \\

 & \multicolumn{1}{c}{[deg]} & \multicolumn{1}{c}{[deg]} &\multicolumn{1}{c}{\IE} &\multicolumn{1}{c}{\YE} &\multicolumn{1}{c}{\JE} &\multicolumn{1}{c}{\HE} &  &  &  &  &\IE  &\YE  &\JE  &\HE  &\multicolumn{1}{c}{}   &\multicolumn{1}{c}{DR3} \\
\hline \noalign{\vskip 1pt}

$2694793520643016275$ & 269.47935 & $64.30163$ & $20.322 ^ {+0.004} _ {-0.003}$ & $17.500 ^ {+0.011} _ {-0.009}$ & $21.724 ^ {+1.247} _ {-0.683}$ & $16.912 ^ {+0.014} _ {-0.014}$ & 0.04 & 14.88 & $-2.94$ & 0000 & 287 & 102 & 0 & 72 & M9 & 1439948156047469696 \\
$2690347233671769139$ & 269.03472 & $67.17691$ & $19.841 ^ {+0.002} _ {-0.002}$ & $17.579 ^ {+0.005} _ {-0.005}$ & $17.391 ^ {+0.005} _ {-0.005}$ & $17.305 ^ {+0.005} _ {-0.004}$ & 0.08 & 16.43 & $-3.01$ & 0000 & 450 & 217 & 225 & 251 & M7 & 1633596239873107328 \\
$2662576498671790183$ & 266.25765 & $67.17902$ & $20.112 ^ {+0.003} _ {-0.003}$ & $17.302 ^ {+0.005} _ {-0.004}$ & $16.993 ^ {+0.004} _ {-0.004}$ & $16.892 ^ {+0.003} _ {-0.003}$ & 0.08 & 18.27 & $-2.76$ & 2222 & 350 & 240 & 246 & 338 & M9: & 1634299343198545664 \\
$2718940376640395469$ & 271.89404 & $64.03955$ & $20.236 ^ {+0.003} _ {-0.004}$ & $18.167 ^ {+0.007} _ {-0.007}$ & $17.978 ^ {+0.005} _ {-0.004}$ & $17.941 ^ {+0.004} _ {-0.005}$ & 0.04 & 14.77 & $-2.88$ & 0000 & 316 & 154 & 234 & 236 & M8 & 2161076944616819200 \\
$2712636910639406462$ & 271.26369 & $63.94065$ & $20.353 ^ {+0.009} _ {-0.008}$ & $18.162 ^ {+0.008} _ {-0.009}$ & $18.304 ^ {+0.348} _ {-0.241}$ & $17.836 ^ {+0.005} _ {-0.006}$ & 0.01 & 12.47 & $-3.01$ & 0000 & 125 & 131 & 4 & 191 & M8: & 2161096735825572352 \\

\hline
\end{tabular}
} 
\tablefoot{Columns: $e$, ellipticity; $r_\mathrm{K}$, Kron radius; and $m$, \texttt{MUMAX\_MINUS\_MAG}. The entire table is available only in electronic form at the CDS.}
\end{sidewaystable}

\renewcommand{\arraystretch}{1.3}
\begin{sidewaystable}
\caption{Photometric T dwarf candidates}
\label{tab.tdwarfs}
\centering
\begin{tabular}{@{\hskip 2mm}c@{\hskip 2mm}r@{\hskip 2mm}r@{\hskip 2mm}r@{\hskip 2mm}r@{\hskip 2mm}r@{\hskip 2mm}r@{\hskip 2mm}r@{\hskip 2mm}r@{\hskip 2mm}l@{\hskip 4mm}l}
\hline \hline\noalign{\vskip 1pt}
Obj & \multicolumn{1}{c}{\Euclid ID} & \multicolumn{1}{c}{RA} & \multicolumn{1}{c}{Dec} & \multicolumn{4}{c}{\Euclid magnitude} & C & SpT & Comments \\
 &  & \multicolumn{1}{c}{[deg]} & \multicolumn{1}{c}{[deg]} & \multicolumn{1}{c}{\IE} & \multicolumn{1}{c}{\YE} & \multicolumn{1}{c}{\JE} & \multicolumn{1}{c}{\HE} & & & \\
\hline\noalign{\vskip 1pt}

a & $2748094058670347269$ & 274.80941 & $67.03473$ & $23.319 ^ {+0.029} _ {-0.031}$ & $19.378 ^ {+0.008} _ {-0.009}$ & $18.980 ^ {+0.007} _ {-0.006}$ & $18.693 ^ {+0.006} _ {-0.005}$ & C1 & 0.0 & T0, Spec. confirmed \\
b & $2710066793674540980$ & 271.00668 & $67.45410$ & $24.606 ^ {+0.080} _ {-0.075}$ & $20.322 ^ {+0.016} _ {-0.016}$ & $19.884 ^ {+0.011} _ {-0.011}$ & $19.548 ^ {+0.009} _ {-0.010}$ & C1 & T2 & \dots \\
c & $2657163304658383990$ & 265.71633 & $65.83840$ & $24.402 ^ {+0.074} _ {-0.071}$ & $20.498 ^ {+0.015} _ {-0.015}$ & $20.101 ^ {+0.013} _ {-0.012}$ & $19.817 ^ {+0.007} _ {-0.009}$ & C1 & T4p & Also in Dominguez-Tagle et al. 2025 \\
d & $-571056342502790814$ & 57.10563 & $-50.27908$ & $25.053 ^ {+0.143} _ {-0.137}$ & $20.852 ^ {+0.021} _ {-0.022}$ & $20.487 ^ {+0.017} _ {-0.016}$ & $20.694 ^ {+0.018} _ {-0.016}$ & C2 & T5: & \dots \\
e & $-519971822278279190$ & 51.99718 & $-27.82792$ & $24.843 ^ {+0.139} _ {-0.124}$ & $20.073 ^ {+0.016} _ {-0.017}$ & $19.512 ^ {+0.010} _ {-0.010}$ & $19.906 ^ {+0.013} _ {-0.013}$ & C1 & T6 &\dots \\
f & $2664850113649936423$ & 266.48501 & $64.99364$ & $24.684 ^ {+0.089} _ {-0.086}$ & $20.436 ^ {+0.019} _ {-0.017}$ & $19.967 ^ {+0.015} _ {-0.016}$ & $20.461 ^ {+0.017} _ {-0.016}$ & C1 & T6: & Also in Mace et al. 2013 \\
g & $-597913643476826162$ & 59.79136 & $-47.68262$ & $24.845 ^ {+0.119} _ {-0.104}$ & $20.207 ^ {+0.013} _ {-0.013}$ & $19.764 ^ {+0.007} _ {-0.008}$ & $20.168 ^ {+0.011} _ {-0.010}$ & C1 & T7 & Also in Zhang et al. 2024 \\ 

\dots  & $-642403482473544180$ & 64.24035 & $-47.35442$ & $25.265 ^ {+0.197} _ {-0.178}$ & $20.478 ^ {+0.017} _ {-0.017}$ & $19.965 ^ {+0.012} _ {-0.012}$ & $20.558 ^ {+0.016} _ {-0.018}$ & C2 & \dots &  \dots                                        \\
\dots  & $-627596072497929093$ & 62.75961 & $-49.79291$ & $25.809 ^ {+0.290} _ {-0.239}$ & $21.846 ^ {+0.042} _ {-0.043}$ & $21.468 ^ {+0.028} _ {-0.028}$ & $21.407 ^ {+0.030} _ {-0.028}$ & C3 & \dots &     \dots                                     \\
\dots  & $-603476608509828998$ & 60.34766 & $-50.98290$ & $24.947 ^ {+0.168} _ {-0.136}$ & $20.786 ^ {+0.018} _ {-0.017}$ & $20.991 ^ {+0.018} _ {-0.016}$ & $21.053 ^ {+0.017} _ {-0.019}$ & C4 & \dots & \dots                                         \\
\dots  & $-596778559467349804$ & 59.67786 & $-46.73498$ & $24.598 ^ {+0.105} _ {-0.086}$ & $20.458 ^ {+0.016} _ {-0.014}$ & $20.171 ^ {+0.257} _ {-0.186}$ & $19.877 ^ {+0.011} _ {-0.010}$ & C3 & \dots & \dots                                         \\
\dots  & $-538097404291216597$ & 53.80974 & $-29.12166$ & $25.158 ^ {+0.158} _ {-0.133}$ & $21.296 ^ {+0.024} _ {-0.023}$ & $20.959 ^ {+0.016} _ {-0.015}$ & $20.836 ^ {+0.015} _ {-0.014}$ & C4 & \dots & \dots                                         \\
\dots  & $-509106704279650699$ & 50.91067 & $-27.96507$ & $25.415 ^ {+0.280} _ {-0.216}$ & $21.010 ^ {+0.022} _ {-0.022}$ & $20.937 ^ {+1.297} _ {-0.707}$ & $20.849 ^ {+0.018} _ {-0.018}$ & C3 & \dots & \dots                                         \\

\hline
\end{tabular}
\tablefoot{Three objects are also identified in the literature (\citealp{carlos25}, \citealp{2013ApJS..205....6M}, and \citealp{2024A&A...686A.171Z}), as noted in the comments. Column C lists spectral classes. Spectral types are from \citet{carlos25}.}
\end{sidewaystable}






\renewcommand{\arraystretch}{1.3}
\begin{sidewaystable}
\caption{Photometric ultracool dwarf candidates in the Q1 data release}
\label{table:catalogphot}
\centering
\begin{tabular}{r@{\hskip 2mm}r@{\hskip 2mm}r@{\hskip 2mm}r@{\hskip 2mm}r@{\hskip 2mm}r@{\hskip 2mm}r@{\hskip 2mm}r@{\hskip 2mm}r@{\hskip 2mm}r@{\hskip 2mm}r@{\hskip 2mm}r@{\hskip 2mm}r@{\hskip 2mm}r@{\hskip 2mm}r@{\hskip 2mm}r@{\hskip 2mm}r}
\hline\hline \noalign{\vskip 1pt}
\multicolumn{1}{c}{\Euclid ID} & \multicolumn{1}{c}{RA} & \multicolumn{1}{c}{Dec} & \multicolumn{4}{c}{\Euclid magnitude} & \multicolumn{1}{c}{$e$} &  \multicolumn{4}{c}{S/N} &\multicolumn{1}{c}{flags} & \multicolumn{1}{c}{F$_\mathrm{dq}$} & \multicolumn{1}{c}{C}&\multicolumn{1}{c}{SpT}&\multicolumn{1}{c}{Pec}  \\
 & \multicolumn{1}{c}{[deg]} & \multicolumn{1}{c}{[deg]} &\multicolumn{1}{c}{\IE} &\multicolumn{1}{c}{\YE} &\multicolumn{1}{c}{\JE} &\multicolumn{1}{c}{\HE}& &\IE  &\YE  &\JE  &\HE  &\multicolumn{1}{c}{}   &&&&\\
\hline \noalign{\vskip 1pt}
\hline

$-670049939478214513$ & 67.00499 & $-47.82145$ & $23.266 ^ {+0.037} _ {-0.034}$ & $20.272 ^ {+0.018} _ {-0.017}$ & $19.954 ^ {+0.014} _ {-0.015}$ & $19.856 ^ {+0.013} _ {-0.013}$ & 0.06 & 31 & 66 & 72 & 84 & 0000 & $0$ & C1 & L1: & 0   \\
$-669812280475874160$ & 66.98123 & $-47.58742$ & $24.871 ^ {+0.185} _ {-0.171}$ & $21.925 ^ {+0.134} _ {-0.118}$ & $21.712 ^ {+0.037} _ {-0.038}$ & $21.549 ^ {+0.031} _ {-0.033}$ & 0.14 & 6 & 9 & 30 & 34 & 0000 & $0$ & C3 & \dots & 0   \\
$-669674147474105903$ & 66.96741 & $-47.41059$ & $24.399 ^ {+0.111} _ {-0.097}$ & $21.565 ^ {+0.034} _ {-0.031}$ & $21.232 ^ {+0.024} _ {-0.026}$ & $21.052 ^ {+0.022} _ {-0.020}$ & 0.10 & 11 & 32 & 42 & 53 & 0000 & $0$ & O & \dots & 0   \\
$-669289912474470806$ & 66.92899 & $-47.44708$ & $23.739 ^ {+0.045} _ {-0.046}$ & $20.942 ^ {+0.022} _ {-0.023}$ & $20.595 ^ {+0.014} _ {-0.016}$ & $20.447 ^ {+0.014} _ {-0.013}$ & 0.16 & 24 & 49 & 69 & 83 & 0000 & $0$ & C3 & \dots & 0   \\
$-669256064476668171$ & 66.92561 & $-47.66682$ & $23.587 ^ {+0.038} _ {-0.038}$ & $20.489 ^ {+0.016} _ {-0.015}$ & $19.944 ^ {+0.012} _ {-0.012}$ & $19.557 ^ {+0.009} _ {-0.008}$ & 0.08 & 29 & 69 & 92 & 122 & 0000 & $0$ & C1 & L2: & 0   \\
$-669217677477463960$ & 66.92177 & $-47.74640$ & $25.587 ^ {+0.249} _ {-0.208}$ & $22.805 ^ {+0.087} _ {-0.081}$ & $22.502 ^ {+0.065} _ {-0.062}$ & $22.425 ^ {+0.059} _ {-0.057}$ & 0.13 & 5 & 13 & 18 & 19 & 0000 & $0$ & C2 & \dots & 0   \\
$-669177240477553807$ & 66.91772 & $-47.75538$ & $24.239 ^ {+0.078} _ {-0.066}$ & $21.701 ^ {+0.046} _ {-0.040}$ & $21.489 ^ {+0.033} _ {-0.030}$ & $21.400 ^ {+0.030} _ {-0.026}$ & 0.05 & 15 & 25 & 36 & 37 & 0000 & $0$ & C4 & \dots & 0   \\
$-668967766476426796$ & 66.89678 & $-47.64268$ & $25.559 ^ {+0.256} _ {-0.206}$ & $23.042 ^ {+0.100} _ {-0.105}$ & $22.658 ^ {+0.070} _ {-0.063}$ & $22.500 ^ {+0.055} _ {-0.054}$ & 0.14 & 5 & 11 & 16 & 19 & 0000 & $0$ & C3 & \dots & 0   \\


\hline
\end{tabular}
\tablefoot{
Columns: $e$, ellipticity; F$_\mathrm{dq}$ is a detection quality flag; C is a spectroscopic class; and SpT, spectral type determined by \citet{carlos25}. Pec is assigned a value of 1 for objects showing a significant mismatch between their spectral type and the expected $\YE-\HE$ colour. Flags are also available for each band separately (for saturation, bad pixels, etc.). The entire table is available only in electronic form at the CDS.}
\end{sidewaystable}

\end{appendix}

\label{LastPage}
\end{document}